\documentclass[12pt]{article}

\usepackage{amsfonts, amsmath, amssymb}
\usepackage{cancel,slashed}
\usepackage{bbm}
\usepackage[dvips]{graphicx}
\usepackage{cite}

\newcommand{\bmat}{\left(\begin{array}}
\newcommand{\emat}{\end{array}\right)}

\def\p{\partial}
\def\a{\alpha}

\def\b{\beta}
\def\g{\gamma}
\def\d{\delta}
\def\th{\theta}
\def\om{\omega}

\def\vphi{\varphi}
\def\-{\hphantom{-}}

\def\s2{\frac{1}{\sqrt2}}

\def\oh{\frac{1}{2}}
\def\beq{\begin{equation}}
\def\eeq{\end{equation}}
\def\beqa{\begin{eqnarray}}
\def\eeqa{\end{eqnarray}}

\def\re{{\rm Re \,}}
\def\tr{{\rm tr \,}}

\def\T{{\rm T}}

\def\ca{{\mathcal A}}

\def\Dsl{\,\raise.15ex\hbox{/}\mkern-13.5mu D} 

\def\e{\epsilon}

\def\CK {{\cal K}}

\def\CR {{\cal R}}
\def\CN {{\cal N}}

\def\CL {{\cal L}}
\def\CA {{\cal A}}

\def\CD {{\cal D}}
\def\CO {{\cal O}}
\def\CP {{\cal P}}
\def\CQ {{\cal Q}}
\def\re{\mbox{Re }}

\def\tr{\mbox{Tr}}
\def\str{\mbox{STr}}

\def\be{\begin{equation}}
\def\ee{\end{equation}}
\def\bea{\begin{eqnarray}}
\def\eea{\end{eqnarray}}
\def\raw{\rightarrow}

\newcommand{\ul}{\underline}

\def\IC{\mathbb{C}}
\def\IN{\mathbb{N}}

\def\IR{\mathbb{R}}

\def\Id{{\mathbb{I}}}

\def\T{{\bf T}}

\def\oh{\frac{1}{2}}
\def\a{{\alpha}}
\def\b{{\beta}}
\def\d{{\delta}}
\def\eps{{\epsilon}}
\def\th{{\theta}}

\def\lam{{\lambda}}

\def\om{{\omega}}
\def\sig{{\sigma}}

\def\g{{\gamma}}
\def\G{{\Gamma}}
\def\vphi{{\varphi}}
\def\p{{\partial}}
\def\rd{{\rm d}}

\begin{document}
\pagestyle{plain}

\makeatletter
\@addtoreset{equation}{section}
\makeatother
\renewcommand{\theequation}{\thesection.\arabic{equation}}

\rightline{ IFT-UAM/CSIC-11-20}
\vspace{0.3cm}
\begin{center}
\LARGE{ Flux and  Instanton Effects  in Local  \\
 F-theory  Models and Hierarchical Fermion Masses
\\[6mm]}
\large{L. Aparicio$^{1,2}$, A. Font$^3$, L.E. Ib\'a\~nez$^{1,2}$ and F. Marchesano$^2$ \\[6mm]}
\small{
${}^1$ Departamento de F\'{\i}sica Te\'orica, 
Universidad Aut\'onoma de Madrid, 
28049 Madrid, Spain  \\[2mm] 
${}^2$ Instituto de F\'{\i}sica Te\'orica UAM-CSIC, Cantoblanco, 28049 Madrid, Spain \\[2mm] 
${}^3$  Departamento de F\'{\i}sica, Centro de F\'{\i}sica Te\'orica y Computacional \\[-0.3em]
 Facultad de Ciencias, Universidad Central de Venezuela\\[-0.3em]
 A.P. 20513, Caracas 1020-A, Venezuela
\\[4mm]} 
\small{\bf Abstract} \\[5mm]
\end{center}
\begin{center}
\begin{minipage}[h]{16cm} 
We study the deformation induced by fluxes and
instanton effects on Yukawa couplings
involving 7-brane intersections 
in local F-theory constructions. 
In the absence of non-perturbative effects,
holomorphic Yukawa couplings do not
depend on open string fluxes. 
On the other hand instanton effects 
(or gaugino condensation on distant 7-branes) 
do induce corrections to the Yukawas. 
The leading order effect may also be captured 
by the presence of closed string (1,2) IASD fluxes,
which give rise to a non-commutative structure. 
We check that even in the presence of these
non-perturbative effects the holomorphic Yukawas remain
independent of magnetic  fluxes.
 Although fermion mass hierarchies 
may be obtained from these non-perturbative effects,
they would give identical Yukawa couplings for D-quark and
Lepton masses in $SU(5)$ F-theory GUT's, in contradiction with experiment.
We point out that this problem may be solved by appropriately
normalizing the wavefunctions.  We show in a simple toy model
how  the presence of hypercharge flux may then  be responsible for 
the difference between D-quarks and Lepton masses 
in local $SU(5)$ GUT's.

\end{minipage}
\end{center}
\newpage
\setcounter{page}{1}
\pagestyle{plain}
\renewcommand{\thefootnote}{\arabic{footnote}}
\setcounter{footnote}{0}


\tableofcontents

\section{Introduction}

If string theory underlies the physical world \cite{thebook} 
it should be able to describe
the observed structure of fermion masses and mixings, governed by the 
Yukawa couplings. In  type IIB orientifold compactifications Yukawas at
tree level may in principle be computed as overlap integrals of three different 
wavefunctions over the compact dimensions.  In practice those wavefunctions 
are only known in some simple examples, like toroidal orientifolds. 
For instance, in ${\bf T^6}/{\bf Z_2}\times {\bf Z_2}$ orientifolds  with 
magnetized D7-branes one can construct semirealistic models  \cite{ms04} in which the Yukawas can be explicitly 
computed by integration of three overlapping wavefunctions.
In these examples the wavefunctions are given by certain Jacobi $\vartheta$-functions
and the resulting mass matrices have rank one, corresponding to a single massive
quark/lepton generation \cite{magnus}. This is a good starting point, and one
hopes that further effects like string instanton may yield masses for the lightest generations \cite{ag06}.

More generally we would like to be able to compute Yukawa couplings 
in more complicated curved geometries which may  arguably be needed 
to obtain more realistic models. This may be less difficult than it sounds.
In the context of intersecting D7-brane models bifundamental 
matter fields reside at pairs of D7-branes intersecting in Riemann curves, and 
Yukawa couplings appear locally at those points where three of these curves intersect.
Thus one might hope to be able to compute the corresponding Yukawa
coupling in terms of just local information around the triple intersection  point.

This possibility is particularly attractive in the context of local F-theory 
GUT models \cite{dw1,bhv1,bhv2,dw2}, in which there is a 4-cycle $S$ on which the GUT  
7-branes wrap and the matter fields reside at the so-called matter curves $\Sigma_i \subset S$ 
at which the GUT symmetry is enhanced (see \cite{weigandRev,heckmanRev} for recent reviews). 
In the $SU(5)$ case the symmetry is enhanced to
$SU(6)$ at curves with 5-plets and to $SO(10)$ at curves with 10-plets.
In addition, these curves intersect at points at which the symmetry is further
enhanced, and which give rise to Yukawa couplings among fields in the matter curves.
In particular the  ${\bf 10}\times {\bf {\overline {5}}}\times {\bf {\overline 5}_H}$  Yukawa
appears at points with enhanced $SO(12)$ symmetry and the 
${\bf 10}\times {\bf { {10}}}\times {\bf { 5}_H}$ appears at 
points of enhanced $E_6$ symmetry. Locally one may describe these couplings in terms
of three intersecting curves where the internal wavefunctions of the zero modes are peaked, 
and so compute the corresponding Yukawa couplings in terms of the overlapping integral 
of the three wavefunctions over $S$ \cite{bhv1,bhv2,fi1,hv08,hktw,fi09,cp09,cchv09,hktw2}. 
Since the wavefunctions are localized along the matter curves, the Yukawa coupling
is expected to depend only on the data around the neighborhood
of the intersection point, and not much on the full structure of the compact space.
This is an F-theory realization of the bottom-up idea for the embedding of the 
Standard Model in string theory \cite{aiqu}.

It turns out that, assuming that there is only one intersecting point for each of the two 
types  of $SU(5)$ couplings, the resulting Yukawa matrices have rank equal to one \cite{hv08}.\footnote{For 
different approaches to fermion hierarchies in F-theory and type IIB models see \cite{DudasPalti,Ross,Krippendorf }.}
Hence only one generation gets massive, similarly to 
the toroidal orientifolds mentioned above.
It was first thought that the dependence of the Yukawa couplings on the worldvolume fluxes on the 7-branes (i.e., those 
required for obtaining chirality from these settings), could correct this result and so give masses to the rest of 
quarks and leptons. However it was soon realized that open string fluxes by themselves are not enough, since 
they do not modify the rank of the Yukawa matrices \cite{cchv09},\cite{cp09},\cite{fi09}.  In particular, one can see that the F-term zero mode equations become 
independent of the worldvolume fluxes in a certain {\it holomorphic gauge} \cite{fi09}  and that, as a consequence, the holomorphic Yukawa couplings remain 
flux independent.  

\begin{figure}[ht]
\begin{center}
\begin{tabular}{lcr}
\includegraphics[width=7.5cm]{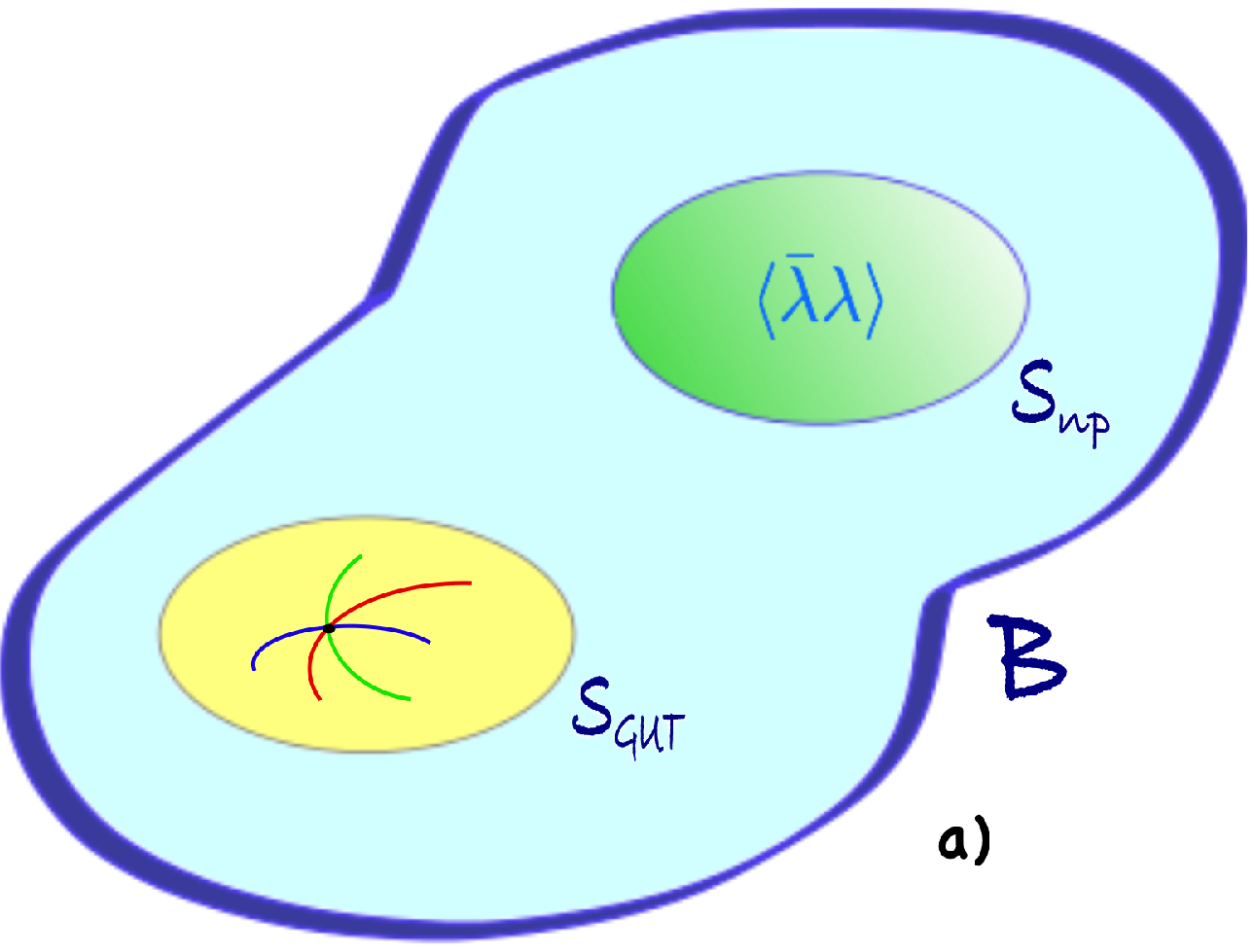}
& \quad &
\includegraphics[width=7.5cm]{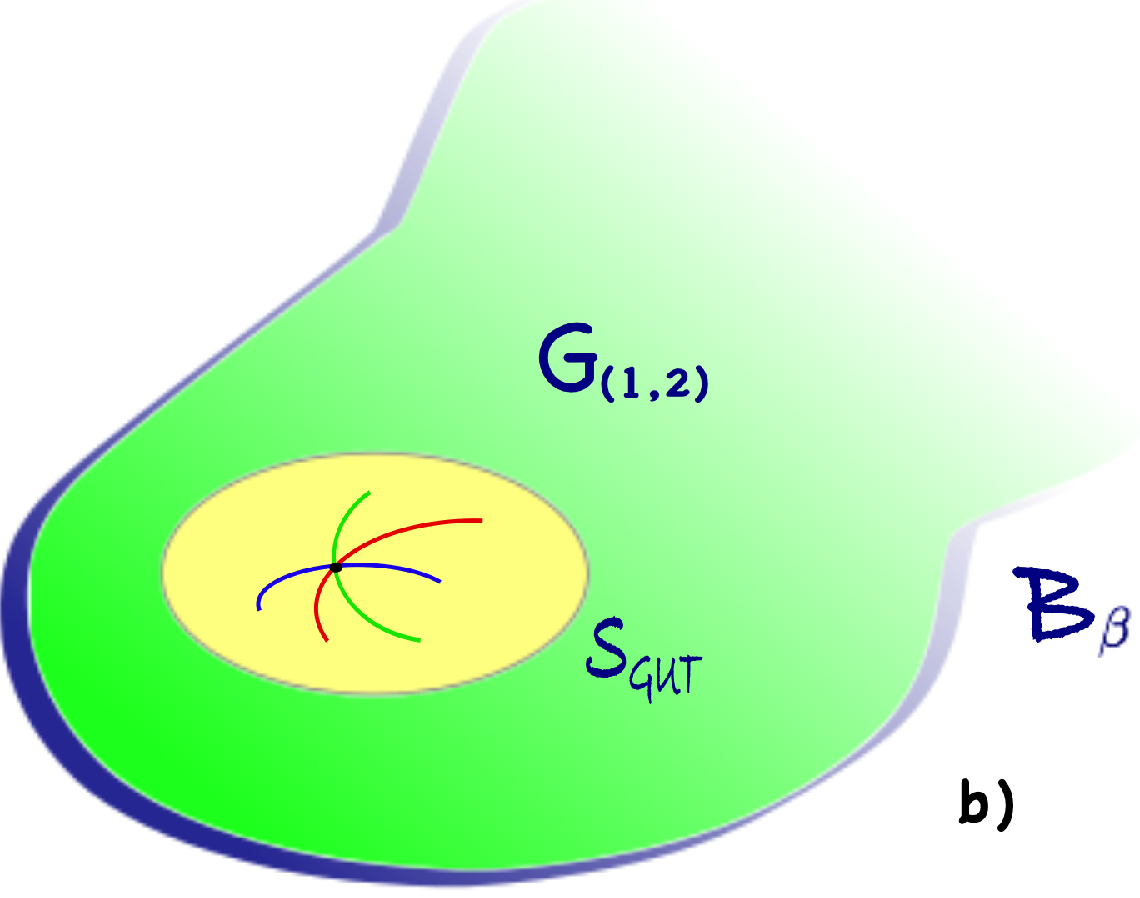}
\end{tabular}
\end{center}
\caption{\small{Sources of corrections to 7-brane Yukawas. Figure $a)$ represents the setup considered in \cite{mm09}, where the Yukawas on a 7-brane stack wrapping the four-cycle $S_{GUT}$ are modified by the gaugino condensate on  7-branes on the distant four-cycle $S_{\rm np}$. Following \cite{km07}, one may identify this setup with the one in figure $b)$, where the non-perturbative sector has been replaced by a $\beta$-deformation of the previous background. This new background contains IASD (1,2) background fluxes that induce a non-commutative deformation on $S_{GUT}$, in the sense of \cite{cchv09}. This $\beta$-deformation is usually not-well defined around $S_{\rm np}$, and so typically the new three-fold $B_\beta$ can only be defined locally.}}
\label{backr}
\end{figure}

Two possible sources of corrections to the holomorphic Yukawas were then put forward.
It was first found that a non-commutative deformation of the 7-brane gauge theory can
induce corrections to the Yukawas such that the rank of the mass matrix is modified \cite{cchv09}. Such 
deformation can be generated by placing D7-branes on type IIB backgrounds with
closed string IASD fluxes of the $(1,2)$-type, often referred to as $\b$-deformed 
backgrounds.\footnote{Such type IIB backgrounds are rather exotic, in the sense that they contain
at least one harmonic 1-form and that D3-branes develop a non-trivial superpotential at tree level. 
As a result, their F-theory lift does not correspond to a Calabi-Yau four-fold compactification. In fact,
to date no compact example of $\beta$-deformed type IIB background has been found.}
The other possible source is the influence of non-perturbative (instanton or gaugino condensate) 
effects on distant 4-cycles in the compact manifold \cite{mm09}, see figure \ref{backr}. 
Although these two proposals look quite 
different they lead to similar physics and it was pointed out that they should be equivalent, the 
reason being that instanton and gaugino condensate effects source IASD $(1,2)$ fluxes on the theory \cite{bdkkm10} (see also \cite{km07}).  
However, a detailed comparison of both kind of effects for the dynamics of local F-theory models is still lacking.

The purpose of this paper is threefold.
 On the one hand we revisit these two different sources of corrections for the Yukawa couplings in local F-theory
  GUT constructions and study their relationship. After a detailed study of the local equations of motion, we show  
  that there is an explicit Seiberg-Witten map which relates the non-commutative and non-perturbative equations
at leading order in the perturbation.

Secondly, we study these  corrections in detail for the simplest 
rank two enhancing example allowing for Yukawa couplings, a $U(3)$ model
 sufficient to capture the main ingredients of the group theoretically
more involved situation in GUT's constructions.
We  obtain  the local wavefunctions to first order in the perturbation and 
compute  the corresponding holomorphic Yukawas for the case of
constant magnetic fluxes and a perturbation linear in the local coordinates.
The result shows an interesting hierarchical structure
but is again independent of the worldvolume fluxes. This is a general
property of the setting, and not an artifact of any particular model.

Finally, we emphasize a problematic phenomenological 
consequence of this persistent flux independence of the holomorphic couplings, 
what we dub  {\it the  $Y(D)=Y(L)$ problem}.  Indeed, 
in a $SU(5)$ GUT the D-quark and lepton Yukawas 
are identical, namely $Y^{ij}(D)=Y^{ij}(L)$ with $i,j$ family labels. Although the related prediction for the 
heaviest generation is consistent with data, those for the lightest generations 
are definitely wrong.  In F-theory models, the hope is that such mass difference
for the lightest generations can be understood in terms of the hypercharge
worldvolume flux, which is the only ingredient breaking the GUT symmetry
down to the MSSM \cite{hv08}. Now, if the Yukawa couplings are flux independent even after 
non-perturbative corrections, that possibility disappears. All is however not lost.
We point out that the wavefunctions used should be normalized
and it is this fact which brings back again the flux dependence for the
physical (not the holomorphic) Yukawas.  
Using this example as a toy model for $SU(5)$
we find that indeed this structure of flux dependence may explain
the hierarchy of masses and mixings of the three quark-lepton
generations.  Wavefunction normalization would thus be the only difference
between D-quark  and lepton masses  at a fundamental level (in addition to the different low-energy
running).

The detailed structure of this paper is as follows. In section \ref{general} we describe the
computation of wavefunctions and Yukawas for local F-theory models in the
absence of any non-perturbative or non-commutative deformation, illustrating the general
computation by means of an explicit $U(3)$ local toy model. In section \ref{npei7} we 
discuss how the previous scheme is modified in the presence of non-perturbative
effects, and describe three different approaches that lead to the same set of corrected
Yukawa couplings. In particular, we consider two different approaches based on the
computation of zero mode wavefunctions for 8d fields living on the GUT 7-brane worldvolume,
one of them related to a commutative 8d gauge theory and the other to a non-commutative
one. The corresponding zero mode equations and wavefunctions look different, but are elegantly 
related by a Seiberg-Witten map, as described in section \ref{corrwf}. Both approaches being 
equivalent, in section \ref{ncyuk} we focus on the non-commutative formalism, in which
the Yukawa independence on the worldvolume fluxes is manifest. There we compute explicitly the 
corrected zero modes and Yukawas for the $U(3)$ model previously introduced, and show that
the Yukawas have a clear hierarchical structure. Finally, in section \ref{norms} we discuss the 
phenomenological implications of this type of Yukawa structure, and more precisely how to solve
the $Y^{ij}(D)=Y^{ij}(L)$ problem. We conclude in section \ref{conclusions} and mention some important 
remarks regarding the generalization of this scheme to concrete F-theory GUT constructions. 

Several technical details and computations have been left for the appendices. In appendix \ref{apD9KK}
we provide a dual description of intersecting D7-brane models in terms of magnetized D9-branes. This 
dual description allows to easily compute not only the spectrum of wavefunctions for the chiral modes of 
the configuration, but also for their massive replicas, as we show for the $U(3)$ model. In appendix
\ref{betawf} we compute the corrected zero mode wavefunctions for the $U(3)$ model in the commutative
formalism, using a slightly different strategy from section \ref{ncyuk}. Finally, appendix \ref{apncmap} 
shows the equivalence of the commutative and non-commutative formalisms at the level of the superpotential.

\section{Wavefunctions and Yukawa couplings in local F-theory models \label{general}}

In this section we describe the standard computation of wavefunctions and Yukawas for local F-theory models in the absence of any non-perturbative or non-commutative deformation, illustrating the general computation by means of an explicit $U(3)$ toy model. While the discussion below is fully carried out  in the context of intersecting and magnetized 7-branes, one may provide a dual description of such system in the more familiar context of magnetized D9-branes, as we show in appendix \ref{apD9KK}.

\subsection{Local F-theory models from intersecting 7-branes}

Following \cite{bhv1,bhv2,dw1,dw2} (see also \cite{collinucci09,bgjw09,mss2,Cordova,mss,gkw09,grimm,dp10}), one may construct a local F-theory model from a set of 7-branes wrapping a compact divisor $S$ of the threefold base $B$ of an elliptically-fibered Calabi-Yau fourfold. The gauge group $G_S$ on such 7-branes is specified by the singularity type of the elliptic fiber on top of the 4-cycle $S$. More precisely, $G_S$ depends on the fiber singularity in the bulk of $S$, as such singularity may be enhanced to a higher type on certain complex submanifolds of $S$. In particular, an enhancement in a curve $\Sigma \subset S$ happens whenever $\Sigma = S \cap S'$ is the intersection locus of $S$ with another divisor $S'$ of $B$ where a different set of 7-branes is wrapped. We can then associate two other gauge groups $G_{S'}$ and $G_{\Sigma}$ to $S'$ and $\Sigma$, respectively, and it is easy to see that $G_S \times G_{S'} \subset G_{\Sigma}$. As in the case of intersecting D7-branes, the intersection locus $\Sigma = S \cap S'$ hosts matter fields charged under the gauge group $G_S \times G_{S'}$. Similarly, an enhancement on a point $p \in S$ happens whenever $p$ is the intersection locus of two or more of these matter curves. Of particular interest for GUT model building are the triple intersections of matter curves, since they give rise to Yukawa couplings between three matter fields charged under $G_S$, which is in turn identified with the GUT gauge group.

The dynamics governing the above construction can be encoded in the 8d effective action found in \cite{bhv1} which, upon dimensional reduction on the 4-cycle $S$, provides the dynamics of the 4d degrees of freedom.\footnote{Alternatively, one may derive such dynamics from a 8d SYM Lagrangian \cite{cp09}.} In particular, the Yukawa couplings between 4d chiral fields arise from the superpotential
\be
W\, =\, m_*^4 \int_{S} \tr \left( F \wedge \Phi \right)
\label{supo7}
\ee
where $m_*^4$ is the F-theory characteristic scale, $F = dA - i A \wedge A$ is the field strength of the gauge vector boson $A$ arising from a stack of 7-branes, and $\Phi$ is a (2,0)-form on the 4-cycle $S$ describing its transverse geometrical deformations. Locally, we can take both $A$ and $\Phi$ to transform in the adjoint of the non-Abelian gauge group $G_p \supset G_S$ associated to the enhanced singularity at the Yukawa point $p$. This initial gauge group is broken by the fact that $\Phi$ and $A$ have a non-trivial profile, and so the actual gauge group is the commutant of $H$ in $G_p$, with $H$ the subgroup generated by $\langle \Phi \rangle$ and $\langle A \rangle$. 

Assuming that $[\langle \Phi \rangle, \langle A \rangle] = 0$, we can write $H = H_{\Phi} \times H_F$, and consider the effect of $\langle \Phi \rangle$ and $\langle A \rangle$ separately. On one hand, the effect of $\langle \Phi \rangle$ is to describe the system of intersecting divisors considered above, so that $G_\Phi = [H_{\Phi}, G_p] = G_S \times \prod_i G_i$, with $G_i$ the gauge groups associated to 7-branes wrapping the divisors $S_i$ intersecting $S$ on $\Sigma_i$. In particular, for a generic point of $S$ the rank of $\langle \Phi \rangle$ is given by ${\rm rank\, } \langle \Phi \rangle = {\rm rank\, } G_p - {\rm rank\, } G_S$, while it decreases to ${\rm rank\, } G_p - {\rm rank\, } G_{\Sigma_i}$  on top of the matter curve $\Sigma_i$ and vanishes at $p$. On the other hand, the effect of $\langle A \rangle$ is to provide a 4d chiral spectrum and to further break the GUT gauge group $G_S$ down to the subgroup $[H_F, G_S]$, as it is usual in  compactifications with magnetized D-branes \cite{magnus,japan,ConlonWF,DiVecchia,cm09}. Hence, one may obtain a 4d MSSM spectrum from the above construction by first engineering the appropriate GUT 4d chiral spectrum via $\langle \Phi \rangle$ and an $\langle A \rangle$ which commutes with $G_S$, and then turn on an extra component of $\langle A \rangle$ along the hypercharge generator in order to break $G_S \raw G_{MSSM}$ \cite{bhv2}.

\subsection{Zero and massive modes at the intersection}\label{zerozero}

While the above scheme provides a general strategy to construct MSSM-like F-theory models, how close a particular construction is to the MSSM crucially depends on the spectrum of 4d zero modes localized at $S$ and of the couplings between them. Again, such information is encoded in the superpotential (\ref{supo7}) which, together with the D-term for $S$,
\be
D\, =\, \int_S \omega \wedge F + \frac{1}{2}  [\Phi, \bar{\Phi}]
\label{FI7}
\ee
(where $\omega$ stands for the fundamental form of $S$) specify the spectrum of 4d zero and massive modes as a set of internal wavefunctions along $S$, and the couplings between these 4d modes in terms of overlapping integrals of such wavefunctions.\footnote{This D-term  receives $\a'$ corrections, and it is also modified by the presence of warping. See \cite{mms10}.}

Indeed, variating $A$ and $\Phi$ in the superpotential (\ref{supo7}), one obtains the F-term equations
\begin{subequations}
\label{Fterm7}
\begin{align}
\label{Fterm7A}
\bar{\p}_A \Phi  & =  0\\
\label{Fterm7phi}
F^{(0,2)}  & =  0
\end{align}
\end{subequations}
where $\bar{\p}_A = (\p_{\bar{x}} - i A_{\bar{x}})\, d\bar{x} + (\p_{\bar{y}} - i A_{\bar{y}})\, d\bar{y}$ is the anti-holomorphic piece of the covariant derivative operator $D_A= \p_A + \bar{\p}_A$ on the 4-cycle $S$, of local complex coordinates $(x,y)$. In addition, from (\ref{FI7}) we obtain the D-term equation 
\be
\omega \wedge F + \frac{1}{2} [\Phi, \bar{\Phi}]\, =\, 0
\label{Dterm7}
\ee
where in this local coordinate system $\omega$ can be described as
\be
\omega = \frac{i}{2}\left( dx\wedge d\bar{x} +  dy\wedge d\bar{y}\right).
\label{kahlerform}
\ee

At the level of the background $\langle A \rangle$ and $\langle \Phi \rangle$, the equations (\ref{Fterm7}) and (\ref{Dterm7}) reduce to the usual supersymmetry conditions for the 7-brane embedding. As $[\langle \Phi \rangle, \langle A \rangle] = 0$, eq.(\ref{Fterm7A}) implies that $\langle \Phi_{xy} \rangle$ is holomorphic on $(x,y)$, consistently with the fact that $S$ and $S_i$ are all holomorphic 4-cycles. If in addition $\langle \Phi_{xy} \rangle$ lives in the Cartan subalgebra of $G_p$ then $[\langle \Phi \rangle, \langle \bar{\Phi} \rangle ] = 0$ and so eqs.(\ref{Fterm7phi}) and (\ref{Dterm7}) imply that $\langle F \rangle$ is a primitive $(1,1)$-form on $S$, similarly to the case of D7-branes in type IIB Calabi-Yau compactifications.

One may, in addition, also obtain the equation of motion for the 7-brane bosonic fluctuations from the above BPS equations. Indeed, by defining
\be
\Phi_{xy}\, =\, \langle \Phi_{xy} \rangle + \varphi_{xy} \quad \quad A_{\bar{m}}\, =\, \langle A_{\bar{m}} \rangle + a_{\bar{m}}
\label{fluctuation}
\ee
and expanding eqs.(\ref{Fterm7}) and (\ref{Dterm7}) to first order in the fluctuations $(\varphi, a_{\bar{x}}, a_{\bar{y}})$ one obtains
\begin{subequations}
\label{fluct7}
\begin{align}
\label{Ffluct7A}
\bar{\p}_{\langle A\rangle} \varphi + i [\langle \Phi\rangle , a]   & =  0\\
\label{Ffluct7phi}
\bar{\p}_{\langle A \rangle} a  & =  0 \\
\label{Dfluct}
\omega \wedge \p_{\langle A \rangle} a - \frac{1}{2} [\langle \bar{\Phi} \rangle, \varphi]  & = 0
\end{align}
\end{subequations}
where $a = a_{\bar{x}} d\bar{x} + a_{\bar{y}} d\bar{y}$ and $\varphi = \varphi_{xy} dx \wedge dy$. These are indeed the zero mode equations of motion for the bosonic fluctuations as obtained from the 8d action derived in \cite{bhv1}, and which pair up with the zero mode fermionic fluctuations in 4d $\CN=1$ chiral multiplets as $(a_{\bar{m}}, \psi_{\bar{m}})$ and $(\varphi_{xy}, \chi_{xy})$. The equation of motion for the latter degrees of freedom can be obtained from the part of the 8d action bilinear in fermions, and read \cite{bhv1,fi09}
\begin{subequations}
\label{ferm7}
\begin{align}
\label{Fferm7A}
\bar{\p}_{A} \chi + i [\Phi , \psi]  -2i\sqrt{2}\, \omega \wedge \p_A \eta  & =  0\\
\label{Fferm7phi}
\bar{\p}_{A} \psi - i\sqrt{2}\, [\bar{\Phi}, \eta] & =  0 \\
\label{Dferm}
\omega \wedge \p_{A} \psi - \frac{1}{2} [\bar{\Phi}, \chi]  & = 0
\end{align}
\end{subequations}
where for simplicity we have replaced $\langle \Phi \rangle \raw \Phi$ and $\langle A \rangle \raw A$, and have included the fermionic degree of freedom within the gauge multiplet $(A_\mu, \eta)$. The latter set of equations can be expressed in matrix notation as
\be
{\bf D_A} \Psi\, =\, 0
\label{Dirac9}
\ee
where 
\be
{\bf D_A}\, =\, 
\left(
\begin{array}{cccc}
0 & D_x & D_y & D_z \\
-D_x & 0 & -D_{\bar{z}} & D_{\bar{y}} \\
-D_y & D_{\bar{z}} & 0 & -D_{\bar{x}} \\
-D_z & -D_{\bar{y}} & D_{\bar{x}} & 0
\end{array}
\right)
\quad \quad
\Psi\, =\, \left(
\begin{array}{c}
\psi_{\bar{0}} \\ \psi_{\bar{x}} \\ \psi_{\bar{y}} \\ \psi_{\bar{z}}
\end{array}
\right) \, \equiv\, 
\left(
\begin{array}{c}
- \sqrt{2}\, \eta \\ \psi_{\bar{x}} \\ \psi_{\bar{y}} \\ \chi_{xy}
\end{array}
\right)\label{matrixDirac}
\ee
where $D_m = \p_m - i [A_m, \cdot]$, $m=x,y,z$ is the covariant derivative. In order to define $D_{\bar{z}}$ we are identifying $A_{\bar{z}} =   \Phi_{xy}$ and imposing that all fields are $z$-independent, so that ${D}_{\bar{z}} = -i [\Phi_{xy}, \cdot]$. As discussed in appendix \ref{apD9KK}, these identifications arise from relating a system of intersecting D7-branes with a system of magnetized D9-branes by T-duality. In such D9-brane picture eq.(\ref{Dirac9}) is nothing but the standard Dirac equation for the fermionic zero modes, ${\bf D_A}$ being the usual Dirac operator.

Interestingly, the latter point allows to write down the eigenmode equation for the 7-brane massive modes in a rather simple way. Indeed, by analogy with the D9-brane picture we have that a fermionic mode of mass $m_\rho$ must satisfy
\be
{\bf D_A}^\dag {\bf D_A}\, \Psi\, =\, |m_\rho|^2 \Psi
\label{eigenferm}
\ee
where ${\bf D_A}^\dag$ is given by (\ref{Ddag}).

\subsection{A $U(3)$ toy model \label{toy}}

In order to illustrate all the above features of F-theory local model building, one may consider a simple toy model made up of three intersecting D7-branes. In particular, let us consider a $U(3)$ gauge theory  on a four-cycle $S$ of local holomorphic coordinates $(x,y)$, and such that the transverse position field $\Phi$ has the vev
\be
\langle \Phi_{xy} \rangle\, =\, 
\frac{m^2_{\Phi_z}}{3}
\left(
\begin{array}{ccc}
1 \\ & 1 \\ & & 1
\end{array}
\right)
{\Phi_0}
+
\frac{m^2_{\Phi_x}}{3}
\left(
\begin{array}{ccc}
-2 \\ & 1 \\ & & 1
\end{array}
\right)
x
+
\frac{m^2_{\Phi_y}}{3}
\left(
\begin{array}{ccc}
1 \\ & 1 \\ & & -2
\end{array}
\right)
y
\label{vevPhi}
\ee
where $m_{\Phi_x}$, $m_{\Phi_y}$ and $m_{\Phi_z}$ are mass scales introduced so that $\Phi_{xy}$ has the usual dimensions of $L^{-1}$. In the following we will assume that 
\be
m_{\Phi_x}^2 = m_{\Phi_y}^2 = m_{\Phi_z}^2 \equiv m_{\Phi}^2\quad \quad  {\rm where} \quad \quad  m_\Phi^2  
= (2\pi)^{3/2} m_*^2
\label{mPhi}
\ee
 leaving the more general case for appendix \ref{apD9KK}. The relation between $m_*$ and $m_{\Phi}$ will be motivated in chapter 3,
 around eq.(\ref{formulilla}). 
 
 From (\ref{vevPhi}) it is easy to see that the initial gauge group is broken as $U(3) \raw U(1)^3$ by the effect of $\langle \Phi \rangle$ alone, and there is then a rank two enhancement $U(1) \raw U(3)$ at the point $p = \{(x,y,z) = (0,0,\Phi_0/3)\}$ where the three D7-branes intersect. 
Such rank two enhancement being generic in the local F-theory GUT setup, one would expect this toy model to capture most of the subtleties involved in computing Yukawa couplings arising from triple intersections of matter curves.

From the geometric viewpoint, the presence of $\langle \Phi \rangle$ indicates that each of the three D7-branes of this model wraps a different four-cycle, algebraically specified by
\begin{subequations}
\label{4cycles}
\begin{align}
S_\a \ :  &\quad 3z + 2x - y - \Phi_0 \, =\, 0\\
S_\b \ :  &\quad 3z - x - y - \Phi_0 \, =\, 0\\
S_\g \ :  &\quad 3z - x + 2y - \Phi_0 \, =\, 0
\end{align}
\end{subequations}
that intersect in the following two-cycles of $S =  \{z = \Phi_0/3\}$
\begin{subequations}
\label{2cycles}
\begin{align}
S_\a \cap S_\b \ :  &\quad \Sigma_a \, =\, \{x=0\} \\
S_\b \cap S_\g \ :  &\quad \Sigma_b \, =\, \{y=0\} \\
S_\g \cap S_\a \ :  &\quad \Sigma_c \, =\, \{x=y\}
\end{align}
\end{subequations}
From the viewpoint of the initial $U(3)$ gauge theory, each of these curves represent a different sector for the fluctuations of a $U(3)$ adjoint field, like the bosonic fields $(\phi, a_{\bar{x}}, a_{\bar{y}})$ or the fermionic fields in the vector $\Psi$ in (\ref{matrixDirac}). In particular, left-handed 4d chiral fermions in the bifundamental will arise from $U(3)$ off-diagonal fluctuations of $\Psi$, that we label as 
\be
{\psi}_{\bar{m}}\,=\, 
\left(
\begin{array}{ccc}
0 & a^+_{\bar{m}} & c^-_{\bar{m}} \\
a^-_{\bar{m}} & 0 & b^+_{\bar{m}} \\
c^+_{\bar{m}} & b^-_{\bar{m}} & 0
\end{array}
\right) \quad \quad \bar{m} = \bar{0}, \bar{x}, \bar{y}, \bar{z}
\label{sectors}
\ee
while their CPT conjugates will be contained in the off-diagonal entries of ${\psi}_m$.

As mentioned above, one extra ingredient necessary to obtain a 4d chiral model is the presence of a non-trivial background worldvolume flux $\langle F \rangle$, usually chosen so that $[\langle \Phi_{xy} \rangle, \langle F \rangle] = 0$. In our $U(3)$ model, a convenient choice is given by 
\be
\langle F\rangle \, =\,  i \left( M_x\, dx \wedge d\bar{x} + M_y\, dy \wedge d\bar{y}\right) 
\frac{1}{3}
\left(
\begin{array}{ccc}
1 \\ & -2 \\ & & 1
\end{array}
\right)
\label{fluxD7}
\ee
so that the background D-term equation (\ref{Dterm7}) is satisfied by imposing  $M_y + M_x = 0$ pointwise. For simplicity, in the following we will assume that $M_x$ and $M_y$ are constant but otherwise arbitrary, see \cite{fi09,cchv09,cp09} for the more general case.

In order to derive the chiral spectrum wavefunctions of this toy model let us consider eq.(\ref{eigenferm}). In general we have that
\be
{\bf D_A}^\dag {\bf D_A}\, =\, - \Delta \Id_4\, - i 
\left(
\begin{array}{cccc}
\sig_{+++} & 0 & 0 & 0\\
0 & \sig_{+--} & F_{y\bar{x}} & F_{z\bar{x}} \\
0 & F_{x\bar{y}} & \sig_{-+-} &  F_{z\bar{y}} \\
0 & F_{x\bar{z}} & F_{y\bar{z}} & \sig_{--+}
\end{array}
\right) 
\label{Lapfer}
\ee
where we have defined  $F_{n\bar{z}} \equiv  D_n \Phi_{xy} $ and\footnote{In our conventions the anticommutator is given by $\{A, B\} \equiv \oh (AB + BA)$.} 
\be
\Delta\, =\, \{D_x, D_{\bar{x}}\} + \{D_y, D_{\bar{y}}\} + \{D_z, D_{\bar{z}}\}
\label{Lapsym}
\ee
\be
\sig_{\eps_x\eps_y\eps_z}\, =\, \oh \left( \eps_x F_{x\bar{x}} + \eps_y F_{y\bar{y}} + \eps_z F_{z\bar{z}}\right)
\label{Lapsig}
\ee
Finally, $F_{n\bar{m}} \sim [F_{n\bar{m}}, \cdot]$ acts in the adjoint on the $U(3)$ gauge indices of $\Psi$, which implies that the worldvolume fluxes $F_{n\bar{m}}$ are felt differently by each matter curve. Indeed, for the $a^{\pm}$ sector in (\ref{sectors}) we have
\be
{\bf D_A}^\dag {\bf D_A}\, =\, -(\Delta_{a^\pm} \pm  M_{xy})\Id_4\, \pm 
\left(
\begin{array}{cccc}
2M_{xy} & 0 & 0 & 0\\
0 & M_x & 0 & -im^2_\Phi\\
0 & 0 & M_y &  0 \\
0 & im_\Phi^2 & 0 & 0
\end{array}
\right)
\label{Lapa}
\ee
where ${M}_{xy} \equiv \oh({M}_x+{M}_y)$. For the $b^\pm$ sector we have instead
\be
{\bf D_A}^\dag {\bf D_A}\, =\, -(\Delta_{b^\pm}\mp M_{xy}) \Id_4\, \pm 
\left(
\begin{array}{cccc}
-2M_{xy} & 0 & 0 & 0\\
0 & -M_x & 0 & 0 \\
0 & 0 & -M_y &  im_\Phi^2 \\
0 & 0 & -im_\Phi^2 & 0
\end{array}
\right)
\label{Lapb}
\ee
and, finally, for the $c^\pm$ sector we have
\be
{\bf D_A}^\dag {\bf D_A}\, =\, -\Delta_{c^\pm} \Id_4\, \pm  m_\Phi^2
\left(
\begin{array}{cccc}
0 & 0 & 0 & 0\\
0 & 0 & 0 & i  \\
0 & 0 & 0 &  -i \\
0 & -i & i & 0
\end{array}
\right)
\label{Lapc}
\ee

Given these expressions and the fact that $M_x$, $M_y$ and $m_\Phi$ are constant it is easy to find the spectrum of eigenvectors of ${\bf D_A}^\dag {\bf D_A}$ in terms of the eigenfunctions $-\Delta \psi_\rho = \rho^2 \psi_\rho$ of the Laplacian. Indeed, in the case of the sector $a^\pm$ we find that the eigenvalues and eigenvectors of the squared Dirac operator are given by
\begin{subequations}
\label{speca}
\begin{align}
\label{speca1}
 & \small{|m_\rho|^2 = \rho^2 \pm  M_{xy}, \,
\Psi =
\left(
\begin{array}{c}
1 \\ 0 \\ 0 \\ 0
\end{array}
\right) \psi_\rho}\ ; \quad
\small{|m_\rho|^2 = \rho^2 \pm  (M_y - M_{xy}), \,
\Psi =
\left(
\begin{array}{c}
0 \\ 0 \\ 1 \\ 0
\end{array}
\right) \psi_\rho} \\
\label{speca2}
& \small{|m_\rho|^2 = \rho^2 \pm  (\lam_a^+ - M_{xy}), \,
\Psi =
\left(
\begin{array}{c}
0\\ \frac{\lam_a^+}{m_\Phi^2}  \\ 0 \\ i
\end{array}
\right) \psi_\rho}\ ; 
\quad 
\small{|m_\rho|^2 = \rho^2 \pm  (\lam_a^- - M_{xy}), \,
\Psi =
\left(
\begin{array}{c}
0\\ \frac{\lam_a^- }{m_\Phi^2} \\ 0 \\ i
\end{array}
\right) \psi_\rho}
\end{align}
\end{subequations}
where
\be
\lam_a^{\pm}\, =\, \frac{M_x}{2} \pm \sqrt{\left(\frac{M_x}{2}\right)^2 + m_\Phi^4}
\label{eigenlam}
\ee
The precise expression for $\psi_\rho$ does in principle depend on which sector we consider, as the Laplacian (\ref{Lapsym}) depends non-trivially on the gauge potential $A$, which acts differently on $a^\pm$. By taking a real gauge 
\be
\langle A\rangle^{\rm real}\, =\, \frac{i}{2} \left[ M_x \left(xd\bar{x} -\bar{x}dx\right) + M_y \left(y d\bar{y} - \bar{y}dy\right) \right]
\frac{1}{3}
\left(
\begin{array}{ccc}
1 \\ & -2 \\ & & 1
\end{array}
\right)
\ee
is easy to see that the modes satisfying the zero mode equation (\ref{Dirac9}) for the sectors $a^\pm$ are given by
\be
\small{\Psi_{0, a^\pm}^{\rm real}\, =\, 
\left(
\begin{array}{c}
0\\ \frac{\lam_a^\mp}{m_\Phi^2} \\ 0 \\ i
\end{array}
\right) e^{- \sqrt{\left( \frac{M_x}{2}\right)^2 + m_\Phi^4}\,  |x|^2} e^{\mp \frac{M_y}{2}   |y|^2} f_{a^\pm}(y)}
\label{zmreala}
\ee
with $f_{a^{\pm}}$ an arbitrary holomorphic function on the intersection coordinate $y$. It is then easy to see that the zero modes on the $a^{\pm}$ sector will only converge for $\pm M_y > 0$, reproducing the usual behavior of magnetized D-brane systems \cite{magnus}. 

For our purposes, however, it is more convenient to express the zero mode wavefunctions in the holomorphic gauge introduced in \cite{fi09}, in which only the holomorphic components of $\langle A \rangle$ are non-vanishing. In the model at hand, such gauge reads
\be
\langle A\rangle^{\rm hol} \, =\, \left( \bar{A_x}dx + \bar{A_y}dy \right)
\frac{1}{3}
\left(
\begin{array}{ccc}
1 \\ & -2 \\ & & 1
\end{array}
\right) \quad\quad
\begin{array}{c}
\bar{A_x}\, =\, -i M_x \bar{x} \\
\bar{A_y}\, =\, -i M_y \bar{y}
\end{array}
\label{holg}
\ee
and it is easy to see that the corresponding zero mode wavefunctions are given by
\be
\small{\Psi_{0, a^\pm}^{\rm hol}\, =\, 
\left(
\begin{array}{c}
0\\ -i\frac{\lam_a^\mp}{m_\Phi^2} \\ 0 \\ 1
\end{array}
\right) \psi_{0, a^\pm}, \quad \quad \psi_{0, a^\pm}\, =\, e^{\pm \lam_a^{\mp}  |x|^2}  f_{a^\pm}(y)}
\label{zmhola}
\ee
In fact, from the Laplace eigenfunction $\psi_{0, a^\pm}$ one may easily construct all the other eigenfunctions of the Laplace operator $\Delta_{a^\pm}$, and so the full spectrum of massive modes in this sector. Indeed, following the discussion in appendix \ref{apD9KK} we have that all the eigenfunctions of $\Delta_{a^\pm}$ are of the form
\be
\psi_{mnl, a^\pm}\, =\, (\CD_x)^m (\CD_y)^n (\CD_{\bar{z}})^l\, \psi_{0, a^\pm}
\label{KKtower}
\ee
with the operators $\CD_m$ defined in (\ref{rotatedsa:ap}).

A similar discussion can be carried out for the sectors $b^{\pm}$ and $c^{\pm}$. 
Leaving the details for appendix \ref{apD9KK}, we obtain that the zero mode wavefunctions in the holomorphic gauge for these sectors are 
\be
\small{\Psi_{0, b^\pm}^{\rm hol}\, =\, 
\left(
\begin{array}{c}
0\\ 0 \\ i\frac{\lam_b^\mp}{m_\Phi^2} \\ 1
\end{array}
\right) \psi_{0, b^\pm}, \quad \quad \psi_{0, b^\pm}\, =\, e^{\pm  \lam_b^{\mp}  |y|^2}  f_{b^\pm}(x)}
\label{zmholb}
\ee
and
\be
\small{\Psi_{0, c^\pm}^{\rm hol}\, =\, 
\left(
\begin{array}{c}
0\\ i\frac{\lam_c}{m_\Phi^2} \\ -i\frac{\lam_c}{m_\Phi^2} \\ 1
\end{array}
\right) \psi_{0, c^\pm}, \quad \quad \psi_{0, c^\pm}\, =\, \g_c m_* e^{  \lam_c  |x-y|^2} }
\label{zmholc}
\ee
where
\be
\lam_b^\pm\, =\, - \frac{{M}_{y}}{2} \pm \sqrt{\left(\frac{{M}_{y}}{2}\right)^2 + m_\Phi^{4}} \quad ;\quad 
\lam_c\, =\, - \frac{m_\Phi^2}{\sqrt2} 
\label{simlams}
\ee
and with similar expressions to (\ref{KKtower}) for their massive replicas. In the following we will consider our zero and massive mode wavefunctions in the holomorphic gauge, avoiding any superscript that indicates so. We will, in addition, assume that $M_x < 0 < M_y$, so that the sectors of interest for computing zero mode Yukawa couplings are 
$a^+$, $b^+$ and $c^+$. Finally, in (\ref{zmholc}) we have introduced a normalization factor to be fixed later.

\subsection{Yukawa couplings \label{yuky}}

The superpotential (\ref{supo7}) gives rise to Yukawa couplings among the 4-dimensional charged fields since it includes the trilinear term
\be
W_{\rm Yuk}\, =\, - im_*^4 \int_{S} \tr \left( A \wedge A \wedge \Phi \right)
\label{supy}
\ee
that induces Yukawa couplings between the zero (and massive) modes of $A$ and $\Phi$. In particular, in the above setup where charged massless matter resides at curves where 7-branes intersect, the Yukawa couplings $Y_{abc}^{ijk}$ are generated at the intersection of three matter curves $\Sigma_a$, $\Sigma_b$ and $\Sigma_c$, whose zero modes are respectively indexed by $i,j,k$.

To describe the Yukawa couplings it is useful to define the vector
\be
\vec{\psi}
\, =\,
\left(
\begin{array}{c}
\psi_{\bar{x}} \\ \psi_{\bar{y}} \\ \chi_{xy}
\end{array}
\right)\, =\, \vec{\psi}_\a \mathfrak{t}_\a
\label{defvec}
\ee
which is a subvector of $\Psi$ in (\ref{matrixDirac}). Here $\mathfrak{t}_\a$ is a generator of the Lie algebra $\mathfrak{g}_p$ of the enhanced group $G_p$ at the Yukawa point $p$, with the normalization $ \tr\, \mathfrak{t}_\a \mathfrak{t}_\b^\dag =  \delta_{\a\b}$. More precisely, $\mathfrak{t}_\a$ is the generator associated to a root $\a$ of $\mathfrak{g}_p$, which  in turn corresponds to a matter curve $\Sigma_\a$ going through that point (see below for an example). The components of $\vec{\psi}_\a$ are scalar wavefunctions describing localized modes at such curve, and in particular its zero modes. As each curve may host several zero modes, we will label each zero mode vector by $\vec{\psi}_\a^{\, i}$, $i$ being the family index.

Recall that the $\psi_{\bar{m}}$ are the superpartners of the fluctuations $a_{\bar{m}}$ of $A$, whereas $\chi_{xy}$ belongs to the same multiplet as the fluctuations $\varphi_{xy}$ of $\Phi$. Notice that, as implicit in (\ref{supy}), the fermion $\eta$ in the gauge multiplet does not contribute to the Yukawa couplings, which is why such degree of freedom does not enter in the definition of $\vec{\psi}$. In fact, as noticed in  \cite{bhv1,fi09} and shown explicitly in the $U(3)$ model above, matter curve zero modes do not have a non-trivial component along $\eta$. As shown in appendix \ref{betawf}, the same applies to the zero modes that arise in the presence of a non-perturbative deformation.

Inserting the zero modes in $W_{\rm Yuk}$ gives the couplings
\be
Y_{abc}^{ijk}\, =\, m_* f_{abc} \int_{S}\, {\rm det\, } (\vec{\psi}_a^{\, i} , \vec{\psi}_b^{\, j}, \vec{\psi}_c^{\, k}) 
\, {\rm d }{\rm vol}_S
\label{yukawa7}
\ee
where $f_{abc} = -i \tr\, ([\mathfrak{t}_a, \mathfrak{t}_b] \mathfrak{t}_c)$ and the integration measure is given by ${\rm d vol}_S = 2 \om^2 = \, {\rm d} x \wedge {\rm d}y \wedge {\rm d}  \bar x \wedge {\rm d} \bar y$.

In the following we will compute the $Y_{abc}^{ijk}$ for the $U(3)$ toy model. Since the couplings
are gauge invariant we can conveniently work in the holomorphic gauge in which the zero modes take a simpler form. Turning on 7-brane fluxes $M_x < 0 < M_y$, there will be normalizable zero modes in the $a^+$ and $b^+$ sectors, which couple to those in the $c^+$ sector. Indeed, given the $U(3)$ structure displayed in (\ref{sectors}) we see that
\be
\mathfrak{t}_{a^+}\, =\,
\left(
\begin{array}{ccc}
 0 & 1 & 0 \\
 0 & 0 & 0 \\
 0 & 0 & 0 
\end{array}
\right) \quad \quad 
\mathfrak{t}_{b^+}\, =\,
\left(
\begin{array}{ccc}
 0 & 0 & 0 \\
 0 & 0 & 1 \\
 0 & 0 & 0 
\end{array}
\right) \quad \quad 
\mathfrak{t}_{c^+}\, =\,
\left(
\begin{array}{ccc}
 0 & 0 & 0 \\
 0 & 0 & 0 \\
 1 & 0 & 0 
\end{array}
\right)
\label{tmats}
\ee
and so $\tr([\mathfrak{t}_{a^+}, \mathfrak{t}_{b^+}] \mathfrak{t}_{c^+}) =1$.  We will take the {\it Higgs} to
arise from the non-chiral sector which is curve $\Sigma_c$, while the chiral families will arise from the curves $\Sigma_a$ and $\Sigma_b$, and will be indexed by $i$ and $j$ respectively. Finally, the Yukawa couplings will be denoted by $Y^{ij}$.

From the results of subsection \ref{toy} and appendix \ref{apD9KK} we see that the vectors (\ref{defvec}) for the $U(3)$ model read
\be
\vec{\psi}_{a^+}^i
\, =\,
\left(\!\!
\begin{array}{c}
-\displaystyle{\frac{i\lam_a}{m_\Phi^2}} \\ 0 \\ 1
\end{array}
\!\! \right)\chi_{a^+}^i
\qquad ; \qquad
\vec{\psi}_{b^+}^j
\, =\,
\left(\!\!
\begin{array}{c}
0\\ \displaystyle{\frac{i\lam_b}{m_\Phi^2}} \\ 1
\end{array}
\!\! \right)\chi_{b^+}^j
\qquad ; \qquad
\vec{\psi}_{c^+}
\, =\,
\left(\!\!
\begin{array}{c}
\displaystyle{\frac{i\lam_c}{m_\Phi^2}} \\[3mm] -\displaystyle{\frac{i\lam_c}{m_\Phi^2}}  \\ 1
\end{array}
\!\! \right)\chi_{c^+}
\label{simvecs}
\ee
where $\lam_a = \lam_a^-$, $\lam_b = \lam_b^-$ and $\lam_c$ are defined in (\ref{eigenlam}) and (\ref{simlams}), and the scalar wavefunctions $\chi$ are given by
\be
\chi_{a^+}^i\, =\, e^{ \lam_a  |x|^2}  f_i(y) \qquad ; \qquad
\chi_{b^+}^j\, =\, e^{ \lam_b  |y|^2}  g_j(x) \qquad ; \qquad
\chi_{c^+}\, =\, \g_c m_* e^{ \lam_c  |x-y|^2}  
\label{simchis}
\ee
For the different zero modes we will follow \cite{hv08} and take a basis in which 
$f_i(y)= \g_{ai} m_*^{4-i} y^{3-i}$ and $g_j(x)= \g_{bj} m_*^{4-j} x^{3-j}$, $i,j=1,2,3$, 
mimicking the physical case with   three families of quarks and leptons. The normalization factors $\g_{ai}$ and
$\g_{bj}$ will be fixed later.

Substituting in (\ref{yukawa7}) readily gives the couplings
\be
Y^{ij}= - i \g_c \frac{m_*^2}{m_\Phi^4} [\lam_a \lam_b + \lam_c(\lam_a + \lam_b)]
\int_S e^{ \lam_a  |x|^2 + \lam_b  |y|^2 + \lam_c  |x-y|^2}  f_i(y) g_j(x) \, {\rm d}{\rm vol}_S
\label{simy}
\ee
Notice that the exponential and the measure of the integral are invariant under the diagonal $U(1)$ rotation 
$x \to e^{i\a} x$ and $y \to e^{i\a} y$. Therefore, the only non-vanishing coupling is 
$Y^{33}$ because $f_3$ and $g_3$ are constant. Even though we are working with a local model for $S$, to evaluate the integral in (\ref{simy}) we extend $|x|$ and $|y|$ to infinite radius. This is justified because the  exponentials are localized on the matter curves and the error due to extending the Gaussian integrals is negligible. An elementary 
calculation then gives the exact result
\be
\int_S e^{\lam_a  |x|^2 + \lam_b  |y|^2 + \lam_c  |x-y|^2}  \, {\rm d}{\rm vol}_S =
\pi^2 [\lam_a \lam_b + \lam_c(\lam_a + \lam_b)]^{-1}
\label{exres}
\ee
Hence, the only non-vanishing Yukawa is given by
\be
Y^{33}= -i\pi^2 \frac{m_*^4 }{m_\Phi^4} \g_{a3} \g_{b3} \g_c 
\label{y33}
\ee
With normalization $\g_{a3}= \g_{b3}= \g_c=1$, the coupling is completely independent of the worldvolume flux. 
Moreover, this result holds without imposing the D-term BPS condition (\ref{Dterm7}) on the background, 
as already noticed in the addendum of \cite{fi09}.
In \cite{cchv09}, Yukawa independence on 7-brane worldvolume fluxes was derived from an exact residue formula.

\section{Non-perturbative effects on intersecting 7-branes\label{npei7}}

As shown above for the $U(3)$ model and more generally in \cite{cchv09}, Yukawa couplings do not depend on 7-brane worldvolume fluxes,\footnote{That is, provided that the latter satisfy the F-term BPS conditions (\ref{Fterm7phi}) at the level of background.} and this result has drastic consequences from the viewpoint of the fermion mass matrices. Namely, if all Yukawa couplings arise from a single triple intersection, the Yukawa matrices derived from (\ref{yukawa7}) will have rank one for any choice of worldvolume flux, and so only one family of quarks and leptons will receive a non-trivial mass in such F-theory construction \cite{cchv09}. While this is a promising starting point to generate the observed hierarchical structure of fermion masses, one still needs an extra ingredient beyond the intersecting 7-brane setup that slightly perturbs the Yukawas away from this rank-one result. 

As pointed out in \cite{mm09}, such extra contribution to the Yukawa couplings will in general arise from non-perturbative effects on a 7-brane far away from the GUT 4-cycle $S$. Indeed, if we consider a distant 7-brane whose 4d gauge theory undergoes a gaugino condensation, then a non-perturbative superpotential will be generated for the GUT 7-brane fields, perturbing the previous tree-level superpotential. In particular, there will a non-trivial contribution to the tree-level Yukawa couplings, so that we will instead have
\be
Y^{ijk}_{\rm total}\, =\, Y^{ijk}_{\rm tree} + Y^{ijk}_{\rm np}
\label{totalyuk}
\ee
where $Y^{ijk}_{\rm tree}$ corresponds to the tree-level contribution (\ref{yukawa7}), while $Y^{ijk}_{\rm np}$ stands for the new set of Yukawa couplings that arise at the non-perturbative level. 
In general $|Y^{ijk}_{\rm np}| << |Y^{ijk}_{\rm tree}|$, and the non-perturbative couplings will provide a slight deviation from the tree-level rank-one result. Finally, the same scenario applies if instead of a gaugino condensate on a 7-brane one considers the effect of an Euclidean 3-brane on the same 4-cycle.\footnote{See \cite{ag06} for an earlier proposal along this lines for the rank-one intersecting D6-brane model of \cite{yukint}.}

Remarkably, as shown in \cite{mm09} such non-perturbative contribution can be computed rather precisely in the case of intersecting 7-branes. In fact, there is not only one, but rather several approaches that one may use in order to compute (\ref{totalyuk}). The purpose of this section is to introduce each one of these approaches separately and show that, at least in the approximation scheme that we will discuss, all lead to the same result. 

The first and more conventional approach consists in computing the non-perturbative effect at the level of the 4d effective action, in terms of a non-perturbative superpotential $W_{\rm np}$ generated for the 4d massless and massive fields of the GUT 7-brane. The second approach consists in treating such non-perturbative superpotential as a 8d deformation of (\ref{supo7}). Notice that, before dimensional reduction to 4d, the superpotential (\ref{supo7}) can be understood as a functional of the 8d fields $A$ and $\Phi$. In this 8d approach, the non-perturbative effect is also understood as a functional $W_\b$ of $(A, \Phi)$, that adds up to the functional (\ref{supo7}) and modifies the wavefunction and Yukawa computation of section \ref{general}. Finally, the third approach is a variant of the 8d approach, in the sense that the analysis is also performed at the level of 8d fields $(\hat{A}, \hat{\Phi})$. The non-perturbative effect, however, is now seen as a non-commutative deformation of the functional (\ref{supo7}), in the sense of \cite{cchv09}. 

Besides describing these different approaches, in this section we will discuss how the (commutative) 8d approach reproduces the results of the more standard effective 4d approach. The matching between commutative and non-commutative 8d approaches will be postponed to section \ref{corrwf} and appendix \ref{apncmap}. In particular, in subsection \ref{ncmap} we will provide a dictionary between those wavefunctions computed in the non-commutative 8d formalism (see section \ref{ncyuk}) and those computed in the commutative 8d approach (see appendix \ref{betawf}). As all these approaches lead to the same physics, the reader who is just interested in the final result for the Yukawa couplings may safely skip section \ref{corrwf}  and proceed to section \ref{ncyuk}, where such Yukawas are computed explicitly for the U(3) model.

\subsection{4d approach\label{4dapp}}

In general, when computing non-perturbative effects in a string compactification, one does so at the level of the 4d effective theory. In particular, for the 7-brane setup considered above one would first compute the gauge kinetic function $f_{7_{\rm np}}$ of the stack of $n$ 7-branes undergoing a gaugino condensation, and then use the standard 4d expression
\be
W_{\rm np}^{\rm 4d}\, =\, \mu^3 \, e^{-f_{7_{\rm np}}/n}
\label{4dsuponp}
\ee
to compute the gaugino condensate contribution to the 4d effective superpotential. Here $\mu \sim m_*$ is the UV scale at which $f_{7{\rm np}}$ is defined. From the IR viewpoint, $f_{7{\rm np}}$ should be understood as a  holomorphic function of the 4d chiral multiplets of the theory, which arise either from the bulk or from the 7-brane sectors of the compactification. More precisely, such 7-brane kinetic function is of the form
\be
f_{7_{\rm np}}\, =\, T_{\rm np} + \, 
f^{\rm 1-loop}_{7_{\rm np}}\left( B_i, C_j \right)
\ee
where the first contribution amounts to the gauge kinetic function $f_{7_{\rm np}}$ computed at tree-level, and is given by  the complexified K\"ahler modulus  $T_{\rm np} = {\rm Vol\, }({S_{\rm np}}) + i \int_{S_{\rm np}} C_4$ corresponding to the 4-cycle $S_{\rm np}$ wrapped by the gaugino condensing 7-branes. The second contribution arises from threshold effects, and is given by a holomorphic function $f^{\rm 1-loop}_{7_{\rm np}}$ of the bulk/closed string fields $\{B_i\}$, and of the 4d fields $\{C_j\}$ arising from the remaining 7-brane sectors of the compactification. The latter set of fields can be divided as $\{C_j\} = \{L_j, H_k, X_l\}$ where $\{L_j\}$ are massless and $\{H_k\}$ massive fields arising from 7-brane intersections and $\{X_l\}$ massive fields spread out along the whole 7-brane worldvolume.\footnote{We are assuming that no chiral or massless fields arise from this sector.} The massive fields $\{H_k\}$ and $\{X_l\}$ are usually integrated out and thus not considered in the expression for $f^{\rm 1-loop}_{7_{\rm np}}$, but we will see that including them is crucial for our analysis.

From this 4d viewpoint, the main problem is to find $f^{\rm 1-loop}_{7_{\rm np}}$ as a function of massless and massive 4d fields. This is however implicit in the expression
\be
f^{\rm 1-loop}_{7_{\rm np}}\, =\, - n\, {\rm log \, } \ca \, - \frac{1}{8\pi^2} \int_S \str ({\rm log\, } h\, F \wedge F)
\label{f1loopa}
\ee
derived in \cite{mm09}. Here $h$ is the divisor function of the 4-cycle $S_{\rm np} = \{h = 0\}$ where the non-perturbative effect is taking place, and ${\cal A}$ is a function of the bulk/closed string fields $B_i$ which will not play any role in the following discussion and can be replaced by their vev $\langle B_i \rangle$. While ${\rm log\, } h$ is a scalar bulk quantity, when plugged into the expression (\ref{f1loopa}) one should follow the prescription of \cite{Myers} and consider its non-Abelian pull-back into $S$. That is
\be
{\rm log\, } h  \, =\, {\rm log\, } h|_{S} + m_\Phi^{-2}\, \Phi^m [\CL_m {\rm log\, } h]_S  +\, m_\Phi^{-4}\, \Phi^m \Phi^n [\CL_m \CL_n {\rm log\, } h]_S  + \dots  
\label{napb}
\ee
with $\CL_m \equiv \CL_{X_m}$ the Lie derivative along a vector $X_m$ transverse to $S$. Since $h$ is holomorphic so will be $X_m$ and so, in the local coordinate system used above, we should take $X_m = z$. Also, if as we assume that $S_{\rm np}$ is distant from our GUT 4-cycle $S$, and in particular that they do not intersect, then $h|_S$ will be a holomorphic function of $S$ with no zeroes or poles, hence a constant. This implies that
\be
- f^{\rm 1-loop}_{7_{\rm np}}\, =\, n\, {\rm log \, } \ca \, + N_{\rm D3}\, {\rm log\, } h|_{S} +  \frac{m_\Phi^{-2}}{8\pi^2} \int_S [\p_z{\rm log\, } h]_{S} \, \tr\, (\Phi^z\, F \wedge F)  +\dots
\label{f1loopb}
\ee
where $N_{D3} = (8\pi^2)^{-1} \int_S \tr (F \wedge F) \in \IN$ stands for the D3-brane charge induced by the presence of $F$ and we are not displaying higher orders in $m_\Phi^{-2}$. Clearly, the dependence of $f^{\rm 1-loop}_{7_{\rm np}}$ on the 7-brane fields $\{C_j\}$ arises only from the third term of the rhs of (\ref{f1loopb}), and is still implicit in the integral over the GUT 4-cycle $S$. In order to extract such dependence one must insert the internal wavefunctions for the fields $\{C_j\}$ in the term $\tr (\Phi^z\, F \wedge F)$, and then perform the integral over $S$ in order to obtain the different 4d couplings.

Once done so, it is straightforward to compute the non-perturbative contribution to the full 4d superpotential. Indeed, inserting (\ref{f1loopb}) into (\ref{4dsuponp}) we obtain
\bea\nonumber
W_{\rm np}^{\rm 4d} & = & \mu^3   (\ca e^{-T_{\rm np}/n} h^{N_{\rm D3}/n}|_{S} )\, {\rm exp } \left[ \frac{(m_\Phi^2 n)^{-1}}{8\pi^2} \int_S [\p_z{\rm log\, } h]_{S} \, \tr\, (\Phi^z\, F \wedge F)  +\dots\right] \\
& = & \mu^3 \eps \left(1 +  \frac{(m_\Phi^2 n)^{-1}}{8\pi^2} \int_S [\p_z{\rm log\, } h]_{S} \, \tr\, (\Phi^z\, F \wedge F)  +\dots \right)
\eea
where we have defined $\eps = \ca\, e^{-T_{\rm np}/n} h |_{S} ^{N_{\rm D3}/n}$. Upon further defining $\theta = \frac{\mu^3/4\pi^2n}{m_*^4 m_\Phi^2} \p_z {\rm log\, } h |_S$ and up to a constant term we have
\be
W_{\rm np}^{\rm 4d}\, =\, m_*^4 \frac{\eps}{2} \int_S {\theta} \,  \tr \left( \Phi_{xy} F \wedge F\right)
\label{finalsuponp4d}
\ee
where we have identified $\Phi_{xy} = \Phi^z$ (see next subsection). We can then approximate the total 4d superpotential by
\be
W_{\rm total}^{\rm 4d}\, =\, W_{\rm tree}^{\rm 4d} + W_{\rm np}^{\rm 4d}\, =\, 
m_*^4\left[  \int_S \tr (\Phi_{xy} F) \wedge dx \wedge dy + \frac{\eps}{2} \int_S {\theta} \,  \tr \left( \Phi_{xy} F \wedge F\right)\right]
\label{totalsupo4d}
\ee

Notice that in this approach the total 4d superpotential is obtained by inserting the zero mode wavefunctions computed at tree level (i.e., the ones of section \ref{general}) into (\ref{totalsupo4d}) and then performing the appropriate integral. That is, we are dimensionally reducing (\ref{totalsupo4d}) with tree level wavefunctions and background values in order to obtain new 4d couplings generated non-perturbatively, and from there performing a 4d analysis.  This is in contrast with the 8d philosophy applied in the next subsection, where new internal wavefunctions need to be computed from the very beginning.

Following the 4d approach, notice that the dimensional reduction of (\ref{supo7}) can be written as
\be
W_{\rm tree}^{\rm 4d}\, =\, \sum_{ij} \mu^{ij} \, C_i C_j + \sum_{ijk} h^{ijk}\, C_iC_jC_k
\label{supotree4d}
\ee
where 
\begin{subequations}
\label{tree4d}
\begin{align}
\mu^{ij} & =\, m_*^2 \int_{S} \tr \left(\vec{\psi}^{\, i} \cdot {\bf \bar{\p}_A} \cdot \vec{\psi}^{\, j}  \right)\, {\rm d vol}_S 
\label{masstree4d}
\\
h^{ijk} & =\,  m_*\, f_{abc}\int_{S}\,  {\rm det\, } (\vec{\psi}^i_a , \vec{\psi}^j_b, \vec{\psi}^k_c) \, {\rm d vol}_S
\label{yuktree4d}
\end{align}
\end{subequations}
Here the vectors $\vec{\psi}$ are defined as in (\ref{defvec}), and the operator ${\bf \bar{\p}_A}$ is the corresponding submatrix of the operator ${\bf D_A}$ in (\ref{matrixDirac}), namely
\be
{\bf \bar{\p}_A}\, =\, 
\left(
\begin{array}{ccc}
 0 & -D_{\bar{z}} & D_{\bar{y}} \\
 D_{\bar{z}} & 0 & -D_{\bar{x}} \\
 -D_{\bar{y}} & D_{\bar{x}} & 0
\end{array}
\right)
\label{mtxD3}
\ee
Also, in (\ref{yuktree4d}) $\cdot$ stands for the usual multiplication of such vectors and matrices. In this sense, it is understood that in (\ref{masstree4d}) $\vec{\psi}$ is given by (\ref{defvec}) when placed at the right of $\cdot$ and by $\vec{\psi} = (\psi_{\bar{x}}, \psi_{\bar{y}}, \chi_{xy})$ when placed at its left. Finally, recall that each of the components of $\vec{\psi}$ is a matrix itself, and that for eigenmodes localized at the matter curve $\Sigma_\a$ we can write $\vec{\psi} = \vec{\psi}_\a \mathfrak{t}_\a$, with $\mathfrak{t}_\a$ a generator of the enhanced group $G_{\Sigma_\a}$. 

Focusing on such matter curve $\alpha$, from eq.(\ref{Dirac9KK}) we deduce that the wavefunctions there localized must satisfy the equation
\be
i\,{\bf \bar{\p}_A}  \vec{\psi}_{H_k^\pm}\, =\, m_k\, \vec{\psi}^{\, \dag}_{H_k^\mp}	
\label{Dirac7KK}
\ee
where $H_k^{\a^+}$ are the replicas of mass $m_k$ of the left-handed chiral multiplets $L^{\a^+}_j$,  $H_k^{\a^-}$ the massive chiral multiplet transforming in the conjugate gauge representation and $\vec{\psi}_{H_k^\pm}$ their corresponding wavefunctions. Finally, $\vec{\psi}^{\, \dag}$ is defined as
\be
\vec{\psi}^{\, \dag}\, =\,
\left(
\begin{array}{c}
 \psi_{\bar{x}}^\dag \\ \psi_{\bar{y}}^\dag \\ \chi_{xy}^\dag
\end{array}
\right) 
\ee
A direct consequence of (\ref{Dirac7KK}) is that, by normalizing our wavefunctions so that
\begin{subequations}
\label{norm}
\begin{align}
 \langle \vec{\psi}_{H_k^\pm} | \vec{\psi}_{H_l^\pm} \rangle & =\,  m_*^2 \int_S \tr \,( \vec{\psi}_{H_k^\pm} \cdot \vec{\psi}_{H_l^\pm}^{\, \dag})\, {\rm d vol}_S\, =\,  \CN_k \d_{kl}
 \\
 \langle \vec{\psi}_{L_i^+} | \vec{\psi}_{L_j^+} \rangle & =\,  m_*^2 \int_S \tr \,( \vec{\psi}_{L_i^+} \cdot \vec{\psi}_{L_j^+}^{\, \dag}) \, {\rm d vol}_S\, =\,  \CN_i\d_{ij}
\end{align}
\end{subequations}
we obtain upon dimensional reduction of (\ref{supo7}) the 4d superpotential
\be
W_{\rm tree}^{\rm 4d}\, =\, - i\, \sum_{\a, k} \CN_k m_k\, H_k^{\a^+}  H_k^{\a^-} \, +\, {\rm Yukawas}
\label{treediag}
\ee
with a very simple diagonal structure for the mass terms. 
Now, when adding the effect of $W_{\rm np}^{\rm 4d}$, it is easy to see that such diagonal structure will be broken, and that in order to restore it we must redefine our fields. Indeed, the dimensional reduction of (\ref{finalsuponp4d}) gives
\be
W_{\rm np}^{\rm 4d}\, =\, \eps \sum_{ij} \mu^{ij}_{\rm np} \, C_i C_j + \eps \sum_{ijk} h^{ijk}_{\rm np}\, C_iC_jC_k
\label{suponp4d}
\ee
where now
\begin{subequations}
\label{np4d}
\begin{align}
\mu^{ij}_{\rm np} & =\, m_*^2 \, d_{abc} \int_{S}  {\rm det }  \left(\vec{\psi}_a^{\th\, i} ,  [\vec{F}_{y}^{\, \th}]_b ,   [D_x \vec{\psi}^{\, \th\, j}]_c  \right) - {\rm det }  \left(\vec{\psi}_a^{\, \th\, i} , [\vec{F}_{x}^{\, \th}]_b ,   [D_y\vec{\psi}^{\, \th\, j}]_c  \right) \, {\rm d vol}_S 
\label{massnp4d}
\\
h^{ijk}_{\rm np}& =\, m_* \, d_{abc} \int_{S}\,  {\rm det\, } \left(\vec{\psi}_a^{\, \th\, i} , [D_y \vec{\psi}^{\, \th\, j} ]_b , [D_x \vec{\psi}^{\, \th\, k}]_c\right) -  {\rm det\, } \left(\vec{\psi}_a^{\, \th\, i} , [D_x \vec{\psi}^{\, \th\, j} ]_b , [D_y \vec{\psi}^{\, \th\, k}]_c\right) \, {\rm d vol}_S
\label{yuknp4d}
\end{align}
\end{subequations}
Here $d_{abc} = \str(\mathfrak{t}_a,\mathfrak{t}_b,\mathfrak{t}_c)$, and $D_{x,y}$ act on each component of the vector $\vec{\psi}^{\, \th}$, defined as
\be
\vec{\psi}^{\, \th}\,=\, \Theta \cdot \vec{\psi}\, =\, 
\left(
\begin{array}{c}
\psi_{\bar{x}} \\ \psi_{\bar{y}} \\ \theta\, \chi_{xy}
\end{array}
\right) \quad \quad \quad
\Theta\, =\,
\left(
\begin{array}{ccc}
 1 & 0 & 0 \\
 0 & 1 & 0 \\
 0 & 0 &\theta
\end{array}
\right)
\label{T}
\ee
Finally, we have also defined the background vectors
\be
\vec{F}_x^{\, \th}\, =\,
\left(
\begin{array}{c}
\langle F_{x\bar{x}} \rangle \\  \langle F_{x\bar{y}} \rangle \\ \p_x (\theta \langle \Phi_{xy} \rangle)
\end{array}
\right) \quad \quad \quad
\vec{F}_y^{\, \th}\, =\,
\left(
\begin{array}{c}
\langle F_{y\bar{x}} \rangle \\  \langle F_{y\bar{y}} \rangle \\ \p_y (\theta \langle \Phi_{xy} \rangle)
\end{array}
\right)
\label{Fvec}
\ee
that, similarly to $\vec{\psi}$ and $\vec{\psi}^{\, \th}$, can be decomposed as $\vec{F}^{\, \th}_m = [\vec{F}^{\, \th}_m]_\a \mathfrak{t}_\a$, with $\mathfrak{t}_\a \in \mathfrak{g}_p$. This time, however, the generators $\mathfrak{t}_\a$ will belong to the Cartan subalgebra of $\mathfrak{g}_p$, as a consequence of the intersecting 7-brane setup of section \ref{general}.

It is useful to rewrite the mass term (\ref{massnp4d}) in the form
\be
\label{2pointnp}
\mu^{ij}_{\rm np} \, =\, m_*^2 \int_{S} \tr \left(\vec{\psi}^{\, \th\, i} \cdot {\bf \tilde{K}}\cdot \vec{\psi}^{\, \th\, j}  \right)\, {\rm d vol}_S 
\ee
where the operator ${\bf \tilde{K}}$ is given by
\be
{\bf \tilde{K}}\, =\,
\left(
\begin{array}{ccc}
 0 & -\CK_{\bar{z}} & \CK_{\bar{y}} \\
 \CK_{\bar{z}} & 0 & -\CK_{\bar{x}} \\
 -\CK_{\bar{y}} & \CK_{\bar{x}} & 0
\end{array}
\right) \quad \quad \quad
\CK_{\bar{m}}\, =\, \{F_{y\bar{m}}^{\, \th}, D_x\cdot\} -  \{F_{x\bar{m}}^{\, \th}, D_y \cdot\}
\label{tildeK}
\ee
and where $F_{x\bar{m}}^{\, \th} \, =\, (\vec{F}_x^{\, \th})_{\bar{m}}$, $F_{y\bar{m}}^\th \, =\, (\vec{F}_y^{\, \th})_{\bar{m}}$. That is
\be
F_{l\bar{n}}^{\, \th} = \langle F_{l\bar{n}} \rangle, \quad \quad 
F_{l\bar{z}}^{\, \th} \, =\, \p_x(\theta \langle \Phi_{xy} \rangle) \quad \quad 
l,n \, =\, x, y
\label{defFth}
\ee

Comparing  (\ref{2pointnp}) to the tree-level 2-point function (\ref{masstree4d}), we have performed the replacement ${\bf \bar{\p}_A} \raw \Theta {\bf \tilde{K}} \Theta$. This new operator does not need to be diagonal on the eigenvectors of the Laplace operator ${\bf \bar{\p}_A}^\dag {\bf \bar{\p}_A}$, as it was the case for ${\bf \bar{\p}_A}$. Recall from section \ref{general} and appendix \ref{apD9KK} that the operators $D_{x, y}$ act as creation operators on the zero mode wavefunctions $\vec{\psi}^{L}_{\a^+}$ for the chiral fields $L^{\a^+}$, and so $D_{x,y} \vec{\psi}^{L}_{\a^+}$ corresponds to the wavefunction of a massive mode. In particular, if in (\ref{2pointnp}) we substitute $\vec{\psi}^j \raw \vec{\psi^{L}}$ the integral will in general not vanish, producing non-vanishing couplings $\mu^{jk}_{\rm np} H_{k}^{\a^-} L_{j}^{\a^+}$ for some $H_k^{\a^-}$. This will result in a  4d effective superpotential of the form
\be
i W_{\rm total}^{\rm 4d} \, =\, 
 \oh \sum_{\a, k}  \left( 
\begin{array}{ccc}
L^{\a^+}_j & H^{\a^+}_k & H^{\a^-}_k
\end{array}
\right)
\left(
\begin{array}{ccc}
0 & 0 & i \eps \mu^{jk}_{\rm np} \\
0 & 0 & \CN_k m_k \\
i \eps \mu^{jk}_{\rm np} & \CN_k m_k & 0
\end{array}
\right)
\left(
\begin{array}{c}
L^{\a^+}_j \\ H^{\a^+}_k \\ H^{\a^-}_k
\end{array}
\right) + \dots
\label{mixing}
\ee
and so, in order to recover the diagonal structure of the tree-level superpotential (\ref{treediag}) we must redefine our fields. In particular, we obtain that the new zero mode is given by
\be
\hat{L}_j^{\a^+} \, =\, L^{\a^+}_j - i \eps \sum_k  \mu^{jk}_{\rm np}(\CN_k m_k)^{-1} H^{\a^+}_k 
\label{truezero}
\ee
In addition, we will have to redefine our massive modes as $H_k \raw \hat{H}_k$. This new set  $\{\hat{H}_k\}$ of massive modes can be safely discarded from the superpotential at energies below $m_k$ \cite{fq86}, obtaining an effective superpotential that only depends on the new zero modes $\hat{L}_j$
\be
W^{\rm 4d}_{\rm eff}\, =\, \sum_{ijk} \left( \hat{h}^{ijk} + \eps\, \hat{h}_{\rm np}^{ijk} \right) \hat{L}_i  \hat{L}_j  \hat{L}_k 
\ee
Note however that the holomorphic Yukawa couplings $\hat{h}^{ijk}$ and $\hat{h}^{ijk}_{\rm np}$ are no longer the ones defined in (\ref{yuktree4d}) and (\ref{yuknp4d}), since now we are expressing everything in the hatted basis of new zero modes (\ref{truezero}). The new couplings $\hat{h}^{ijk}$ will then be a linear combination of the previous ones ${h}^{ijk}$, in a way consistent with (\ref{truezero}). Schematically, we will have
\be
\hat{h}^{LLL}\, =\, h^{LLL} - i \eps  \sum_H \frac{\mu_{\rm np}^{LH}}{m_H} h^{LLH} + \CO(\eps^2)
\ee
where $h^{LLL}$ is an unhatted Yukawa coupling involving three tree-level zero modes $L$, and $h^{LLH}$ are Yukawa coupling involving two zero modes and one massive mode. Similar statements apply to $\hat{h}^{ijk}_{\rm np}$ and so we obtain the Yukawa structure
\be
\hat{h}^{LLL}\, =\, h^{LLL} + \eps  \left(h_{\rm np}^{LLL} - i \sum_H \frac{\mu_{\rm np}^{LH}}{m_H} h^{LLH}\right) + \CO(\eps^2)
\label{sqtotalyuk}
\ee
which should be compared with the expression (\ref{totalyuk}) advanced at the beginning of this section. Clearly, we have that $Y_{\rm tree} = h^{LLL}$ and that $Y_{\rm np}$ is suppressed with respect to $Y_{\rm tree}$ by the small parameter $\eps$. The main contribution to $Y_{\rm np}$ is given by the quantity in brackets and, although not clear at this point, there can be non-trivial cancellations between the two factors therein.

In fact, the above sketchy expressions can be made more precise by the following observation. It is easy to convince oneself that the couplings $\hat{h}^{ijk}$ and $\hat{h}^{ijk}_{\rm np}$ may be easily computed from the rhs of (\ref{yuktree4d}), respectively (\ref{yuknp4d}), by simply replacing the wavefunctions $\vec{\psi}^{\, j}$ there by the linear combination of wavefunctions
\be
\vec{\psi}^{\, \hat{L}_j}\, =\, \vec{\psi}^{\, L_j} - i\eps \sum_k \frac{\mu^{jk}_{\rm np}}{\CN_k m_k}\, \vec{\psi}^{\, H_k^+}
\label{newwf}
\ee
that correspond to the corrected zero modes (\ref{truezero}). It is interesting to note that this new zero modes no longer satisfy the classical zero mode equation ${\bf \bar{\p}_A} \vec{\psi} = 0$, but rather
\be
{\bf \bar{\p}_A} {\vec{\psi}}^{\, \hat{L}_j} \, =\,  \eps\,  \sum_k  \mu^{jk}_{\rm np}\, \CN_k^{-1} (\vec{\psi}^{H_k^-})^{\, \dag}
\ee
On the other hand, from (\ref{2pointnp}) and the fact that $\vec{\psi}^{\, H_i^-}_{\a^-}$ is a complete basis of wavefunctions for the sector $\a^-$ one can deduce that
\be
\sum_k  \mu^{jk}_{\rm np}\, \CN_k^{-1}  (\vec{\psi}^{H_k^-})^{\, \dag} \, =\, \Theta {\bf \tilde{K}} \Theta\, \vec{\psi}^{L_j}
\ee
and so
\be
{\bf \bar{\p}_A} {\vec{\psi}}^{\, \hat{L}_j} \, =\, \eps\,  \Theta {\bf \tilde{K}} \Theta\, {\vec{\psi}}^{\, \hat{L}_j} + \CO(\eps^2)
\label{almosteq}
\ee
This last equation will be particularly relevant when comparing the present 4d approach to the 8d approach discussed in the next subsection. 

Finally, note that in deriving (\ref{np4d}) we have used the classical values of $\langle A \rangle$ and $\langle \Phi \rangle$. One can however show that such values are also shifted by the effect of $W_{\rm np}^{\rm 4d}$. This can be seen from the fact that the massive $7$-brane bulk fields $X_l$ have a term $m_X (X_l -\langle X_l \rangle)^2$ generated at tree-level and a linear term $\eps \lam^l X_l$  contained in $W_{\rm np}^{\rm 4d}$, with
\be
\lam^l\, =\, m_*^3 \, d_{abc} \int_{S}  {\rm det }  \left(\vec{\psi}^{X_l}_{\th\, a} ,  [\vec{F}_y^{\, \th}]_b ,  [\vec{F}_x^{\, \th}]_c  \right) -  {\rm det }  \left(\vec{\psi}^{X_l}_{\th\, a} ,  [\vec{F}_x^{\, \th}]_b ,  [\vec{F}_y^{\, \th}]_c  \right)\, {\rm d vol}_S
\label{tadnp}
\ee
Hence, unless $m_X \langle X_l \rangle >> \lam^l$ the vev of $X_l$ will be shifted by the non-perturbative effect by a non-negligible amount compared to $\eps$, and so the same will apply to $\langle A \rangle$ and $\langle \Phi \rangle$. This would not only affect quantities like $\mu_{\rm np}$ and $h^{ijk}_{\rm np}$, but also the tree-level mass terms (\ref{masstree4d}) via a shift in the operator (\ref{mtxD3}). In fact, the latter effect will arise at first order in $\eps$ and so it will correct   (\ref{sqtotalyuk}) non-trivially. While one may compute the effect of (\ref{tadnp}) within the present 4d approach, let us turn our attention to a 8d description of the same physics. As we will see, the latter will provide a systematic approach to compute this and all the non-perturbative effects that we have discussed.

\subsection{8d approach\label{8dapp}}

An interesting point regarding the 4d analysis above is that the two superpotentials $W_{\rm tree}^{\rm 4d}$ and $W_{\rm np}^{\rm 4d}$ have very different origin. On the one hand, $W_{\rm tree}^{\rm 4d}$ is obtained from the dimensional reduction of the 8d field theory on the worldvolume of a stack of 7-branes, and in particular from reducing the functional (\ref{supo7}) that depends on the 8d fields $(A_{\bar{m}}, \Phi_{xy})$. On the other hand, $W_{\rm np}^{\rm 4d}$ arises at the level of the 4d effective action, via the expression (\ref{4dsuponp}). This means that, just like $f_{7{\rm np}}$, $W_{\rm np}^{\rm 4d}$ should be defined as a function of the 4d fields $\{C_i\} = \{L_j, H_k, X_l\}$, rather than $(A_{\bar{m}}, \Phi_{xy})$. 

Nevertheless, as it is clear from eq.(\ref{totalsupo4d}), both superpotentials may be put on equal footing, in the sense that $W_{\rm total}^{\rm 4d} = W_{\rm tree}^{\rm 4d} + W_{\rm np}^{\rm 4d}$ may be expressed as a sum of two functionals that depend on the 8d fields $(A_{\bar{m}}, \Phi_{xy})$. The reason for this is that, when carrying the 4d effective theory analysis, it was necessary to consider the full spectrum of 7-brane massive modes $\{C_i\}$, which contain the same information as the 8d fields $(A_{\bar{m}}, \Phi_{xy})$. As a consequence, rather than expressing (\ref{totalsupo4d}) in terms of the classical 4d fields $\{C_i\}$, one may analyze $W_{\rm total}^{\rm 4d}$ directly in terms of $(A_{\bar{m}}, \Phi_{xy})$ and obtain the same results. Notice that if we variate $W_{\rm total}^{\rm 4d}$ with respect to $(A_{\bar{m}}, \Phi_{xy})$ we will obtain a set of BPS equations that are different from (\ref{Fterm7}), and that this implies a new set of background values and zero mode wavefunctions for the 7-brane fields $(A, \Phi)$, slightly different to those obtained in section \ref{general}. This is indeed what we expect from the results of the 4d approach. The fact that we have new values for $\langle A \rangle$ and $\langle \Phi \rangle$ corresponds to a shift in the 7-brane vacuum induced by the linear terms $\e \lam^l X_l$ in the 4d effective theory, and the fact that we have a new set of zero mode wavefunctions corresponds to the result that in the 4d theory the true zero modes are given by (\ref{truezero}).

In the following we will analyze $W_{\rm total}^{\rm 4d}$ from this 8d point of view, in which the main objects are given by the 8d fields $(A_{\bar{m}}, \Phi_{xy})$. Since now we have a description of the perturbative + non-perturbative dynamics in terms of the 8d functional (\ref{totalsupo4d}), one may wonder if there could be an underlying 8d field theory from which $W_{\rm total}^{\rm 4d}$ could be derived, as it is the case for $W_{\rm tree}^{\rm 4d}$. While this latter point is more speculative, it fits nicely with the set of ideas put forward in \cite{km07}, which analyzed the contribution of non-perturbative effects to 4d effective superpotentials from a higher dimensional viewpoint. Indeed, in the scheme of \cite{km07} (see also \cite{bdkkm10,dm10}) one would trade the 4d non-perturbative field theory effect by a deformation of the 10d classical supergravity background that engineers the same 4d physics, and then extract the non-perturbative physics from a 10d supergravity analysis of this new background.

For instance, in type IIB/F-theory compactifications on warped Calabi-Yau manifolds a gaugino condensing D$7$-brane would be `backreacted' to a so-called $\beta$-deformation of the background which, among other things, involves the introduction of 3-form fluxes of the IASD kind, as in \cite{gp00}. While it is not known how to construct compact examples of these $\b$-deformed backgrounds, it is possible to implement $\beta$-deformations on local Calabi-Yau geometries \cite{gp00,bdkkm10,hmt10,dm10}. In particular, one may do so around a local type IIB/F-theory model of intersecting $7$-branes. Then, in this deformed 10d background $7$-branes should deform their 8d action, and in particular develop a new superpotential that would take into account all the non-perturbative effects computed in the standard 4d approach.
 
As pointed out in \cite{mm09} this is indeed the case, with the new superpotential looking precisely like (\ref{totalsupo4d}). The derivation can be more clearly done in the type IIB orientifold limit of F-theory and it goes as follows. In a $\b$-deformed background a D$7$-brane wrapping a 4-cycle $S$ develops a superpotential of the form \cite{martucci06}
\be
W_{\rm total}\,=\, W_{\rm CY} + W_{\b}\, =\, \frac{\mu_5 }{2\pi} \int_S \tr (\chi_2 \wedge F) + \eps  \frac{\mu_3}{8\pi^2}  \int_S \str (\chi_0 F \wedge F)
\label{totalsupo}
\ee
with  $\mu_p = (2\pi)^{-p} \a'^{-\frac{p+1}{2}}$ and $\chi_2$ a holomorphic $(2,0)$-form locally defined around $S$ and such that
\be
\rd \chi_2\, =\, \Omega
\ee
Just like ${\rm log\, } h$ in (\ref{napb}), $\chi_2$ needs to be Taylor expanded and pulled-back into $S$
\be
 \chi_2 \, =\, \chi_2|_{S} + m_\Phi^{-2} \Phi^z [\CL_z \chi_2]_S  +\, m_\Phi^{-4} (\Phi^z)^2 [\CL_z^2 \chi_2]_S  + \dots  
\label{supotree}
\ee
where now $m_\Phi^{-2} = 2\pi \alpha'$. Since in our local coordinate system $\CL_z \chi_2 = \iota_z \Omega = dx \wedge dy$ and $\CL_z^2 \chi_2 = 0$, we have that
\be
W_{\rm CY}\, =\, \frac{\mu_5}{2\pi} \int_S \tr (\chi_2 \wedge F)\, =\, \frac{\mu_5}{2\pi} \int_S \chi_2|_{S} \wedge \tr F + \frac{\mu_3}{4\pi^2} \int_S \tr (\Phi^z F) \wedge dx \wedge dy 
\label{formulilla}
\ee
This reproduces (\ref{supo7}) by choosing $\chi_2$ such that $\chi_2|_S = 0$, and substituting $\Phi^z \raw \Phi_{xy}$ and $\mu_3/4\pi^2 \raw m_*^4$. Note that from the last identification it follows the second equation in (\ref{mPhi}), which we also expect to be valid in a general F-theory compactification.

On the other hand, the extra piece $W_{\b}$ is specified by a holomorphic 0-form $\chi_0$, which is proportional to $ {\rm log\, } h$, with $h$ some divisor function. Again, $\chi_0|_S$ is a constant and so 
\bea\nonumber
\eps^{-1} W_{\b} & = &  \frac{\mu_3 }{8\pi^2} \int_S \str  (\chi_0 F \wedge F)\, =\, \frac{\mu_3}{8\pi^2} \int_S \chi_0 \tr (F \wedge F) + \frac{\mu_3}{8\pi^2} \int_S \frac{\p_z \chi_0}{m_\Phi^2}\,  \tr \left( \Phi^z F \wedge F\right) + \dots \\
& = & \chi_0|_S\,  \mu_3 N_{D3} + \frac{\mu_3}{8\pi^2} \int_S {\theta} \,  \tr \left( \Phi^z F \wedge F\right) + \dots
\label{suponp}
\eea
where we have defined ${\theta} \equiv m_\Phi^{-2} \p_z\chi_0|_S$ and ignored higher powers of $\Phi^z =\Phi_{xy}$. Hence, up to constant terms and up to linear order in $\Phi_{xy}$ we are left with the superpotential
\be
W_{\rm total}\, =\, m_*^4\left[  \int_S \tr (\Phi_{xy} F) \wedge dx \wedge dy + \frac{\eps}{2} \int_S {\theta} \,  \tr \left( \Phi_{xy} F \wedge F\right)\right]
\label{supototal}
\ee
which indeed reproduces (\ref{totalsupo4d}).

Let us now analyze this functional. Recall that in the 8d approach the prescription is to variate the whole of (\ref{supototal}) with respect to $(A_{\bar{m}}, \Phi_{xy})$. We then obtain the F-term equations 
\begin{subequations}
\label{Fterm7np}
\begin{align}
\label{Fterm7Anp}
D_{\bar{m}} \Phi_{xy} + \eps \left[  \{F_{\bar{m}x}, D_{y} (\theta \Phi_{xy})\} + \{F_{y\bar{m}}, D_{x} (\theta\Phi_{xy})\}  + \{F_{xy}, D_{\bar{m}} (\theta \Phi_{xy})\} \right]  & =  0\\
\label{Fterm7phinp}
F \wedge dx \wedge dy + \eps \theta \frac{1}{2} F \wedge F  & =  0
\end{align}
\end{subequations}
with $\bar{m} = \bar{x}, \bar{y}$ and where we have used the Bianchi identity $D_{[m} F_{np]} = 0$. Clearly, eqs.(\ref{Fterm7np}) reduce to (\ref{Fterm7}) for $\eps = 0$, and become much harder to solve for non-vanishing $\eps$. This problem becomes somewhat easier for small values of $\eps$ (as is the case in our  setup), since then we can apply perturbation theory to solve the new F-term equations. Indeed, one can then perform a perturbative expansion of the fields in the parameter $\eps$
\be
\Phi_{xy}\, =\, \Phi^{(0)}_{xy} + \eps\, \Phi^{(1)}_{xy} +\dots \quad \quad \quad A_{\bar{m}}\, =\, A_{\bar{m}}^{(0)} + \eps\, A_{\bar{m}}^{(1)} + \dots
\label{expeps}
\ee
and then solve (\ref{Fterm7np}) order by order in $\eps$. It is then easy to see that to zeroth order these equations are given by the unperturbed F-term equations (\ref{Fterm7})
\be
(D_{\bar{m}} \Phi_{xy})^{(0)}\, =\, F_{\bar{x}\bar{y}}^{(0)} \, = \,  0
\label{Fterm7np0}
\ee
while to first order we have
\begin{subequations}
\label{Fterm7np1}
\begin{align}
\label{Fterm7Anp1}
(D_{\bar{m}} \Phi_{xy})^{(1)} & =\, - \left(  \{F_{\bar{m}x}, D_{y} (\theta \Phi_{xy})\} + \{F_{y\bar{m}}, D_{x} (\theta\Phi_{xy})\} \right)^{(0)}  \\
\label{Fterm7phinp1}
F^{(1)} \wedge dx \wedge dy & =\, - \theta \frac{1}{2} F^{(0)} \wedge F^{(0)}  
\end{align}
\end{subequations}
One should then solve these equations at the level of the 7-brane background and then at the level of the fluctuations. At the level of the background and at zeroth order in $\eps$ one may use, as in the previous section, the holomorphic gauge $\langle A_{\bar{x}}\rangle^{(0)}  = \langle A_{\bar{y}} \rangle^{(0)}  = 0$, which drastically simplifies eqs.(\ref{Fterm7np1}). Indeed, (\ref{Fterm7Anp1}) then reads
\be
\p_{\bar{m}} \langle \Phi_{xy} \rangle^{(1)} \, =\, \p_{\bar{m}} \langle A_y\p_x (\theta\Phi_{xy}) -  A_x\p_y (\theta\Phi_{xy}) \rangle^{(0)}
\ee
where similarly to the previous section we are assuming that $[\langle A_m \rangle^{(0)}, \langle \Phi_{xy}\rangle^{(0)} ] = 0$. It is then easy to see that a solution to the $\CO(\eps)$ background F-term equations is given by
\begin{subequations}
\label{order1}
\begin{align}
\label{cPhi}
\langle \Phi_{xy}\rangle^{(1)}  & = \,  \langle A_y \rangle^{(0)} \p_x (\theta\langle\Phi_{xy}\rangle^{(0)}) -  \langle A_x\rangle^{(0)} \p_y (\theta\langle\Phi_{xy}\rangle^{(0)}) \\
\label{cAx}
\langle A_{\bar{x}} \rangle^{(1)} &=\, \oh \theta \langle A_y F_{x\bar{x}} - A_x F_{y\bar{x}}\rangle^{(0)} \\
\label{cAy}
\langle A_{\bar{y}} \rangle^{(1)} &=\, \oh \theta \langle A_y F_{x\bar{y}} - A_x F_{y\bar{y}}\rangle^{(0)} 
\end{align}
\end{subequations}
These corrections to the $7$-brane background fields $\langle A_{\bar{m}} \rangle$, $\langle \Phi_{xy} \rangle$ correspond to the shifts induced by the operators (\ref{tadnp}) derived in the 4d approach, as one may explicitly check. Notice that (\ref{order1}) is no longer compatible with the holomorphic gauge and that $\langle \Phi \rangle^{(1)}$ is not holomorphic. We then find that the correction $W_{\rm np}$ to the 7-brane superpotential (\ref{supo7}) spoils the holomorphic structure of the 7-brane background F-term in two different ways. On the one hand $\langle \Phi \rangle$ cannot be holomorphic and on the other hand we cannot maintain the holomorphic gauge $\langle A_{\bar{x}}  \rangle= \langle A_{\bar{y}}  \rangle =0$. These two non-holomorphic features arise already at first order in $\eps$ and, as we will see in section \ref{ncmap}, they can be understood in terms of the holomorphic structure of a related non-commutative theory. 

Similarly, one may solve for the zero mode wavefunctions order by order in $\eps$. First, the expansion (\ref{expeps}) implies that at the level of the fluctuations we have a similar expansion
\be
\vphi_{xy}\, =\, \vphi^{(0)}_{xy} + \eps\, \vphi^{(1)}_{xy} +\dots \quad \quad \quad a_{\bar{m}}\, =\, a_{\bar{m}}^{(0)} + \eps\, a_{\bar{m}}^{(1)} + \dots
\label{expepsfluc}
\ee
Second, the equations of motion for the zeroth order wavefunctions $(\vphi_{xy}^{(0)}, a_{\bar{x}}^{(0)}, a_{\bar{y}}^{(0)})$ are given by expanding (\ref{Fterm7np0}) to first order in fluctuations. These are nothing but eqs.(\ref{Ffluct7A}) and (\ref{Ffluct7phi}) which, together with the D-term equation (\ref{Dfluct}), precisely specify the zero mode equations discussed in section \ref{zerozero}. Thus, as expected, to zeroth order in $\eps$ our wavefunctions match the zero modes of the unperturbed 7-brane superpotential (\ref{supo7}). Finally, the equations for $(\vphi_{xy}^{(1)}, a_{\bar{x}}^{(1)}, a_{\bar{y}}^{(1)})$ are obtained by expanding (\ref{Fterm7np1}) to first order in fluctuations. For instance, from (\ref{Fterm7Anp1}) one obtains 
\bea
\label{FA1fluc}
\p_{\bar{m}} \vphi^{(1)}  - i [a_{\bar{m}}^{(1)}, \langle \Phi_{xy} \rangle^{(0)}] & = &  
  \left(\{F_{x\bar{m}}, D_{y} (\theta \Phi_{xy}) \} - \{F_{y\bar{m}}, D_{x} (\theta\Phi_{xy})\}  \right)^{(0)}\nonumber\\
& & + i [\langle A_{\bar{m}} \rangle^{(1)}, \vphi^{(0)}] +i  [a_{\bar{m}}^{(0)}, \langle \Phi_{xy} \rangle^{(1)}]
\eea
where the first line of the rhs of (\ref{FA1fluc}) is still to be expanded to linear order in fluctuations. That is, we must replace
\bea
\{F_{x\bar{m}}, D_{y} (\theta \Phi_{xy}) \}^{(0)}& \raw & \{ \langle F_{x\bar{m}} \rangle^{(0)}, \p_y  (\theta \vphi^{(0)}) -i  \theta [\langle A_y \rangle^{(0)}, \vphi^{(0)}]\} \\ & & + \{ \p_x a_{\bar{m}}^{(0)} - i[\langle A_x\rangle^{(0)}, a_{\bar{m}}^{(0)}], \p_y (\theta \langle \Phi \rangle^{(0)}) \} \nonumber
\eea
and similarly for $\{F_{y\bar{m}}, D_{x} (\theta\Phi_{xy})\}^{(0)}$. Note that the first line of (\ref{FA1fluc}) matches exactly with two of the three equations in (\ref{almosteq}), the remaining eq.\hspace*{-.05cm} arising from expanding (\ref{Fterm7phinp1}). Hence, we see that the linear combination of wavefunctions (\ref{newwf}) found in the 4d approach correspond to the solutions to the $\b$-deformed zero mode equations.\footnote{Obviously, the same thing will happen when we include the effect of $\langle A_{\bar{m}} \rangle^{(1)}$ and $\langle \Phi_{xy} \rangle^{(1)}$.} From this point of view, one can identify the corrections $(\vphi_{xy}^{(1)}, a_{\bar{x}}^{(1)}, a_{\bar{y}}^{(1)})$ with linear combinations of tree-level wavefunctions for $7$-brane massive modes $\{H_k\}$, a fact that will be used in appendix \ref{betawf} in order to find explicit solutions for (\ref{expepsfluc}).

Iterating the above procedure one can obtain the background and wavefunction corrections $\vec{\psi}^{(n)}\, =\, (\vphi_{xy}^{(n)}, a_{\bar{x}}^{(n)}, a_{\bar{y}}^{(n)})$ to order $\eps^n$. One then computes the corrected Yukawa couplings by inserting back these data into the functional (\ref{supototal}) and integrating over the 4-cycle $S$. It is then easy to see that the corrected Yukawa couplings look like
\be
Y^{ijk}_{abc}\, =\, (Y_{\rm tree}^{(0)})^{ijk}_{abc} + \eps \left( Y_{{\rm tree}}^{(1)} + Y_{{\rm np}}^{(0)}\right)^{ijk}_{abc} + \CO(\eps^2)
\label{expepsyuk}
\ee
in agreement with the previous structure (\ref{sqtotalyuk}). Indeed, $Y_{\rm tree}^{(0)}$ is given by the previous expression (\ref{yukawa7}), with all the three vectors in the determinant being zeroth order wavefunctions $\vec{\psi}^{(0)}$. The  $\CO(\eps)$ contribution $Y_{{\rm tree}}^{(1)}$ arises from a similar expression, but now with two zeroth order wavefunction and one first order correction $\vec{\psi}^{(1)}$. More precisely
\bea\nonumber
(Y_{{\rm tree}}^{(1)})_{abc}^{ijk} & = & m_* f_{abc} \int_{S}\, \left[ {\rm det\, } \left((\vec{\psi}_a^i)^{(1)} , (\vec{\psi}_b^j)^{(0)}, (\vec{\psi}_c^k)^{(0)} \right) + {\rm det\, } \left((\vec{\psi}_a^i)^{(0)} , (\vec{\psi}_b^j)^{(1)}, (\vec{\psi}_c^k)^{(0)} \right) \right. \\ & & \quad \quad \quad \quad 
\left. +\,  {\rm det\, } \left((\vec{\psi}_a^i)^{(0)} , (\vec{\psi}_b^j)^{(0)}, (\vec{\psi}_c^k)^{(1)} \right)\right] \, d{\rm vol}_S
\label{yukawa7np1}
\eea
Finally, the  $\CO(\eps)$ contribution $Y_{{\rm np}}^{(0)}$ arises from inserting zeroth order wavefunctions into $W_{\rm np}$ in (\ref{supototal}). That is
\bea\nonumber
(Y_{{\rm np}}^{(0)})_{abc}^{ijk} & = & m_* d_{abc} \int_{S}\, \theta\,  (\varphi_{xy}^{(0)})_a^i\, \p_{\langle A \rangle} (a^{(0)})_b^j \wedge \p_{\langle A \rangle} (a^{(0)})_c^k  \\ & &
{\rm + \ cyclic \ permutations\ in \ } abc
\label{yukawa7np2}
\eea
where $ \p_{\langle A \rangle} = \p - i \langle A\rangle^{(0)} \wedge$. As in our 4d results, from these expressions we reproduce a Yukawa structure of the form (\ref{totalyuk}). In particular, from $Y_{\rm tree}^{(0)}$ we obtain a rank-one Yukawa matrix with $\CO(1)$ coefficients, and so this contribution will play the role of $Y_{\rm tree}$ in (\ref{totalyuk}). The extra contribution to Yukawa couplings arising from non-perturbative physics will be given by $\eps\left( Y_{{\rm tree}}^{(1)} + Y_{{\rm np}}^{(0)}\right) + \CO(\eps^2)$ and, while with suppressed $\CO(\eps)$ coefficients, will generically raise the total Yukawa matrix to its maximal rank.

\subsection{Non-commutative approach\label{ncapp}}

A quite interesting variation of the 8d approach consist of expressing the $\beta$-deformed $7$-brane superpotential (\ref{totalsupo}) in terms of 8d non-commutative fields $\hat{A}_{\bar{m}}$ and $\hat{\Phi}$. Indeed, as shown in \cite{mm09} and in appendix \ref{apncmap}, by applying a generalized Seiberg-Witten map the superpotential $W_{\rm CY} + W_\b$ is transformed to
\be
\hat W \, =\, m_*^4 \int_S \tr \left( \hat\Phi\circledast \hat F \right)
\label{ncsupo}
\ee
up to $\CO(\e^2)$ terms. This new superpotential is a non-commutative version of $W_{\rm CY}$ in the sense that it has the same structure, but now the $7$-brane fields $\hat{A}$ and $\hat{\Phi}$ should be multiplied according to a non-commutative version of the usual scalar and wedge products. More precisely, two scalar functions $f$ and $g$ will be multiplied by the holomorphic Moyal product, which reads
\be
f * g\, =\, fg + \frac{i}{2} \eps\,  \theta^{ij} \p_if \p_jg + \CO(\e^2)  \quad \quad \quad \theta^{yx} = - \theta^{xy} = \theta
\label{Moyal}
\ee
whenever $\theta$ is a constant. For non-constant $\theta  = \theta(x, y)$ this definition has to be modified, as explained in appendix B of \cite{cchv09},  
where also the non-commutative version $\circledast$ of the ordinary wedge product was discussed for this case.

In fact, the non-commutative superpotential (\ref{ncsupo}) was already proposed in \cite{cchv09} as a way to overcome the Yukawa rank one problem discussed in section \ref{general}. As a proof of concept,  in \cite{cchv09} an explicit example was analyzed, for which the Yukawa couplings were computed using the non-commutative approach. This showed that, indeed, the Yukawa rank one result no longer holds when one considers the non-commutative deformation (\ref{ncsupo}) of the superpotential (\ref{supo7}).

While perhaps less intuitive, the non-commutative approach has a number of advantages for the computation of wavefunctions and Yukawa couplings. In particular, as we will discuss in section \ref{ncyuk}, it allows to realize that even if now the Yukawa matrices have maximal rank, still they do not depend on the worldvolume magnetic fluxes $\hat{F}$ on the $7$-branes. As discussed in the introduction, this could still pose a severe phenomenological handicap for local F-theory models. 

Just like for its commutative counterpart (\ref{supo7}) we may compute the F-term equations for (\ref{ncsupo}). Following \cite{cchv09} they read
\begin{subequations}
\label{ncFterm7}
\begin{align}
\label{ncFterm7A}
\bar{\p}_{A\, \circledast} \hat{\Phi}  & =  \bar{\p} \hat{\Phi} - i [\hat{A}, \hat{\Phi}]_* \, =\, 0\\
\label{ncFterm7phi}
\hat{F}^{(0,2)}  & =  \bar{\p} \hat{A} - i \hat{A} \circledast \hat{A}\, =\, 0 
\end{align}
\end{subequations}
where $\hat{A} = \hat{A}_{\bar{x}} d\bar{x} + \hat{A}_{\bar{y}} d\bar{y}$ is a $(0,1)$-form. As in section \ref{general}, these equations greatly simplify if we take a non-commutative version of the holomorphic gauge of \cite{fi09}, namely setting $\langle \hat{A}_{\bar{m}} \rangle = 0$ for $\bar{m} = \bar{x},\bar{y}$. Indeed, we then have that at the level of the background they amount to set $\langle \hat{\Phi}_{xy} \rangle$ holomorphic. In addition, defining the non-commutative fields fluctuations as
\be
\hat{\Phi}_{xy}\, =\, \langle \hat{\Phi}_{xy} \rangle + \hat{\varphi}_{xy} \quad \quad \hat{A}_{\bar{m}}\, =\, \langle \hat{A}_{\bar{m}} \rangle + \hat{a}_{\bar{m}}
\label{ncfluctuation}
\ee
and expanding (\ref{ncFterm7}) to first order in fluctuations we find the wavefunction equations
\begin{subequations}
\label{ncfluct7}
\begin{align}
\label{ncFfluct7A}
\bar{\p}_{\bar{m}} \hat{\varphi}_{xy} - i [\hat{a}_{\bar{m}}, \langle \hat{\Phi}_{xy} \rangle]   + \eps \sum_{ij} \{\p_i \hat{a}_{\bar{m}}, \p_j (\theta^{ij} \langle \hat{\Phi}_{xy} \rangle) \} & =  \CO(\e^2)\\
\label{ncFfluct7phi}
\bar{\p}_{\bar{x}} \hat{a}_{\bar{y}} -  \bar{\p}_{\bar{y}} \hat{a}_{\bar{x}}  & =  \CO(\e^2)
\end{align}
\end{subequations}
where now all products are commutative, $i,j = x,y$ and again $\theta^{yx} = - \theta^{xy} = \theta$. 

Besides modifying the wavefunctions, the non-commutative superpotential (\ref{ncsupo}) also induces $\theta$ depending corrections to the Yukawa
couplings. In particular, $\hat W$ includes the trilinear term
\be
\hat W_{\rm Yuk}\, =\, - im_*^4 \int_{S} \tr \left(\hat A \circledast \hat A \circledast \hat \Phi \right)
\label{ncsupy}
\ee
which has the $\eps$ expansion
\be
\hat W_{\rm Yuk} = \hat W_0 + \eps \hat W_1 +  \CO(\e^2)\ .
\label{exphatw}
\ee
To  zeroth order in $\e$ such trilinear term reads
\be
\hat W_0=-im_*^4 \int_{S} \tr \left(\hat A \wedge \hat A \wedge \hat \Phi \right)
\ee
while the first order correction turns out to be
\be
\hat W_1\, =\, m_*^4 d_{abc} \int_{S} \th \, \p \hat A_a \wedge \p \hat A_b  \, \hat \Phi_{c xy} 
+ {\rm cyclic \ permutations \ in \ } a, \, b, \, c
\label{ncsupy3}
\ee 
where a surface term has been dropped. 
The $\CO(\e)$ corrections to the Yukawa couplings due to $\hat W_1$ are the non-commutative counterpart of the non-perturbative
 couplings of equation (\ref{yukawa7np2}). There will be in addition an $\CO(\e)$ correction analogous to (\ref{yukawa7np1}), arising
 from inserting the the zero mode wavefunctions solving (\ref{ncfluct7}) within $\hat W_0$, and then expanding to first order in $\e$.

We will continue the discussion of non-commutative zero modes and Yukawa couplings in section \ref{ncyuk}, to which the reader may safely jump if not interested in the relation between commutative and non-commutative formalisms to be discussed in the next section. In particular, in subsection \ref{ncmap} we will relate the two set of equations (\ref{FA1fluc}) and (\ref{ncFfluct7A}) by a Seiberg-Witten map and show that, up to $\CO(\e^2)$ corrections, solving one of them is equivalent to solving the other. 

\section{Corrected wavefunctions\label{corrwf}}

As discussed in the previous section, the effect of a non-perturbative superpotential for the chiral matter fields of a local F-theory model can be understood in terms of a deformation of their internal wavefunctions. Namely, the new wavefunctions read
 \be
 \Psi\, =\, \Psi_0 + \sum_\lam c_\lam \Psi_\lam
 \label{expwf}
 \ee
where $\Psi_0$ is the wavefunction for the chiral field before the non-perturbative effect has been taken into account, $\{\Psi_\lam\}$ is an appropriate basis of wavefunctions and $c_\lam$ are complex coefficients that vanish when the strength $\e$ of the non-perturbative effect does.

In this section we describe the corrected wavefunction $\Psi$ in terms of the 8d approach of section \ref{8dapp}. That is, the new wavefunction $\Psi$ will be the solution of a new set of zero mode equations, which arise from deforming the previous F-term equations (\ref{Fterm7}) to the more complicated ones (\ref{Fterm7np}). While computing $\Psi$ exactly is quite involved, one may simplify the problem by expanding $\Psi$ in the small parameter $\eps$ and then applying perturbation theory. In the following we will describe this perturbative strategy for the deformed F-terms of section \ref{8dapp},  and obtain the equations for the $\CO(\e)$ correction to $\Psi_0$  in this case. In fact, the same perturbative strategy may be applied to the non-commutative F-term equations of section \ref{ncapp}, where the new zero mode equations are given by (\ref{ncfluct7}). We will do so and show that, up to $\CO(\e^2)$ corrections, the commutative and non-commutative corrected wavefunctions are related by a simple Seiberg-Witten map. 

As both kind of wavefunctions are equivalent, in section \ref{ncyuk} we will focus on solving for corrected wavefunctions within the non-commutative formalism, leaving the computation in the commutative formalism for appendix \ref{betawf}. It will become clear from the latter analysis that a basis of wavefunctions $\{\Psi_\lam\}$ suitable to solve (\ref{expwf}) is given by the tower of unperturbed massive modes at the matter curve (i.e., those wavefunctions satisfying eq.(\ref{eigenferm})) as already suggested by the 4d analysis of section \ref{4dapp}.

\subsection{Wavefunctions and perturbation theory}

Let us consider a chiral zero mode wavefunction $\Psi_0$ in the absence of non-perturbative effect/$\beta$-deformation. By the results of section \ref{general}, $\Psi_0$ must satisfy the equation
\be
{\bf D_A} \Psi_0\, =\, 0
\label{Dirac9b}
\ee
where the matrix operator ${\bf D_A}$ and the wavefunction vector $\Psi$ are defined as in (\ref{matrixDirac}). 

Let us now assume that the system is perturbed such that the zero mode equations are modified to
\be
{\bf D_A} \Psi\, =\, {\bf K} \Psi
\label{corrected}
\ee
where {\bf K} is a linear operator that may contain derivatives and functions but does not depend on $\Psi$, and admits an expansion of the form
\be
{\bf K}\, =\,  \eps {\bf K}^{(1)} + \eps^2 {\bf K}^{(2)} + \dots 
\label{expK}
\ee
with $\eps$ a small parameter. It is then natural to consider an expansion of the form
\be
\Psi\, =\, \Psi^{(0)} + \eps \Psi^{(1)} + \eps^2 \Psi^{(2)} + \dots  
\label{expwf2}
\ee
and to solve (\ref{corrected}) order by order in $\eps$. Clearly, to zeroth order in $\eps$ we just need to impose $\Psi^{(0)} = \Psi_0$, while to $\CO(\eps)$ we have
\be
{\bf D_A} \Psi^{(1)}\, =\, {\bf K}^{(1)} \Psi_0
\label{premaster}
\ee

Note that this perturbative method may be applied to solve any equation of motion of the form (\ref{corrected}).\footnote{See for instance \cite{mms10} for its application to compute chiral wavefunctions in warped backgrounds.} In the following, however, we will focus on those corrections that only deform the F-term equations (\ref{Ffluct7A}) and (\ref{Ffluct7phi}), and leave the D-term equation (\ref{Dfluct}) unchanged. This means that 
\be
{\bf K}\, =\,
\left(
\begin{array}{cc}
0 & 0 \\
0 & {\bf k}
\end{array}
\right)\quad \quad \quad 
{\bf k}\, =\,  \eps {\bf k}^{(1)} + \eps^2 {\bf k}^{(2)} + \dots
\label{K3}
\ee
for some $3 \times 3$ submatrix ${\bf k}$. Eq.(\ref{premaster}) then reduces to the standard D-term equation and the deformed F-term equations
\be
\bar{\p}_{\bf A} \vec{\psi}^{(1)}\, =\, {\bf k}^{(1)} \vec{\psi}_0
\label{premaster3}
\ee
where $\vec{\psi} = \vec{\psi}_0 + \e  \vec{\psi}^{(1)} + \dots$ and $\bar{\p}_{\bf A}$ are defined by (\ref{defvec}) and (\ref{mtxD3}), respectively.

\subsubsection*{Commutative equations}

By the results of section \ref{8dapp}, the wavefunction equations that arise from the $\beta$-deformed F-terms (\ref{Fterm7np}) are indeed of the form (\ref{corrected}) and (\ref{K3}). More precisely we have that
\be
{\bf k}^{(1)}\, =\,
- \Theta ({\bf \tilde{K}} + i {\bf \tilde{A}})^{(0)} \Theta
\label{K12}
\ee
with ${\bf \tilde{K}}$ and $\Theta$ defined as in (\ref{tildeK}) and (\ref{T}), respectively. The operator ${\bf \tilde{A}}$ contains the terms arising from the second line of (\ref{FA1fluc}). Using the background solutions (\ref{order1}), we find that it reads
\be
{\bf \tilde{A}}\, =\, 
\left(
\begin{array}{ccc}
0 & - \CA_{\bar{z}} & \frac{1}{2}\CA_{\bar{y}} \\
\CA_{\bar{z}} & 0 & -\frac{1}{2}\CA_{\bar{x}} \\
- \frac{1}{2}\CA_{\bar{y}} & \frac{1}{2} \CA_{\bar{x}} &0
\end{array}
\right) 
\quad \quad
\CA_{\bar{m}}  
\, =\, 
[\langle F_{y\bar{m}}^{\, \th} A_x - F_{x\bar{m}}^{\, \th} A_y  \rangle,\cdot]
\label{tildeA}
\ee
where $F_{n\bar{m}}^{\, \th}$ is again given by (\ref{defFth}). Finally, the superscript $(0)$ in (\ref{K12}) indicates that within ${\bf \tilde{K}}$ and ${\bf \tilde{A}}$ one should take  the uncorrected vevs $\langle A_{\bar{m}} \rangle^{(0)}$ and $\langle \Phi_{xy} \rangle^{(0)}$. 

That is, we have that the $\CO(\e)$ wavefunction equation is given by (\ref{premaster3}) with
\bea
\label{K1K2}
{\bf k}^{(1)} & = &
- \Theta ({\bf \tilde{K}} + i {\bf \tilde{A}})^{(0)} \Theta \\ \nonumber
&  & -({\bf \tilde{K}} + i{\bf \tilde{A}}) \, = \,
\left(
\begin{array}{ccc}
0 & \CK_{\bar{z}} + i \CA_{\bar{z}} & -(\CK_{\bar{y}} + \frac{i}{2}\CA_{\bar{y}}) \\
-(\CK_{\bar{z}} + i \CA_{\bar{z}}) & 0 & \CK_{\bar{x}} + \frac{i}{2}\CA_{\bar{x}} \\
\CK_{\bar{y}} + \frac{i}{2}\CA_{\bar{y}} & -(\CK_{\bar{x}} +\frac{i}{2}\CA_{\bar{x}}) &0
\end{array}
\right)
\eea
and
\bea
 \nonumber
& & \CK_{\bar{m}} \, = \, \{F_{y\bar{m}}^{\, \th}, D_x\cdot\} -  \{F_{x\bar{m}}^{\, \th}, D_y \cdot\} \quad \quad \CA_{\bar{m}}\, =\, [\langle F_{y\bar{m}}^{\, \th} A_x -F_{x\bar{m}}^{\, \th} A_y \rangle,\cdot]
\\ \nonumber
& & F_{l\bar{n}}^{\, \th} = \langle F_{l\bar{n}} \rangle \quad \quad 
F_{l\bar{z}}^{\, \th} \, =\, \p_l(\theta \langle \Phi_{xy} \rangle) \quad \quad 
l,n \, =\, x, y
\eea

\subsubsection*{Non-commutative equations}

In order to write the analogous equations in the non-commutative formalism it is useful to consider the non-commutative version of the vectors (\ref{defvec}) and (\ref{T})
\be
\hat{\psi}
\, =\,
\left(
\begin{array}{c}
\hat{a}_{\bar{x}} \\ \hat{a}_{\bar{y}} \\ \hat{\vphi}_{xy}
\end{array}
\right)\, =\, \hat{\psi}_\a \mathfrak{t}_\a
\quad \quad \quad
\hat{\psi}^{\, \th}\,=\, \Theta \cdot \hat{\psi}\, =\, 
\left(
\begin{array}{c}
\hat{a}_{\bar{x}} \\ \hat{a}_{\bar{y}} \\ \theta\, \hat{\vphi}_{xy}
\end{array}
\right)
\label{defvecnc}
\ee
So that eqs.(\ref{ncfluct7}) can be rewritten as
\be
\label{Qm}
{\bf \bar{\p}_A} \hat{\psi}\, =\, \eps 
\left(
\begin{array}{ccc}
 0 & \CQ_{\bar{z}}^{(0)} & 0 \\
- \CQ_{\bar{z}}^{(0)} & 0 & 0 \\
 0 & 0 & 0
\end{array}
\right)
\hat{\psi}\, +\, \CO(\e^2) \quad \quad \CQ_{\bar{m}} \, =\,  \{F_{y\bar{m}}^{\, \th}, \p_x \cdot\} - \{F_{x\bar{m}}^{\, \th}, \p_y \cdot\}
\ee
where as before $F_{n\bar{m}}^{\, \th}$ is given by (\ref{defFth}). This again implies a corrected zero mode wavefuction of the form
\be
\hat{\psi} \, =\, \hat{\psi}_0 + \e \hat{\psi}^{(1)} + \dots \quad \quad {\rm with} \quad \quad 
\bar{\p}_{\bf A} \hat{\psi}_0 \, =\, 0\, , \quad
\bar{\p}_{\bf A} \hat{\psi}^{(1)}\, =\, {\bf \hat{k}}^{(1)} \hat{\psi}_0 
\label{premaster3nc}
\ee
and
\be
{\bf \hat{k}}^{(1)} \, =\, 
\left(
\begin{array}{ccc}
 0 & \CQ_{\bar{z}}^{(0)} & 0 \\
- \CQ_{\bar{z}}^{(0)} & 0 & 0 \\
 0 & 0 & 0
\end{array}
\right)
\ee
Finally, note that for $\theta \not= 0$ eq.(\ref{Qm}) is equivalent to
\be
\label{Qmth}
{\bf \bar{\p}_A}^{\, \th}  \hat{\psi}^{\, \th}\, =\, \eps \th
{\bf \hat{k}}^{(1)}
\hat{\psi}^{\, \th}\, +\, \CO(\e^2) 
\ee
with ${\bf \bar{\p}_A}^{\, \th}$ given by 
\be
{\bf \bar{\p}_A}^{\, \th}\, =\, 
\left(
\begin{array}{ccc}
 0 & i [\th \langle\Phi_{xy}\rangle^{(0)}, \cdot] & \p_{\bar{y}} \\
 -i [\th \langle\Phi_{xy}\rangle^{(0)}, \cdot ] & 0 & -\p_{\bar{x}} \\
 -\p_{\bar{y}} & \p_{\bar{x}} & 0
\end{array}
\right)
\label{mtxD3b}
\ee
This alternative expression will be important in proving the equivalence between the commutative and non-commutative zero mode equations, to which we now turn.

\subsection{Relating commutative and non-commutative wavefunctions\label{ncmap}}

As pointed out in \cite{mm09}, the 8d $\beta$-deformed superpotential (\ref{totalsupo}) can be expressed in terms of the non-commutative superpotential $\hat{W}$ in (\ref{ncsupo}), upon applying a Seiberg-Witten map relating the non-commutative 7-brane fields $(\hat{A}, \hat{\Phi})$ to the standard ones $(A, \Phi)$. As shown explicitly in appendix \ref{apSWmap} such SW map is given by
\begin{subequations}
\label{SWmap}
\begin{align}
\label{SWmapA}
\hat A_{\bar{m}}&=\,A_{\bar{m}} + \tilde{A}_{\bar{m}}\, =\, A_{\bar{m}} - \frac\e2\theta^{ij}\{A_i, \p_j A_{\bar{m}} + F_{j \bar{m}}\}+\CO(\eps^2)\\
\label{SWmapphi}
\hat\Phi_{xy} &=\, \Phi_{xy} + \tilde{\Phi}_{xy}\, =\, \Phi_{xy} - \frac\e2 \{A_i, (\p_j+D_j)(\theta^{ij}\Phi_{xy})\} +\CO(\eps^2)
\end{align}
\end{subequations}
where all the products in the rhs are standard ones. As in (\ref{ncfluct7}),  $\theta^{yx} = - \theta^{xy} = \theta$.

One can gain some intuition on the meaning of this Seiberg-Witten map by applying it at the level of the background. We have that $\langle \hat{A}_{\bar{m}} \rangle = \langle {A}_{\bar{m}} \rangle + \langle \tilde{A}_{\bar{m}} \rangle$, where
\begin{subequations}
\label{SWvsncA}
\begin{align}
\langle A_{\bar{m}} \rangle& =\,  \langle A_{\bar{m}} \rangle^{(0)} + \eps  \langle A_{\bar{m}} \rangle^{(1)} + \CO(\eps^2) \\
\langle \tilde{A}_{\bar{m}} \rangle & =\,  \frac\eps2 \theta \langle A_x(\p_yA_{\bar{m}} +F_{y\bar{m}})- A_y(\p_xA_{\bar{m}} +F_{x\bar{m}})\rangle^{(0)} + \CO(\eps^2)
\end{align}
\end{subequations}
and that $\langle \hat{\Phi}_{xy} \rangle = \langle {\Phi}_{xy} \rangle + \langle \tilde{\Phi}_{xy} \rangle$, with
\begin{subequations}
\label{SWvsncPhi}
\begin{align}
\langle \Phi_{xy} \rangle& =\,  \langle \Phi_{xy} \rangle^{(0)} + \eps  \langle \Phi_{xy} \rangle^{(1)} + \CO(\eps^2) \\
\langle \tilde{\Phi}_{xy} \rangle & =\,  \eps \langle A_x\p_y(\theta \Phi_{xy}) - A_y\p_x(\th \Phi_{xy})\rangle^{(0)} + \CO(\eps^2)
\end{align}
\end{subequations}
Both in (\ref{SWvsncA}) and in (\ref{SWvsncPhi}) we have used our previous assumption that $\langle A_{m} \rangle^{(0)}$, $\langle A_{\bar{m}} \rangle^{(0)}$ and $\langle \Phi_{xy} \rangle^{(0)}$ commute. In particular, taking the holomorphic gauge $\langle A_{\bar{m}} \rangle^{(0)} = 0$     and the corresponding first order perturbations $\langle A_{\bar{m}} \rangle^{(1)}$ and $\langle \Phi_{xy} \rangle^{(1)}$ found in eqs.(\ref{order1}), we find that $\langle \tilde{A}_{\bar{m}}\rangle = - \eps \langle {A}_{\bar{m}}\rangle^{(1)}$ and $\langle \tilde{\Phi}_{xy}\rangle = - \eps \langle {\Phi}_{xy}\rangle^{(1)}$. Hence
\begin{subequations}
\label{SWcancel}
\begin{align}
\label{SWcancelA}
\langle \hat A_{\bar{m}} \rangle &= 0+\CO(\eps^2)\\
\label{SWcancelphi}
\langle \hat\Phi_{xy}\rangle &=\, \langle \Phi_{xy}\rangle^{(0)}  +\CO(\eps^2)
\end{align}
\end{subequations}
and so the background values for the non-commutative fields $\hat{A}$, $\hat{\Phi}$ correspond to those of the commutative fields $A$, $\Phi$ at zeroth order in $\eps$, that is before the $\b$-deformation/non-perturbative effect was taken into account. Note in particular that in the non-commutative variables we recover the holomorphic structure (that is, the holomorphic gauge for $\langle \hat{A}\rangle$ and the fact that $\langle \hat{\Phi}\rangle $ is holomorphic) that was lost for $\langle A \rangle$, $\langle\Phi\rangle$ when we included the effect of the $\b$-deformation. In this sense, the SW map (\ref{SWmap}) relates rather directly the holomorphic gauge introduced in \cite{fi09} and its non-commutative version. We see that the non-commutative holomorphic gauge used in \cite{cchv09} is nothing but a deformation of the standard holomorphic gauge such that it makes the latter compatible with the $\b$-deformed equations of motion.

In addition, the map (\ref{SWmap}) between commutative and non-commutative 8d fields implies a well-defined dictionary between the corresponding 7-brane wavefunctions, and in particular between the $\theta$-corrected zero modes of appendix \ref{betawf} and their non-commutative counterparts solving eqs.(\ref{ncfluct7}). Indeed, let us build explicitly such dictionary between commutative and non-commutative zero modes. For this we need to expand (\ref{SWmap}) to first order in the fluctuations $(\hat{a}_{\bar{m}}, \hat{\varphi}_{xy})$ defined in (\ref{ncfluctuation}). We find
\be
\label{changenc}
\hat{a}_{\bar{m}}\, =\, a_{\bar{m}} + \tilde{a}_{\bar{m}} \quad \quad 
\hat{\vphi}_{xy}\, =\, \vphi_{xy} + \tilde{\vphi}_{xy}
\ee
where
\begin{subequations}
\label{tildes}
\begin{align}\nonumber
\tilde{a}_{\bar{m}} & =\, \eps\th \left[ \{\langle A_x\rangle , \oh(\p_y + D_y) {a}_{\bar{m}}\} - \{\langle A_y\rangle, \oh(\p_x + D_x) {a}_{\bar{m}}\}\right]  \\
\label{tildea} 
& =\, \eps\th \left( \{\langle A_x\rangle , \p_y  {a}_{\bar{m}}\} -  \{\langle A_y\rangle, \p_x {a}_{\bar{m}}\}\right)  \\ \nonumber
\tilde{\vphi}_{xy} & = \, \eps \left[ \{\langle A_x\rangle , \oh(\p_y + D_y) (\theta\vphi_{xy})\} -  \{\langle A_y\rangle, \oh(\p_x + D_x)  (\theta\vphi_{xy})\}\right]  \\
& =\, \eps \left( \{\langle A_x\rangle , \p_y  (\theta\vphi_{xy})\} - \{\langle A_y\rangle, \p_x (\theta\vphi_{xy})\} \right) 
\label{tildephi}
\end{align}
\end{subequations}
up to $\CO(\eps^2)$ corrections. This map can be expressed in a more compact way in terms of the vector $\hat{\psi}^\th$ defined in (\ref{defvecnc}). Indeed, we have that
\be
\label{Q}
\hat{\psi}^{\, \th}\, =\, \left(\Id -  \eps\th\, {\CQ}^{(0)} \right)  \vec{\psi}^{\, \th} \quad \quad \quad
{\CQ}\, =\, \{\langle A_y\rangle, \p_x \cdot \} - \{\langle A_x\rangle , \p_y \cdot \} 
\ee
with $\vec{\psi}^{\, \th}$ given by (\ref{T}). 

Let us now see that (\ref{Q}) maps non-commutative zero modes to $\beta$-deformed zero modes. 
 Combining the non-commutative zero mode equation (\ref{Qmth}) with (\ref{Q}) we obtain
\be
\label{QQm}
{\bf\bar{\p}_A}^{\, \th}  \vec{\psi}^{\, \th} \, =\, \eps \theta
\left[ 
{\bf \hat{k}}^{(1)} + {\bf \bar{\p}_A}^{\, \th} \CQ^{(0)}
 \right]
 \vec{\psi}^{\, \th} + \CO(\e^2)
\ee
We may now perform an expansion on the wavefunction $\vec{\psi}^{\, \th}$
\be
\vec{\psi}^{\, \th}\, =\, (\vec{\psi}^{\, \th})^{(0)} + \eps\,  (\vec{\psi}^{\, \th})^{(1)} + \dots
\ee
and write down (\ref{QQm}) order by order in $\eps$. At zeroth order we have that
\be
{\bf\bar{\p}_A}^{\, \th}  (\vec{\psi}^{\, \th})^{(0)} \, =\, 0 \quad\quad \iff \quad \quad {\bf\bar{\p}_A}  \vec{\psi}^{(0)}\, =\, 0
\label{0nc}
\ee
and to first order we obtain
\bea\nonumber
{\bf\bar{\p}_A}^{\, \th}  (\vec{\psi}^{\, \th})^{(1)} & = & 
\theta
\left[ 
{\bf \hat{k}}^{(1)} + [{\bf \bar{\p}_A}^{\, \th}, \CQ^{(0)}]
 \right]
 (\vec{\psi}^{\, \th})^{(0)} \\
& = & \theta
\left(
\begin{array}{ccc}
 0 & \CQ_{\bar{z}} +i \CR_{\bar{z}} & -\CQ_{\bar{y}} \\
 - (\CQ_{\bar{z}} +i \CR_{\bar{z}})  & 0 & \CQ_{\bar{x}}   \\
\CQ_{\bar{y}} &- \CQ_{\bar{x}} & 0
\end{array}
 \right)^{(0)}
 (\vec{\psi}^{\, \th})^{(0)}
 \label{1nc}
\eea
where we have defined
\be
\CR_{\bar{m}}  =   \{\langle A_x \rangle, [F_{y\bar{m}}^{\, \th}, \cdot ]\} - \{\langle A_y \rangle, [F_{x\bar{m}}^{\, \th}, \cdot ]\}  
\ee
and made use of the fact that $[\p_{\bar{n}}, \CQ] = - \CQ_{\bar{n}}$.

To show that (\ref{1nc}) is equivalent to (\ref{premaster3}), (\ref{K1K2}) notice that they can be rewritten as
\be
{\bf\bar{\p}_A} (\vec{\psi}^{\, \th})^{(1)}\, =\, -\th({\bf \tilde{K}} + i {\bf \tilde{A}})^{(0)} (\vec{\psi}^{\, \th})^{(0)}
\label{K1K2c}
\ee
and so one just needs to show that the elements of $-({\bf \tilde{K}} + i {\bf \tilde{A}})^{(0)}$ and of the matrix in (\ref{1nc}) are the same. For this one needs to notice that
\be
\CK_{\bar{m}} + i \CA_{\bar{m}} \, =\, \CQ_{\bar{m}} + i \CR_{\bar{m}}
\ee
where $\CK_{\bar{m}}$ and $\CA_{\bar{m}}$ are defined as in (\ref{K1K2}). Finally, we have that\footnote{In (\ref{Qxy}) we are using that, whenever $[A, B] = 0$ we have the identity
\be
\{A,[B,C]\} - \oh [AB,C]\, =\, \oh (BCA - ACB)
\nonumber
\ee
and that the rhs of this equation vanishes if $A$ and $B$ differ by multiplication of a function.} 
\be
\CK_{\bar{n}} + \frac{i}{2} \CA_{\bar{n}} \, =\, \CQ_{\bar{n}}  \quad \quad n = x, y
\label{Qxy}
\ee
whenever $\langle A_x \rangle^{(0)}$ and $\langle A_y \rangle^{(0)}$ are proportional matrices, which should be the case if the D-term equation (\ref{Dterm7}) is to be satisfied for some choice of K\"ahler parameters.\footnote{Indeed, note that for our assumptions $[\langle A_x \rangle, \langle A_y \rangle] =0$ and $[\langle \Phi\rangle, \langle\bar{\Phi}\rangle] = 0$ the D-term equation (\ref{Dterm7}) reduces to $\omega_{y\bar{y}} \p_{\bar{x}} \langle A_x \rangle + \omega_{x\bar{x}} \p_{\bar{y}} \langle A_y \rangle = 0$ at the level of the background. The matrices $\langle A_x \rangle$ and $\langle A_x \rangle$ are clearly proportional if the D-term equation is satisfied. Changing the K\"ahler moduli of the compactification can be understood as changing the value of $\omega_{m\bar{m}}$, which will not change the fact that  $\langle A_x \rangle^{(0)}$ and $\langle A_y \rangle^{(0)}$ must be proportional.}

We have then shown that, upon the change of variables (\ref{changenc}) the zero mode equations for non-commutative fields $(\hat{a}_{\bar{m}}, \hat{\varphi}_{xy})$ is mapped to the $\b$-deformed zero mode equations for the standard commutative fields $({a}_{\bar{m}}, {\varphi}_{xy})$. One may, in addition, show that the Yukawa couplings deduced in both formalisms match. In particular, this can be deduced from the results of appendix \ref{apSWmap}, where the equivalence of the 8d commutative and non-commutative formalisms is shown at the level of the superpotential. In the following we will focus on the computation of wavefunctions and Yukawa couplings in the non-commutative formalism which, as we will see, is particularly useful for deducing some important features of Yukawa couplings in intersecting 7-brane models.

\section{Wavefunctions and Yukawas in the non-commutative formalism\label{ncyuk}}

In this section we discuss the equations of motion that follow from the non-perturbative superpotential
$\hat W$ in (\ref{ncsupo}). Our main motivation is to use the zero mode solutions to determine the $\theta$ corrections
to the Yukawa couplings due to the trilinear term in $\hat W$. 

In the previous section we have explained that the background values for the non-commutative fields
$\langle \hat A \rangle$ and $\langle \hat\Phi_{xy}\rangle$ are equal to those of the commutative fields at
zeroth order in $\epsilon$. More precisely, it is possible to take the holomorphic gauge in which
$\langle \hat A_{\bar{m}} \rangle=0$, for $\bar m = \bar x, \bar y$, and therefore the equations of motion
for the background imply that $\langle \hat\Phi_{xy}\rangle$ is holomorphic. The equations of motion for the
fluctuations will then take a simpler form. In particular, in the holomorphic gauge the F-term equations
for the fluctuations are given in equations (\ref{ncfluct7}). In addition there is a D-term which is the 
non-commutative extension of (\ref{Dterm7}) \cite{cchv09}. Expanding the non-commutative wedge product
according to the prescription in \cite{cchv09}, and defining the fluctuations as in  (\ref{ncfluctuation}), 
yields the D-term equation    
\be
\omega \wedge \p_{\langle \hat A \rangle} \hat a - \frac12 [\langle \hat{\bar{\Phi}} \rangle, \hat\varphi] = 0
\label{Dnc}
\ee
Notice that this equation does not receive $\CO(\e)$ corrections.

In the following we will use the $U(3)$ toy model in order to carry out explicit calculations in
the non-commutative approach. With the resulting zero modes we will also check that the non-commutative
wave functions are related to the corrected wave functions obtained in appendix \ref{betawf} via the dictionary
established in (\ref{tildes}).

\subsection{Non-commutative zero modes in the $U(3)$ toy model \label{ncu3}}

The background value for $\langle \hat\Phi_{xy}\rangle$ is given exactly by (\ref{vevPhi}). Just as in the
commutative case the non zero vev breaks $U(3)$ to $U(1)^3$, with full enhancement occurring at the point
of intersection of the three curves $\Sigma_a =\{x=0\}$, $\Sigma_b =\{y=0\}$, and $\Sigma_c =\{x=y\}$.
Associated to each curve $\Sigma_\a$ there is a different set of fluctuations for which we use the notation
\be
\hat{\vec{\psi}}_{(\a)} = \left(\!\!
\begin{array}{c}
\hat\psi_{\a\bar x} \\ \hat\psi_{\a\bar y} \\ \hat\chi_{\a}
\end{array}
\!\! \right){\mathfrak t}_\a = \hat{\vec{\psi}}_\a {\mathfrak t}_\a  
\label{hatvec}
\ee 
For simplicity we write $\chi = \chi_{xy}$. We are using the fermionic fluctuations which satisfy
the same equations as their supersymmetric partners $(\hat a_{\a\bar x}, \hat a_{\a\bar y}, \hat \varphi_{\a xy})$.

To obtain a 4d chiral model we turn on a worldvolume flux of the form (\ref{fluxD7}). 
We then choose the holomorphic gauge in which 
\be
\langle \hat A \rangle = -\frac{i}{3} (M_x\bar x  dx + M_y \bar y dy){\rm diag} (1, -2, 1)
\label{ahol}
\ee 
exactly as in  (\ref{holg}). As in the commutative case we take $M_x < 0 < M_y$, so that the normalizable
zero modes appear in the $a^+$, $b^+$ and $c^+$ sectors. The corresponding ${\mathfrak t}_\a$ generators
for each curve are those given in (\ref{tmats}). 

The F-term equation (\ref{ncFfluct7phi}) arising from $\hat F^{(0,2)}=0$ takes the same form in all sectors
and will not be written separately. To first order in $\eps$ it implies that
$\bar\partial_{\bar x} \hat  \psi_{\a\bar y}= \bar\partial_{\bar y} \hat \psi_{\a\bar x}$, which
will indeed be satisfied by the zero modes determined below. 

With  the main ingredients at hand we next proceed to obtain the zero modes in each sector. 
We will work out in more detail the case of $\th$ constant. In section \ref{ncthvar}
we will briefly discuss an example with $\th$ depending on the coordinates.

\subsubsection*{Sector $a^+$ \label{nca}}

Given the generator ${\mathfrak t}_{a^+}$ it is straightforward to evaluate the various commutators and 
anticommutators in the F-term equations (\ref{ncFfluct7A}). For generic $\th$ we find
\bea
\bar\p_{\bar m} \hat\chi \ & - &im_{\Phi}^2 \, x \hat \psi_{\bar m} 
- \frac{\eps m_{\Phi}^2}{6}[\th + \p_x\th(x-2y-2\Phi_0)]\p_y \hat\psi_{\bar m} \nonumber \\
& - &  \frac{\eps m_{\Phi}^2}{6}[2\th - \p_y\th(x-2y-2\Phi_0)] \p_x \hat\psi_{\bar m}     =  \CO(\e^2)
\label{ancfer}
\eea
for $\bar m = \bar x, \bar y$.
We have dropped the subindex $\a=a^+$ that will be reinserted at the end. The D-term equation
derived from (\ref{Dnc}) turns out to be
\be
(\p_x-  M_x \bar x) \hat \psi_{\bar x} +  (\p_y- M_y \bar y) \hat \psi_{\bar y} 
+ i m_{\Phi}^2 \bar x \hat \chi = 0
\label{ancferD}
\ee
Notice that the worldvolume fluxes $M_x$ and $M_y$ only appear in the D-term equation. In practice this feature
greatly simplifies the search of zero mode solutions. More importantly, it makes clear from the beginning
that the Yukawa couplings, which only depend on the F-terms, will turn out to be independent of fluxes \cite{cchv09}.

To find the zero modes we make an Ansatz motivated by the form of the solutions when $\theta=0$
collected in eq.~(\ref{simvecs}). 
In particular we set $ \hat \psi_{\bar y} =0$, which then implies  
$\bar\partial_{\bar y}\hat \psi_{\bar x} =  \bar\partial_{\bar y} \hat\chi=0$.
To solve the remaining equations we have to treat separately the cases of $\th$ constant and $\th$ coordinate
dependent.

When $\th$ is constant we further impose
\be
 \hat \psi_{\bar x} = -\frac{i\lam_a}{m_\Phi^2} \hat \chi \quad ; \quad
\hat \chi= e^{\lam_a  |x|^2} G_a(y, \bar x)
\label{aans}
\ee
In this way the D-term equation is satisfied with $\lam_a = \lam_a^-$, defined in (\ref{eigenlam}).
The root $\lam_a^+$ is discarded because it yields zero modes that are not localized on the curve
$x=0$. It remains to solve the F-term (\ref{ancfer}). Inserting the Ansatz for
$\hat \psi_{\bar x}$ and $\hat \chi$ gives an equation for $G_a(\bar x, y)$ that is easily solved
to first order in $\eps$. The final result can be written as 
\be
\hat{\vec{\psi}}_{a^+}^i
\, =\,
\left(\!\!
\begin{array}{c}
-\displaystyle{\frac{i\lam_a}{m_\Phi^2}} \\ 0 \\ 1
\end{array}
\!\! \right)\hat\chi_{a^+}^i
\quad ; \quad
\hat\chi_{a^+}^i\, =\, e^{\lam_a  |x|^2}
\left(f_i(y) - \frac{i\eps\theta}{6}\lam_a^2 \bar x^2 f_i(y)
 -  \frac{i\eps\th}{6} \lam_a \bar x f_i^\prime(y) \right) 
\label{achinc}
\ee
where $f_i(y)$ is an arbitrary holomorphic function, and $i$ labels the different zero modes.

It is interesting to compare the non-commutative wavefunctions with those obtained in appendix
\ref{betawf} within a commutative approach. In sector $a^+$ we can read $\chi_{a^+}^i$ from 
(\ref{finalwfa}) after applying the operators $\CD_x$ and $\CD_y$ as indicated. We find
\be
\chi_{a^+}^i\, =\, e^{\lam_a  |x|^2}
\left\{f_i(y) - \frac{i\eps\theta}{6} \lam_a^2 \bar x^2 f_i(y)
 -  \frac{i\eps\th}{6} \bar x \left[(\lam_a + M) f_i^\prime(y) + M\lam_a \bar y  f_i(y) \right] \right\} 
\label{achic}
\ee
where we used that $\psi_{0,a^+}= e^{\lam_a  |x|^2}\, f_i(y)$. To first order in $\eps$ we see that
the difference with (\ref{achinc}) is given by
\be
\tilde \chi_{a^+}^i=\hat\chi_{a^+}^i - \chi_{a^+}^i=
\frac{i\eps\th}{6} M\left(\bar x \p_y \chi_{a^+}^i + \bar y \p_x \chi_{a^+}^i \right)
\label{chitildea}  
\ee
For $\th$ constant this is precisely the result expected from the map established in (\ref{tildephi}). 
Finally, notice that $\psi_{a^+ \bar x}^i$ and $\hat\psi_{a^+ \bar x}^i$ are also correctly related.

\subsubsection*{Sector $b^+$ \label{ncb}}

In this sector the F-term equations (\ref{ncFfluct7A}) reduce to
\bea
\bar\partial_{\bar m} \hat\chi \ & + & i m_{\Phi}^2 \, y \hat \psi_{\bar m} 
+ \frac{\eps m_{\Phi}^2}{6}[\th - \p_y\th(2x-y+2\Phi_0)] \partial_x \hat\psi_{\bar m} \nonumber \\ 
& + & \frac{\eps m_{\Phi}^2}{6}[2\th + \p_x\th(2x-y+2\Phi_0)] \partial_y \hat\psi_{\bar m}     =  \CO(\e^2)
\label{bncfer}
\eea
for $\bar m = \bar x, \bar y$. The D-term equation is given by
\be
(\p_x+ M_x \bar x) \hat \psi_{\bar x} +  (\p_y+  M_y \bar y) \hat \psi_{\bar y} 
- i  m_{\Phi}^2 \bar y \hat \chi = 0
\label{bncferD}
\ee
The subindex $\a=b^+$ will be omitted until the final result.
Now it is consistent to set $\hat \psi_{\bar x} =0$, which then requires   
$\bar\partial_{\bar x}\hat \psi_{\bar y} =  \bar\partial_{\bar x} \hat\chi=0$.

Our previous results for $\th$ constant suggest the Ansatz
\be
 \hat \psi_{\bar y} = \frac{i\lam_b}{m_\Phi^2} \hat \chi \quad ; \quad
\hat \chi= e^{\lam_b  |y|^2} G_b(x,\bar y)
\label{bans}
\ee
The D-term equation is then verified with $\lam_b = \lam_b^-$, defined in (\ref{simlams}).
The auxiliary function $G_b$ is determined to first order in $\eps$ substituting the Ansatz in the 
F-term (\ref{bncfer}). In the end we obtain 
\be
\hat{\vec{\psi}}_{b^+}^j
\, =\,
\left(\!\!
\begin{array}{c}
0\\ \displaystyle{\frac{i\lam_b}{m_\Phi^2}} \\ 1 
\end{array}
\!\! \right)\hat\chi_{b^+}^j
\quad ; \quad
\hat\chi_{b^+}^j\, =\, e^{\lam_b  |y|^2}
\left(g_j(x) - \frac{i\eps\theta}{6} \lam_b^2 \bar y^2 g_j(x)
 -  \frac{i\eps\th}{6} \lam_b \bar y g_j^\prime(x) \right) 
\label{bchinc}
\ee
with $j$ indexing different zero modes.

In this sector the commutative wavefunction $\chi_{b^+}^j$ obtained from 
(\ref{finalwfb}) reads
\be
\chi_{b^+}^j\, =\, e^{\lam_b  |y|^2}
\left\{g_j(x) - \frac{i\eps\theta}{6} \lam_b^2 \bar y^2 g_j(x)
 -  \frac{i\eps\th}{6} \bar y \left[(\lam_b + M) g_j^\prime(x) + M\lam_b \bar x g_j(x) \right] \right\} 
\label{bchic}
\ee
where we took $\psi_{0,b^+}= e^{\lam_b  |y|^2}\, g_j(x)$. It is easy to check that $\hat\chi_{b^+}^j$
is recovered after adding $\tilde\chi_{b^+}^j$ computed according to (\ref{tildephi}).

\subsubsection*{Sector $c^+$ \label{ncc}}

Substituting the fluctuations and the vevs in (\ref{ncFfluct7A}) yields the F-term equations
\bea
\bar\p_{\bar m} \hat\chi \ & + & i m_{\Phi}^2 \, (x-y) \hat \psi_{\bar m} 
+ \frac{\eps  m_{\Phi}^2}{6}[\th + \p_y\th(x+y-2\Phi_0)] \p_x \hat\psi_{\bar m} \nonumber \\
& - & \frac{\eps  m_{\Phi}^2}{6}[\th + \p_x\th(x+y-2\Phi_0)] \p_y \hat\psi_{\bar m}     =  \CO(\e^2)
\label{cncfer}
\eea
for $\bar m = \bar x, \bar y$.
As before the subindex $\a=c^+$ is dropped.
The D-term equation takes the simple form
\be
\p_x  \hat \psi_{\bar x} +  \p_y \hat \psi_{\bar y} 
- i  m_{\Phi}^2 (\bar x - \bar y) \hat \chi = 0
\label{cncferD}
\ee
In this case the condition $\hat \psi_{\bar x}= -\hat \psi_{\bar y}$ can be imposed 
consistently with (\ref{ncFfluct7phi}).

For $\th$ constant the Ansatz inspired by known results is now  
\be
 \hat \psi_{\bar x}= -\hat \psi_{\bar y} = \frac{i\lam_c}{m_\Phi^2} \hat \chi \quad ; \quad
\hat \chi= e^{\lam_c  |x-y|^2} G_c(\bar x, \bar y)
\label{cans}
\ee
Inserting  the Ansatz in the equations shows that it works with $\lam_c$ given in (\ref{simlams}).
Summarizing we obtain
\be
\hat{\vec{\psi}}_{c^+}
\, =\,
\left(\!\!
\begin{array}{c}
 \displaystyle{\frac{i\lam_c}{m_\Phi^2}}\\[2mm]
 -\displaystyle{\frac{i\lam_c}{m_\Phi^2}} \\ 1 
\end{array}
\!\! \right)\hat\chi_{c^+}
\quad ; \quad
\hat\chi_{c^+}\, =\, \g_c m_* \, e^{\lam_c  |x-y|^2}
\left(1 - \frac{i\eps\theta}{6} \lam_c^2 (\bar x - \bar y)^2 \right) 
\label{cchinc}
\ee
To determine $G_c$ we assumed that it is constant at lowest order in $\eps$. The reason is that
at $\Sigma_c$ there is only one set of zero modes corresponding to the Higgs. The constant is taken to 
be $\g_c m_*$, where $\g_c$ is an adimensional normalization factor.

The commutative wavefunction $\chi_{c^+}$ computed from (\ref{finalwfc}), using 
$\psi_{0,c^+}= \g_c m_* e^{\lam_c  |x-y|^2}$, is given by
\be
\chi_{c^+}\, =\, \g_c m_* e^{\lam_c  |x-y|^2}
\left[1 - \frac{2i\eps\theta}{3} \lam_c^2 (\bar x - \bar y)^2 
 \left(\frac14 + \frac{M}{2\lam_c} \right) \right] 
\label{cchic}
\ee
There is again perfect matching with $\hat\chi_{c^+}$ after the map (\ref{tildephi}).

\subsubsection{Zero modes with $\th$ coordinate dependent \label{ncthvar}}

As we will see, with $\th$ constant it is not possible to generate a realistic
pattern for the Yukawa couplings. This motivates us to consider a more general case
with $\th$ coordinate dependent. We have been able to find the zero modes in closed form
when $\th$ is the linear function
\be
\th= \th_0 + \th_1 x + \th_2 y
\label{thfun}
\ee
where the $\th_\ell$ are constants. In this case it is no longer consistent to make
an Ansatz such as (\ref{cans}) in which the non-zero $\hat \psi_{\a\bar m}$ are proportional to $\hat \chi_\a$.

To illustrate the strategy that works for the linear $\th$ let us focus in the $a^+$ sector. 
Again it is allowed to take $\hat \psi_{a^+\bar y}=0$. The new Ansatz for the non-trivial
fluctuations consists of first setting
\be
\hat \psi_{a^+\bar x}=  -\frac{i\lam_a}{m_\Phi^2} e^{\lam_a  |x|^2} H_a(y,x,\bar x)
\label{newaansA}
\ee 
and then solving for $\hat \chi_{a^+}$ from the D-term equation (\ref{ancferD}).
The constant $\lam_a$ is again equal to the $\lam_a^-$ defined in (\ref{eigenlam}). It thus follows that
\be
\hat \chi_{a^+}=  e^{\lam_a  |x|^2} \left(H_a + \frac{\lam_a}{m_\Phi^4 \bar x} \p_x H_a\right)
\label{newaansB}
\ee 
We also impose the condition that $\hat \chi_{a^+} \to f_i(y)$ when $x \to 0$. To determine the function
$H_a$ we substitute the above $\hat \psi_{a^+\bar x}$ and $\hat \chi_{a^+}$ into the F-term (\ref{ancfer})
and solve to first order in $\eps$. In this way we find
\bea
H_a & = & f_i(y) + \eps\th_0 \bar x \a_0 + \eps\th_1\left(\bar x \a_1 + 
\frac{i\lam_a^2}{3(\lam_a^2 + m_\Phi^4)}f_i^\prime(y)\right) + \eps\th_2 \bar x \a_2 \nonumber \\
\a_0 & = &  - \frac{i}{6} \lam_a^2 \bar x f_i(y) -  \frac{i}{6} \lam_a f_i^\prime(y)
\label{hasol}\\
\a_1 & = & \frac{i\lam_a^2}{3(\lam_a^2 + 2m_\Phi^4)}
(\lam_a - m_\Phi^4 x\bar x )f_i(y) 
-\frac{i\lam_a}{3}\left(\frac{m_\Phi^4}{(\lam_a^2+ m_\Phi^4)}x- y - \Phi_0 \right)f_i^\prime(y) \nonumber \\ 
\a_2 & = &   - \frac{i\lam_a}{6} y f_i^\prime(y)  - \frac{i \lam_a^2}{6}
\left(\frac{\lam_a - m_\Phi^4 x\bar x}{(\lam_a^2+ 2m_\Phi^4)} + 2y\bar x + \Phi_0 \bar x \right)f_i(y)
\nonumber
\eea
Notice that for $\theta$ constant we recover our previous result (\ref{achinc}). We remark that the
F-term equations are satisfied to $\CO(\e)$ for any $\lam_a$. Hence, we expect the $\lam_a$ dependence
to drop out completely in the computation of Yukawa couplings.

The wavefunctions in the $b^+$ sector can be found in a similar fashion. We start with 
$\hat \psi_{b^+\bar x}=0$, together with
\be
\hat \psi_{b^+\bar y}=  \frac{i\lam_b}{m_\Phi^2} e^{\lam_b  |y|^2} H_b(x,y,\bar y)
\label{newbans}
\ee 
where $\lam_b$ is equal to $\lam_b^-$ defined in (\ref{simlams}).
The corresponding $\hat \chi_{b^+}$ is such that the D-term equation (\ref{bncferD}) is verified and is
required to satisfy $\hat \chi_{b^+} \to g_j(x)$ when $y \to 0$.
The function $H_b$ is determined from the F-term eq.(\ref{bncfer}). To order $\eps$ it is given by
\be
H_b  =  g_j(x) + \eps\th_0 \bar y \b_0 + \eps\th_1 \bar y \b_1 
+ \eps\th_2 \left(\bar y \b_2 + \frac{i\lam_b^2}{3(\lam_b^2 + m_\Phi^4)}g_j^\prime(x)\right) 
\label{hbsol}
\ee
The $\b_\ell$ are obtained from the $\a_\ell$ in eq.(\ref{hasol}) as $\b_0=\a_0$, $\b_1=\a_2$,
and $\b_2 = \a_1$, upon the exchanges $x \to y$, $\lam_a \to \lam_b$, and $f_i(y) \to g_j(x)$.
The F-term equations do not constrain the value of $\lam_b$.

In the $c^+$ sector we take
\be
\hat \psi_{c^+\bar x}=  \frac{i\lam_c}{m_\Phi^2} e^{\lam_c  |x-y|^2} H_c(x,\bar x, y,\bar y) = 
- \hat \psi_{c^+\bar y}
\label{newcans}
\ee 
where $\lam_c=-m_\Phi^2/\sqrt2$. {} From the D-term equation (\ref{cncferD}) 
$\hat \chi_{c^+}$ is then determined to be
\be
\hat \chi_{c^+}=  e^{\lam_c  |x-y|^2} \left(H_c 
+ \frac{\lam_c}{m_\Phi^4}\frac{ \p_x H_c - \p_y H_c}{\bar x - \bar y}\right)
\label{newcansB}
\ee
For $H_c$ we make the Ansatz
\be
H_c  =  m_* \g_c \left[ 1 + \eps(\bar x-\bar y)(\th_0 \nu_0 + \th_1\nu_1 + \th_2\nu_2)\right] 
\label{hcsol}
\ee
and deduce the $\nu_\ell$ substituting in the F-term eqs.(\ref{cncfer}). This procedure yields
\bea
\nu_0 & = &  - \frac{i}{6} \lam_c^2 (\bar x-\bar y)  
\label{nusols}\\
\nu_1 & = & \frac{i\lam_c^2}{12(\lam_c^2 + m_\Phi^4)} \left\{ 2\lam_c   
-(\bar x-\bar y)\left[(2\lam_c^2 + 3m_\Phi^4)x + (2\lam_c^2 + m_\Phi^4)y -2(\lam_c^2 + m_\Phi^4)\Phi_0\right] \right\}    
\nonumber \\
\nu_2 & = & -\frac{i\lam_c^2}{12(\lam_c^2 + m_\Phi^4)} \left\{ 2\lam_c   
+(\bar x-\bar y)\left[(2\lam_c^2 + 3m_\Phi^4)x + (2\lam_c^2 + m_\Phi^4)y -2(\lam_c^2 + m_\Phi^4)\Phi_0\right] \right\}    
\nonumber
\eea
These solutions can be simplified inserting the actual value of $\lam_c$. However, the F-term equations are satisfied
for generic $\lam_c$.

\subsection{Non-commutative Yukawa couplings \label{ncyuks}}

The starting point is the $\eps$ expansion of the trilinear superpotential $\hat W_{\rm Yuk}$
given in eq.(\ref{exphatw}).
To compute the couplings we further need to substitute the zero modes collected in the vector
$\hat{\vec{\psi}}_\a$. {}From $\hat W_0$ there is a coupling $Y_0$ that is given by 
(\ref{yukawa7}) replacing $\vec{\psi}_\a$ by $\hat{\vec{\psi}}_\a$. {} From $\hat W_1$ in (\ref{ncsupy3})
we obtain a contribution
\bea
(Y_1)_{abc}^{ijk} \!\! & = & \!\!\! -m_* d_{abc} 
\int_{S} \th  \left(\p_x \hat \psi_{a\bar x}^i \,\p_y \hat \psi_{b \bar y}^j 
-\p_y \hat \psi_{a\bar x}^i \,\p_x \hat \psi_{b \bar y}^j  
- \p_x \hat \psi_{a\bar y}^i \,\p_y \hat \psi_{b \bar x}^j 
+ \p_y \hat \psi_{a\bar y}^i \,\p_x \hat \psi_{b \bar x}^j \right)\hat\chi_c^k \, d{\rm vol}_S
\nonumber \\
& + & {\rm cyclic \ permutations \ in \ } a, \, b, \, c
\label{ncy1}
\eea 
Both pieces $Y_0$ and $Y_1$ have an $\eps$ expansion since the zero modes can be written as  
\mbox{$\hat{\vec{\psi}} = \hat{\vec{\psi}}^{(0)} + \eps \hat{\vec{\psi}}^{(1)} + \CO(\e^2)$}.
Notice that at zeroth order the non-commutative and  commutative wavefunctions do coincide and in fact
$\hat{\vec{\psi}}^{(0)} =  \vec{\psi}^{(0)}$. Taking into account the expansion of $\hat W_{\rm Yuk}$ we see that 
indeed the Yukawa couplings have the schematic structure
\be
Y = Y_0^{(0)} + \eps\left( Y_0^{(1)}+ Y_1^{(0)}\right) +  \CO(\e^2) 
\ee
where we have omitted indices for simplicity.
Here $Y_0^{(0)}$ and $Y_0^{(1)}$ both originate from $Y_0$.
More concretely,  $Y_0^{(0)}$ is computed from (\ref{yukawa7}) replacing $\vec{\psi}_\a^i$
by $(\vec{\psi}_\a^i)^{(0)}$, whereas  $Y_0^{(1)}$ is given by (\ref{yukawa7np1}) 
with  $\vec{\psi}_\a^i$ replaced by $\hat{\vec{\psi}}_\a^i$. 
On the other hand, the $\CO(\e)$ contribution $Y_1^{(0)}$ is computed inserting the uncorrected
wavefunctions $(\vec{\psi}_\a^i)^{(0)}$ in (\ref{ncy1}).

\subsubsection*{Corrected Yukawa couplings in the $U(3)$ model \label{ncyuku3}}

The calculation of the Yukawa couplings corrected to $\CO(\e)$ can be implemented in the $U(3)$ model
using the wavefunctions determined in section \ref{ncu3}. 
As before, we take the {\it Higgs} to  arise from the curve $\Sigma_c$, whereas the {\it quark and lepton 
families}  come from the curves $\Sigma_a$ and $\Sigma_b$, and are indexed by $i$ and $j$ 
respectively. The full Yukawa couplings are denoted by $Y^{ij}$, and the two types of
contributions by $Y_0^{ij}$ and $Y_1^{ij}$. For the holomorphic functions appearing in 
the wavefunctions we again adopt the basis of \cite{hv08} in which 
$f_i(y)= \g_{ai} m_*^{4-i} y^{3-i}$ and $g_j(x)= \g_{bj} m_*^{4-j} x^{3-j}$, with $i,j=1,2,3$. 
The normalization factors $\g_{ai}$ and $\g_{bj}$ will be specified later.

We will first determine the couplings when $\th$ is constant.
The calculation simplifies considerably because the wavefunctions $\hat\psi_{\a \bar m}$ are either
zero or are proportional to $\hat \chi_\a$ as shown in equations (\ref{aans}), (\ref{bans}) and (\ref{cans}).
From section \ref{yuky} we already know that $(Y_0^{(0)})^{ij}$ is zero for $i\not=3, j\not=3$, and
$(Y_0^{(0)})^{33}=-i\pi^2 \g_{a3}\g_{b3} \g_c m_*^4/m_\Phi^4$. We will then concentrate on the pieces  
$Y_0^{(1)}$ and $Y_1^{(0)}$ for which we can write down explicit expressions using the properties
of the wavefunctions derived in section \ref{ncu3}.

Inserting the known wavefunctions in (\ref{yukawa7np1}) we obtain the  $\CO(\e)$ contribution from $\hat W_0$ 
\bea
(Y_0^{(1)})^{ij} & = & -\frac{\th m_*^2 \g_c}{6 m_\Phi^4}(\lam_a\lam_b + \lam_a\lam_c + \lam_b\lam_c)\, I_0^{ij} 
\nonumber \\
I_0^{ij} & = & \int_S \bigg\{f_i(y) g_j(x)\left[\lam_c^2(\bar x- \bar y)^2 
+ \lam_a^2{\bar x}^2 +\lam_b^2{\bar y}^2\right]  \label{yuk01} \\[2mm]
& + & \left[ \lam_a \bar x f_i^\prime(y) g_j(x) + \lam_b \bar y f_i(y) g_j^\prime(x) \right]\bigg\}
 e^{\lam_a  |x|^2 + \lam_b  |y|^2 + \lam_c  |x-y|^2} \, d{\rm vol}_S
\nonumber
\eea 
Substituting in (\ref{ncy1}) yields the $\CO(\e)$ contribution from $\hat W_1$  
\bea
(Y_1^{(0)})^{ij} & = & -\frac{m_*^2 \g_c}{2 m_\Phi^4}I_1^{ij} 
\nonumber \\
I_1^{ij} & = & \int_S \th \bigg\{ 
\lam_c^2 (\bar x- \bar y)\left[\left(\lam_a^2 \bar x - \lam_b^2 \bar y\right)f_i(y) g_j(x)  
+\lam_a f_i^\prime(y) g_j(x) - \lam_b f_i(y) g_j^\prime(x) \right]
\nonumber \\[2mm]
& + &  \lam_a  \lam_b \left[\lam_a \lam_b \bar x \bar y f_i(y) g_j(x) 
- f_i^\prime(y)  g_j^\prime(x) \right] 
\bigg\} e^{\lam_a  |x|^2 + \lam_b  |y|^2 + \lam_c  |x-y|^2} \, d{\rm vol}_S
\label{yuk10}
\eea 
This formula is also valid when $\th$ is coordinate dependent. Here we have used that in the $U(3)$ model
$d_{a^+ b^+ c^+} =  \str(\mathfrak{t}_{a^+},\mathfrak{t}_{b+},\mathfrak{t}_{c^+}) = \frac12$.

The coupling $Y^{33}$ does not receive $\CO(\e)$ corrections. 
Indeed, the integrals appearing in $(Y_0^{(1)})^{33}$ and $(Y_1^{(0)})^{33}$ vanish because the product
of wavefunctions in the integrand is not invariant under the diagonal $U(1)$ rotation 
$x \to e^{i\a} x$ and $y \to e^{i\a} y$. By the same token we also conclude that for constant $\th$ only the 
couplings $Y^{22}$, $Y^{31}$, and  $Y^{13}$ can be different from zero. 

To evaluate the integrals we extend $|x|$ and $|y|$ to infinity as we did in absence of $\theta$ corrections.
After a simple calculation we obtain the couplings
\be
Y^{22}= \frac{\eps \pi^2 \th m_*^6}{3 m_\Phi^4}\g_{a2}\g_{b2} \g_c \quad ; \quad
Y^{31} = -\frac{\eps \pi^2 \th m_*^6}{3 m_\Phi^4}\g_{a3}\g_{b1} \g_c  \quad ; \quad
Y^{31} = -\frac{\eps \pi^2 \th m_*^6}{3 m_\Phi^4}\g_{a1}\g_{b3} \g_c 
\label{thconstresu}
\ee
We find exactly the same couplings following the commutative approach,
using the wavefunctions computed in appendix B.
Observe that, up to normalization, the couplings are independent of worldvolume flux,
in agreement with the general result of \cite{cchv09}.

We have also found the couplings for perturbations linear in the local coordinates, i.e.
\be
\th  \ = \ 3i(\th_0 \ +\  \th_1 x \  + \ \th_2 y)
\label{lineartheta}
\ee
where the $3i$ factor is added to simplify the results.
Inserting the wavefunctions 
given in section \ref{ncthvar} and evaluating the integrals leads to the Yukawa matrix
\be
\frac{Y}{Y^{33}} = 
\left(
\begin{array}{ccc}
\CO(\e^2)   &   \CO(\e^2)   &   \eps m_*^2 \frac { \g_{a1}}{\g_{a3}}(\th_0+ \theta_1 \Phi_0)\\
\CO(\e^2)   &  \eps m_*^2 \frac { \g_{a2} \g_{b2} } {\g_{a3}\g_{b3}}  [(\theta_1+\theta_2)\Phi_0-\theta_0]  
&  \eps m_* \frac {  \g_{a2}}{\g_{a3}}\theta_2 \\
\eps m_*^2 \frac { \g_{b1}}{\g_{b3}}(\theta_0+\theta_2\Phi_0) &  \eps m_*  \frac { \g_{b2}}{\g_{b3}}\theta_1 &  1
\end{array}
\right) 
\label{yukasmm}
\ee
As expected, up to normalization the couplings are independent of worldvolume fluxes.
We have obtained the same results using the residue techniques of \cite{cchv09} where a similar
calculation was performed but only in the case $\th_0=0$ and $\th_1=-\th_2$.
In the next section we will discuss  possible  phenomenological implications of this type of  Yukawa structure.

\section{Flux dependence, D-terms and Yukawa couplings\label{norms}}

\subsection{The $Y(D)=Y(L)$ problem}

It is interesting to see to what extent a simplified model like the $U(3)$ example above  with constant fluxes 
and linearly dependent perturbations can display a hierarchical structure which could be useful in
a more realistic local $SU(5)$ F-theory GUT. Setting $\g_{ai}=\g_{bi}=1$, one observes that the
Yukawa matrix eq.(\ref{yukasmm}) has eigenvalues of order $1,\epsilon , \epsilon^2$ and hence 
has a promising structure to generate such mass hierarchies.  However the Yukawa couplings 
computed above are flux independent. In an  $SU(5)$ local GUT the Yukawa couplings of 
D-quark and leptons come from  couplings ${\bf 10}\times {\bf {\overline 5}}\times {\bf {\overline 5}_H}$
and before  the addition of hypercharge fluxes one has identical Yukawa couplings for 
D-quarks and leptons of all three generations, i.e., $Y^{ij}(D)=Y^{ij}(L)$.
Since  eq.(\ref{yukasmm}) is independent of  fluxes this equality will persist even after the addition
of hypercharge fluxes.  These equalities are however not consistent with the measured
quark and lepton masses of the first two generations.\footnote{After renormalization 
group corrections  one finds $m_b(M_Z)/m_\tau(M_Z)\simeq$
$(\alpha_3(M_Z)/\alpha_3(M_*))^{\gamma /2b_3}\simeq (\alpha_3(M_Z)/\alpha_3(M_*))^{8/9}\simeq 2.5$,
in reasonable agreement with experiment.
However for the lightest two generations such a prediction fails. 
Experimental results are better described by boundary conditions 
$m_\mu \simeq 3m_s$ and $m_e \simeq m_d/3 $ at the unification scale.
In standard GUT's getting these boundary conditions  requires higher dimensional Higgs 
representations in a ${\bf 45}$ of  $SU(5)$ \cite{GJ} or else non-renormalizable interactions  \cite{EG}  involving the
adjoint Higgs, i.e. ${\bf 10}\times \Phi _{{\bf 24}}^n\times {\bf {\overline 5}}\times {\bf {\overline 5}}_H$,
both options using  additional,  poorly motivated,  flavour symmetries.}
This  seems to be a  general problem of 
Yukawa couplings in  local F-theory GUT models, and is ultimately connected to the existence of the
holomorphic gauge discussed in previous sections. Using this holomorphic gauge the fluxes
disappear from the local F-term equations and the resulting holomorphic Yukawas are flux independent.

\subsection{Normalization and flux dependence in physical couplings}

There is however a relevant point we did not take into account yet. 
The wave functions we have used to compute the Yukawa couplings were not
normalized as they should if one is to compare with physical quantities. 
To see the effect of normalization 
it will be sufficient for our purposes to compute the normalization of the
wave functions in the absence of the perturbations. 
 Let us consider the $a$ sector and to simplify let us  impose the BPS condition on the fluxes, i.e. $M_x=M$
and $M_y=-M$. We take $M<0$ so that the normalizable zero modes are in the $a^+$ sector. 
We need to use the wave functions in real gauge as given in eq.(\ref{zmreala}), i.e.
\be
\vec{\psi}_{a\, i}^{ {\rm real}}
\, =\,
\left(\!\!
\begin{array}{c}
-\displaystyle{\frac{i\lam_a}{m_\Phi^2}} \\ 0 \\ 1
\end{array}
\!\! \right)\
\chi^{{\rm real}}_{a\, i} 
\label{realveca}
\ee
where $\lam_a=\lam_a^-$ is defined in eq. (2.24) 
and the real scalar wavefunction is 
\be
\chi^{{\rm real}}_{a\, i} = e^{- \sqrt{\left( \frac{M}{2}\right)^2 + m_\Phi^4}\,  |x|^2} e^{-\frac{|M|}{2} |y|^2} f_i(y) \ .
\label{chireala}
\ee
In the real gauge an exponential suppression along the  $y$ coordinate is made explicit. 
The wave function is only sensitive to local physics and 
one can  normalize
the wave function without the addition  of any volume dependent cut-off.
To normalize the states we perform the integration
\be
||\chi^{{\rm real}}_{a \, i}||^2 = 
 \int_{S}\, |\chi^{{\rm real}}_{a \, i}|^2
\, d{\rm vol}_S\ =\  
|\g_i|^2 \frac{\pi^2 (3-i) !} {2 \sqrt{\left( \frac{M}{2}\right)^2 + m_\Phi^4} \, \big(|M|/m_*^2\big)^{4-i}} \ ,
\label{chinorm}
\ee
where we have extended  the integration 
along  $|x|$ and $|y|$ to infinite radius as in the calculation of Yukawa couplings. 
The normalization condition amounts to imposing 
\be
\langle \vec{\psi}_{a \, i}^{\rm real} | \vec{\psi}_{a \, j}^{\rm real} \rangle  =
\,  m_*^2 \int_S \tr \,( \vec{\psi}_{a \, i}^{\rm real} \cdot \vec{\psi}_{a\, j}^{\, \dag\, {\rm real}})\, {\rm d vol}_S\, =
\, \d_{ij}
\label{normcond}
\ee 
where the $\d_{ij}$ structure arises because the exponentials in the wavefunctions, as well as the measure, are invariant under the 
diagonal $U(1)$ rotation $x \to e^{i\theta} x$ and $y \to e^{i\theta} y$.
Upon normalization we get
\be
\g_{ai}^2\ =\ \frac {( |M|/m_*^2)^{3-i}}{(3-i)!} \times  \CN_{a}^{-1}
\ \ ,\ \ 
\CN_{a}\ =\  \frac {\pi^2m_*^4}{2m_{\Phi}^4} (1+\sqrt{1+ \frac {4m_{\Phi}^4}{|M|^2}}) \ ,
\label{gamillas}
\ee
where $\CN_{a}$ is generation independent. Note that in the dilute flux limit 
$m_*,m_{\Phi}\gg |M|$ one has $ \CN_{a}\simeq  m_*^4 / (m_{\Phi}^2|M|)$.
If the flux scales like $|M| \simeq 1/R^2$ this correspond to a metric for the matter fields
$K_{ii}\simeq 1/R^2$. This  is analogous to the  modulus behavior of of the metrics of perturbative intersecting 
D7-brane models, see e.g. refs. \cite{cremades,aparicio}.

Similar results are obtained for the matter fields in curve $b$. In the case of the curve $c$ with the 
flux choice in section 2 there is  no flux along the curve. In this case the exponential
damping along the curve  is missing and one has to take a volume dependent cut-off in order to normalize the
wave function. However, we will not need to do so because in the Yukawa coupling ratios in eq.(\ref{yukasmm}) 
the normalization of the c curve cancels out.
Plugging these values for $\g_{ai},\g_{bi}$ in eq.(\ref{yukasmm}) one obtains a matrix
\be
\frac{Y}{Y^{33}} = 
\left(
\begin{array}{ccc}
\CO(\epsilon ^2)    &   \CO(\epsilon ^2)    &   \frac{ |M| }{\sqrt{2}} \eps (\th_0+ \theta_1 \Phi_0)\\
\CO(\epsilon ^2)    &   |M| \eps [(\theta_1+\theta_2)\Phi_0-\theta_0]  
&  |M|^{1/2}  \eps \theta_2 \\
\frac{|M| }{\sqrt{2}}\eps (\theta_0+\theta_2\Phi_0) &    |M|^{1/2} \eps \theta_1 &  1
\end{array}
\right) \ . 
\label{yukasmf}
\ee
As we can see, the physical Yukawa couplings now  depend on the fluxes. Interestingly enough  these 
physical couplings only depend on the dimensionless combinations $(\epsilon\theta_0 |M|), (\epsilon\theta_{1,2}\Phi_0 |M|)$ and $(\epsilon\theta_{1,2}|M|^{1/2}) $.
In other words, there is an invariance under the rescalings
\be
(\epsilon\theta_0, \epsilon\theta_{1,2} ,   \Phi_0 , M)\ \rightarrow \   (\lambda ^2\epsilon\theta_0, \lambda \epsilon\theta_{1,2} ,   \lambda\Phi_0 , \lambda ^{-2}M)
\label{scalings}
\ee
and there is  no explicit dependence on the ``stringy'' scales $m_*,m_{\Phi}$. This is as it should since
Yukawa couplings in type IIB string compactifications may be computed in the large volume limit
just in terms of the compactified 10d field theory, without explicit reference to any string theoretical
scale. Note that, had we included $\theta$ corrections to the wave functions in 
the normalization we would have obtained corrections for each entry of higher order in $(\epsilon\theta_0 M), (\epsilon\theta_i\Phi_0|M|)$, etc.

\subsection{A model of quark-lepton hierarchies}

We can take the $U(3)\rightarrow U(1)$ model as a toy model for the  $SO(12)\rightarrow SU(5)$  and 
$E_6\rightarrow SU(5)$  symmetry structure underlying the ${\bf 10}\times {\bf {\overline 5}}\times {\bf {\overline 5}_H}$
and ${\bf 10}\times {\bf {{10}}}\times {\bf { 5}_H}$ Yukawas in a local $SU(5)$ GUT. 
In F-theory $SU(5)$ GUT's one breaks the gauge symmetry down to the SM by the addition of
$F_Y$ fluxes along the hypercharge direction. The matter localized at the curves then feel both the flux in the
curves and the hypercharge flux in the bulk  $S$.  Although the {\it net}  hypercharge  flux on the matter curves associated to
quark and leptons is assumed to vanish (so that the SM family spectrum is not spoiled), the value of the hypercharge flux
density  at the intersection Yukawa point is in general non-vanishing. This effect can then make the physical Yukawa
couplings sensitive to the hypercharge flux.  One can model out this situation by taking the above wave functions 
with constant fluxes and making the replacement
$M =M_0  +YM_1$ 
in the physical Yukawa couplings in eq.(\ref{yukasmf}). Here $M_0$, $M_1$ are the curve flux and the bulk hypercharge flux
respectively, and $Y=1,-4,2,-3,6$ are the hypercharges  of the SM fields $Q_L,U_R,D_R,L,E_R$ respectively, in standard notation.
This may be a good approximation to the extent that the local flux in the vicinity of the Yukawa point may be 
slowly varying. This might be the case recalling that the physical Yukawa coupling and the normalized wave functions are
only sensitive to backgrounds in the vicinity of this Yukawa point. 
Left- and right-handed fermions in a given coupling have different hypercharges $Y_{L,R}$ so that 
one finds that the ratio of physical Yukawa couplings have a hypercharge dependence of the form
{ 
\beq
\frac{Y}{Y^{33}} = 
\left(
\begin{array}{ccc}
  \CO(\epsilon^2)   &   \CO(\epsilon^2)  &    (\theta_0+ \theta_1\Phi_0)\frac {\epsilon}{\sqrt{2}}  |M_R|\\
   \CO(\epsilon^2)    &    \epsilon [(\theta_1+\theta_2)\Phi_0-\theta_0]   |M_LM_R|^{1/2}    &  \epsilon \theta_2  |M_R|^{1/2} \\
(\theta_0+\theta_2\Phi_0)  \frac {\epsilon}{\sqrt{2}}  |M_L|\ &  \epsilon \theta_1 |M_L|^{1/2} &  1
\end{array}
\right)  \ ,
\label{yukasmmflux} 
\eeq }
where
\be
M_L \ =\ M_0\ +\ Y_L M_1 \ \ ,\ \ 
M_R\ =\ M_0 \ +\ Y_RM_1  .
\ee
Note that the leading QCD low energy running for $D$-quarks cancels out in this  matrix, so that it is easier the
comparison with the leptonic hierarchies.
It is interesting to check numerically whether this kind of structure  is able to describe the observed
hierarchy of fermion masses and their mixing.  Since the above matrix is not hermitian it is simpler 
to compute the eigenvalues and eigenvectors of its product by its adjoint and take the square root.
In  $SU(5)$ GUT's the right handed $D$-quarks  live in a ${\bf {\overline 5}}$ matter curve
whereas the left-handed quarks $Q_L$ live in a ${\bf { {10}}}$. For the leptons the 
opposite happens, left-handed leptons $L$ live in the ${\bf {\overline 5}}$ and the 
right-handed leptons in the ${\bf { {10}}}$. This means that when going from a 
$D$-quark Yukawa matrix to a lepton matrix we have to interchange $Y_R,M_R\leftrightarrow Y_L,M_L$. 
In the case of $U$-quark masses Yukawa couplings come from an intersection 
${\bf 10}\times {\bf { {10}}}\times {\bf { 5}_H}$, both left and right $U$-quarks are in a
${\bf 10}$ and one has to symmetrize the Yukawa coupling. Note also that in this case, as noted in 
\cite{bhv2} the 10-plet matter curve must self-pinch  or rather both 10-plet branches must be related by
some discrete symmetry \cite{hktw} in order to be able to get eventually rank=3 matrices.
We will further assume that there is a single intersection point of matter curves for each of the
two types of Yukawa couplings.

We have looked for values of the six parameters $\epsilon\theta_0, \epsilon\theta_1, \epsilon\theta_2, \Phi_0 , M_0, M_1$  (with all parameters real
for simplicity)  able to reproduce the observed fermion hierarchies and mixings. Note that in a realistic setting  these sets of
parameters are different for the $D/L$ and the $U$ physical Yukawa couplings since the
corresponding intersection points are in general different in a $SU(5)$ local GUT.
We will compare the results with 
the {\it observed } ratios of physical Yukawa couplings all evaluated at a scale of order the electroweak scale.
One has for those  (see e.g. \cite{Fritzsch} )
\begin{equation}
\begin{array}{lcl}
\left(\frac{Y_1}{Y_3}\right)_U = (0.5-1.6)\cdot10^{-5} &  \mathrm{and}  & \left(\frac{Y_2}{Y_3}\right)_U = (3-4)\cdot10^{-3}\\
\left(\frac{Y_1}{Y_3}\right)_D = (0.6-1.8)\cdot10^{-3} & \mathrm{and} & \left(\frac{Y_2}{Y_3}\right)_D = (1-3)\cdot10^{-2}\\
\left(\frac{Y_1}{Y_3}\right)_L = (2.8)\cdot10^{-4} & \mathrm{and} & \left(\frac{Y_2}{Y_3}\right)_L = (5.9)\cdot10^{-2}.
\end{array}
\end{equation}
The experimental CKM mixing matrix with 90\%  CL is (see \cite{PDB})
\begin{equation}
|V_{CKM}|  = \left(\begin{array}{ccc}
0.9741-0.9756 & 0.219-0.226  & 0.0025-0.0048 \\
 0.219-0.226 & 0.9732-0.9748  & 0.038-0.044\\
 0.004-0.014 & 0.037-0.044   & 0.9990-0.9993
\end{array}\right) \qquad 
\end{equation}
As an example take  parameters 
\be
(\epsilon\theta_0, \epsilon\theta_1, \epsilon\theta_2,\Phi_0, M_0, M_1)_{D, L} \  =  \  (-0.066, 0.10, -0.27, 0.5,1.41,-0.31) 
\ee
\be
(\epsilon\theta_0, \epsilon\theta_1, \epsilon\theta_2, \Phi_0, M_0, M_1)_U \  =  \  (-0.033, -0.27, -0.33, -0.1, -1.1, -0.47) \ .
\ee
for the backgrounds at the $D/L$ and $U$ Yukawa intersecting points respectively.
One then obtains mass  ratios
\begin{equation}
 \begin{array}{lcl}
  (m_1,\ m_2,\ m_3)_U & = & (6.9\cdot10^{-5},\ 3.8\cdot10^{-3},\ 1)\\
  (m_1,\ m_2,\ m_3)_D & = & (0.65\cdot10^{-3},\ 2.96\cdot10^{-2},\ 1)\\
  (m_1,\ m_2,\ m_3)_L & = & (4.8\cdot10^{-4},\ 5.3\cdot10^{-2},\ 1)
 \end{array}
\end{equation}
with a CKM mixing matrix
\begin{equation}
V_{CKM}  = \left(\begin{array}{ccc}
0.9834 & 0.1812  & 0.0056\\
 0.1809 & 0.9827  & 0.0382\\
0.0125  & 0.0365   & 0.9992
\end{array}\right)
\end{equation}
The agreement is quite good taking into account the simplicity of the model and the uncertainties.
In particular the first  (smallest) eigenvalues are less reliable since in the computation of the
Yukawa couplings we have neglected effects of order $\epsilon ^2$ which could be relevant for the
masses and mixings of the first generation.  Many other solutions leading to similarly acceptable
results exist.
Note that due to the scale invariance in eq.(\ref{scalings}) 
the same numerical results may be obtained with magnetic fluxes a factor $1/\lambda^2$ smaller
by compensating taking larger values for the rest of the parameters. 
Note also that in the above we have computed ratios of 
physical Yukawa couplings. However, flux effects do also affect the relative size of the third generation 
physical Yukawa couplings. Indeed, from eq.(\ref{gamillas}) one obtains 
\beq
\frac {Y^{33}(L)}{Y^{33}(D)} \ =\  \frac {(\CN_{D} \CN_{Q})^{1/2} }  {(\CN_{L} \CN_{E})^{1/2} } \ .
\eeq
As we saw, in the dilute flux limit with $m_*,m_{\Phi}\gg |M|$ one has $ \CN_{a}\simeq  m_*^4 / (m_{\Phi}^2|M|)$ so 
\be
\frac {Y^{33}(L)}{Y^{33}(D)} \  \simeq \  \frac {( |M_0+Y_LM_1||M_0+Y_EM_1|)^{1/2} }  {( |M_0+Y_DM_1||M_0+Y_QM_1|)^{1/2} }
\ =\  \frac {( |M_0-3M_1||M_0+6M_1|)^{1/2} }  {( |M_0+2M_1||M_0+M_1|)^{1/2} } \ .
\ee
For the above choice of parameters this leads to $Y^{33}(L)/Y^{33}(D)\simeq 1.13$ , so that the (successful) standard
$SU(5)$ prediction $m_b(M_{GUT})=m_{\tau} (M_{GUT})$ is not much distorted.

In summary, although the above estimations are based on the results obtained for a simple $U(3)$ model
with constant magnetic  flux, the lesson seems to be more general.  Non-perturbative effects from distant
7-branes sectors (or, equivalently, local closed string $(1,2)$ fluxes) can give rise to the observed hierarchy of quark and lepton masses. 
On the other hand  turning on hypercharge fluxes on $SU(5)$ models  seems able to explain the difference in masses between the D-quarks and charged leptons of the lightest generations.

\section{Final comments and conclusions\label{conclusions}}

In this paper we have analyzed the dependence of Yukawa couplings on two important ingredients of F-theory GUT models: 7-brane worldvolume fluxes and corrections which are non-perturbative on K\"ahler moduli couplings. The former are essential in building the local F-theory model itself, since they are needed to obtain a 4d chiral spectrum and, in most scenarios, to perform the gauge symmetry breaking $G_{GUT} \raw G_{MSSM}$. The latter can be built independently of the local GUT model, since they take place on extra 7-branes distant from the GUT 4-cycle $S$ (see figure \ref{backr}), but are nevertheless crucial for achieving full moduli stabilization in the type IIB/F-theory context \cite{dk06,revDenef}. 

As shown in \cite{cchv09},\cite{cp09},\cite{fi09}, before such non-perturbative corrections holomorphic Yukawas do not depend on 7-brane worldvolume fluxes,\footnote{For the type IIB limit of F-theory, this can be guessed from the fact that worldvolume fluxes enter into the 7-brane action via their density, which depends on the K\"ahler moduli of the compactification. A dependence of the holomorphic Yukawas on the worldvolume fluxes would then imply a dependence of the superpotential on the K\"ahler moduli, and the latter is forbidden for type IIB compactifications on Calabi-Yau manifolds from general worldsheet arguments \cite{bdlr99}.} which implies that the rank of the matrix of physical Yukawas does not depend on them either. Here we have found that the same result applies in the presence of gaugino condensate/D3-instanton effects, and that only such non-perturbative corrections may modify the number of quark/lepton generations that are massive in a local F-theory model. 

While the above picture is rather simple, it opens up a quite interesting scenario. First, in the context of the $SU(5)$ models described in the introduction, one needs non-perturbative effects in order to give mass to the two lightest generations of quarks and leptons. The resulting Yukawa matrix then displays a clear hierarchical structure, just like (\ref{yukasmm}) for the $U(3)$ toy model that we have analyzed, but is nevertheless independent of the worldvolume fluxes. As the hypercharge flux is the only ingredient that breaks the GUT symmetry down to the MSSM, the D-quark and lepton Yukawas are identical even after the breaking $G_{GUT} \raw G_{MSSM}$, except from the fact that they have different normalization. Taking into account that wavefunction normalization does depend on worldvolume fluxes, we obtain a matrix of physical Yukawas of the form (\ref{yukasmf}). This flux dependence allows to accommodate the observed MSSM mass ratios 
in a simple toy model of quark/lepton hierarchies.

In fact, the above scenario is quite robust, in the sense that it also applies to F-theory models built in supersymmetric backgrounds with IASD (1,2)-fluxes in the three-fold base, dubbed in the literature as $\beta$-deformed backgrounds.\footnote{Actually, such backgrounds have only been built in the type IIB limit. It would be interesting to generalize the concept of $\beta$-deformation to genuine F-theory compactifications.} Indeed, as pointed out in \cite{mm09}, a $\beta$-deformation on the closed string background induces a non-commutative deformation on the GUT 7-brane gauge theory in the sense of \cite{cchv09}, and this deformation is equivalent to the one induced by non-perturbative effects on the 7-brane superpotential. We have elaborated on such equivalence and given an specific dictionary between internal wavefunctions in the non-perturbative and non-commutative scenarios, and verified that the Yukawas in both formalisms are identical to first order in the non-perturbative/non-commutative parameter $\e$. It would be interesting to check whether such equivalence holds to higher orders in $\e$.

For the purposes of this paper the non-commutative formalism has been perhaps more useful than the commutative one, in the sense that the flux independence of the Yukawa couplings is explicit. Nevertheless, the commutative formalism is bound to play a more prominent role when generalizing the computations made here for the $U(3)$ model to more involved F-theory setups. Indeed, as can be deduced from the discussion in section \ref{npei7}, the correction to both the wavefunctions and the Yukawa couplings is not only proportional to the parameter $\e$, but also to the symmetric tensor $d_{abc} = \str(\mathfrak{t}_a, \mathfrak{t}_b, \mathfrak{t}_c)$ of the group of enhanced symmetry $G_p$ at the Yukawa point $p$. While for $U(N)$ Yukawa points $d_{abc}$ is non-zero, the Yukawa points of main interest in F-theory have either $G_p = SO(2N)$ or $E_n$, for which $d_{abc}=0$ and hence the non-perturbative corrections  to Yukawas would seem to vanish identically.

This is however not so, in the following sense. Recall that in order to arrive to the $\beta$-deformed superpotential (\ref{supototal}) in the commutative formalism we had to perform the expansion (\ref{suponp}) on the scalar function $\chi_0$ that defines the non-perturbative effect/$\beta$-deformation. While we based our analysis on the first non-trivial term in the Taylor expansion, we could have considered higher order terms, so that
\be
  \oh \int_S \str  (\chi_0 F \wedge F)\, =\,  \chi_0|_S\,  N_{D3} + \oh \int_S \theta\,  \tr \left( \Phi_{xy} F \wedge F\right) +  \frac{1}{4} \int_S \theta'\,  \str \left( \Phi_{xy}^2 F \wedge F\right) + \dots 
\ee
where again ${\theta} \equiv m_\Phi^{-2} \p_z\chi_0|_S$ and we have defined ${\theta}' \equiv m_\Phi^{-4} \p_z^2\chi_0|_S$. The second term in this expansion vanishes identically whenever $d_{abc} = 0$, but the third one does not. Hence for those cases where, e.g.,  $G_p = SO(2N)$ we will have the total superpotential
\be
W_{\rm total}\, =\, m_*^4\left[  \int_S \tr (\Phi_{xy} F) \wedge dx \wedge dy + \frac{\eps}{4} \int_S {\theta}' \,  \str \left( \Phi_{xy}^2 F \wedge F\right)\right]
\ee
instead of the previous expression (\ref{supototal}).  Naively, for the purpose of computing Yukawa couplings, this new superpotential can be related to the previous one by the replacement $\theta \raw \theta' \langle \Phi_{xy} \rangle$, so one would expect that  the results derived for $U(N)$ groups via (\ref{supototal}) have an analogue for $SO(2N)$ groups in terms of this new superpotential. It is however of obvious interest to generalize the analysis of this paper to include the above superpotential. In particular, it would be interesting to  see whether it can also be understood as a non-commutative deformation of (\ref{supo7}). These and other related issues are currently under investigation. 

\bigskip

\bigskip

\centerline{\bf \large Acknowledgments}

\bigskip

We thank  P.G.~C\'amara, I.~Garc\'{\i}a-Etxebarria, L.~Martucci,  S.~Theisen and A.~Uranga  for useful discussions. 
F.M. thanks the CERN TH group and HKUST-IAS for hospitality during the completion of this paper. 
A.F. thanks the IFT UAM/CSIC, the AEI-Potsdam, as well as the AS ICTP, 
for hospitality and support during progress of this work; 
and CDCH-UCV for a research grant No. PI-03-007127-2008.
This work has been partially supported by the grants FPA 2009-09017, FPA 2009-07908, Consolider-CPAN (CSD2007-00042) from the MICINN, HEPHACOS-S2009/ESP1473 from the C.A. de Madrid and the contract ``UNILHC" PITN-GA-2009-237920 of the European Commission.
F.M. is supported by the MICINN Ram\'on y Cajal programme through the grant RYC-2009-05096.


\appendix

\section{Magnetized D9-branes and KK modes\label{apD9KK}}

The aim of this section is to describe the intersecting 7-brane models discussed in the text from the T-dual vantage point of magnetized D9-branes, which have been extensively studied for the purpose of computing Yukawa couplings \cite{magnus,bblmr05,DiVecchia,akp09,ccd09}. As we will see this D9-brane picture will also allow to describe the spectrum of open string massive modes in a rather natural way. Recall from section \ref{general} that the local intersecting 7-brane configuration considered in the main text are based on a compact divisor $S$ of a threefold $B$, and that Yukawas arise from points of intersection of matter curves $\Sigma_i \subset S$. Near such intersection points we can describe our geometry by the local holomorphic coordinates  $(x,y,z) \in B$, with $(x,y) \in S$, $z$ a transverse coordinate to $S$, and such that the fundamental form of $S$ is given by (\ref{kahlerform}).

To proceed, let us assume that all the 7-branes are in fact D7-branes and that $B$ is elliptically fibered over $S$. We can then perform fiberwise T-duality to a system of magnetized D9-branes on a threefold $\tilde{B}$. While the D7-brane worldvolume fluxes $F_{x\bar{x}}$, etc. remain as magnetic fluxes in the D9-brane picture, the D7-brane field $\Phi$ is translated into the D9-brane magnetic fluxes $F_{x\bar{z}} = D_x\Phi_{xy}$ and $F_{y\bar{z}} = D_y\Phi_{xy}$. Hence, a generic system of intersecting D7-branes, for which $\langle \Phi_{xy} \rangle \not= 0$, translates into a D9-brane system where $F_{x\bar{z}}$ and $F_{y\bar{z}}$ have a non-trivial background value. 

For concreteness, let us consider the $U(3)$ model introduced in subsection \ref{toy}.\footnote{In particular, one may easily embed this $U(3)$ model into a global geometry such as $(\T^2)_x \times (\T^2)_y \times (\T^2)_z$ or orbifolds thereof, so that $S = (\T^2)_x \times (\T^2)_y$ and to arrive to the magnetized D9-brane picture we simply need to perform two T-dualities along $(\T^2)_z$.} After performing two T-dualities along the transverse coordinates $(z,\bar{z})$, one obtains a model with a stack of 3 D9-branes, with a gauge flux $F_{D9}$ given by
\bea
\label{fluxD9a}
\langle F_{D9} \rangle & = & i \left(M_x dx \wedge d\bar{x} + M_y dy \wedge d\bar{y}\right) 
\frac{1}{3}
\left(
\begin{array}{ccc}
1 \\ & -2 \\ & & 1
\end{array}
\right) \\ \nonumber
& & 
+ 
\frac{M_{\Phi_x}}{3}
\left(
\begin{array}{ccc}
-2 \\ & 1 \\ & & 1
\end{array}
\right)
d\bar{x} \wedge d{z} 
+
\frac{M_{\Phi_y}}{3}
\left(
\begin{array}{ccc}
1 \\ & 1 \\ & & -2
\end{array}
\right)
d\bar{y} \wedge d{z} +{\rm c.c.}
\eea
where the first line arises from T-dualizing (\ref{fluxD7}), while the second line is the T-dual version of (\ref{vevPhi}) for $M_{\Phi_x} = m_{\Phi_x}^2$ and  $M_{\Phi_y} = m_{\Phi_y}^2$. In addition, the piece of (\ref{vevPhi}) proportional to $\Phi_0$ will translate into a Wilson line background for $A_{\bar{z}}$, which for simplicity is set to zero in the following.

Similarly to the D7-brane picture, the D9-brane background (\ref{fluxD9a}) breaks the gauge group as $U(3) \raw U(1)^3$. The wavefunctions for open string zero modes can then be computed  by simply solving the Dirac or Laplace equations. One may do so by choosing a gauge for $A$ analogous to (\ref{holg}), which in the D9-brane picture reads
\bea
\label{sholgD9}
\langle A_{D9}\rangle  & = & - \frac{i}{3} \left(M_x \bar{x}dx +  M_y \bar{y}dy \right)
\left(
\begin{array}{ccc}
1 \\ & -2 \\ & & 1
\end{array}
\right)
\\ \nonumber & &
+\frac{2}{3}M_{\Phi_x} \re (\bar{x}dz) 
\left(
\begin{array}{ccc}
-2 \\ & 1 \\ & & 1
\end{array}
\right)
+ \frac{2}{3}M_{\Phi_y} \re(\bar{y}dz)
\left(
\begin{array}{ccc}
1 \\ & 1 \\ & & -2
\end{array}
\right)
\eea
Unlike in the D7-brane case, (\ref{sholgD9}) is not a pure holomorphic gauge because $\langle A_{\bar{z}} \rangle \not= 0$. Note, however, that a purely holomorphic gauge would imply a $z$-dependent $\langle A_{D9} \rangle$, which would prevent to T-dualize this background back to a system of intersecting D7-branes.

Given (\ref{sholgD9}), and a local set of coordinates such that the fundamental form of $\tilde{B}$ reads
\be
\mathcal{J} = \frac{i}{2}\left( dx\wedge d\bar{x} +  dy\wedge d\bar{y} +  dz\wedge d\bar{z} \right).
\label{kahlerformD9}
\ee
it is straightforward to write down the zero mode equation for the D9-brane fermionic degrees of freedom  (see e.g. \cite{magnus,cm09}). Indeed, we may describe them by a negative chirality 10d Majorana-Weyl spinor $\chi$ in the adjoint of $U(3)$, which we decompose as
\beq \chi\, =\, \zeta + \mathcal{B}^* \zeta^* \quad \quad \zeta\,
=\, \chi_4 \otimes \chi_6
\label{splitgaug}
\eeq
with  $\chi_6$ a 6d Weyl spinor of negative chirality,
$\mathcal{B} = \mathcal{B}_4 \otimes \mathcal{B}_6$ a Majorana
matrix and $\chi_4$ is a 4d Weyl spinor of positive chirality satisfying
\begin{equation}
\gamma_{(4)}
\slashed{\p}_{\IR^{1,3}} \mathcal{B}_4^* \chi_4^* = - m_\rho\,
\chi_4
\label{ferm4d}
\end{equation}
where $\g_{(4)}$ is the chirality operator in 4d (see subsection \ref{fconv} for explicit expressions). Then, the 10d Dirac equation $\slashed{D}_{10} \chi = 0$ reduces to the 6d internal KK mode equation
\begin{equation}
 \slashed{D}_{\tilde{B}}  \chi_6 \,
=\,  m_\rho\, \mathcal{B}_6^* \chi_6^*
\label{dirac6d}
\end{equation}
with $\slashed{D}_{\tilde{B}} = \tilde{\g}^m D_m$ the Dirac operator on the internal 6d manifold $\tilde{B}$. Taking the conventions of subsection \ref{fconv}, this last equation can be expressed as
\be
i{\bf D_A} \Psi\, =\, m_\rho \Psi^{\dag}
\label{Dirac9KK}
\ee
with
\beq
\Psi\, =\,
\left(
\begin{array}{c}
\psi^0 \\ \psi^ 1\\ \psi^2 \\ \psi^3
\end{array}
\right) 
\quad\quad\quad\quad
\Psi^{\dag}\, =\,
\left(
\begin{array}{c}
(\psi^0)^\dag \\ (\psi^1)^\dag \\ (\psi^2)^\dag \\ (\psi^3)^\dag
\end{array}
\right) 
\label{fvector}
\eeq
and
\be
{\bf D_A}\, =\, 
\left(
\begin{array}{cccc}
0 & D_x & {D}_y & {D}_z \\
-{D}_x & 0 & -{D}_{\bar{z}} & {D}_{\bar{y}} \\
-{D}_y & {D}_{\bar{z}} & 0 & -{D}_{\bar{x}} \\
-{D}_z & -{D}_{\bar{y}} & {D}_{\bar{x}} & 0
\end{array}
\right)
\label{matrixDirac:ap}
\ee
with $D_m = \p_m - i [A_m, \cdot]$ the usual covariant derivative. Note that in principle (\ref{Dirac9KK}) for $m_\rho = 0$ is not the same zero mode equation as (\ref{Dirac9}), since in (\ref{matrixDirac:ap}) $D_{\bar{z}} = \p_{\bar{z}} - i[A_{\bar{z}}, \cdot]$. However, if we want to T-dualize along the $(z,\bar{z})$ coordinates back to a system of intersecting D7-branes, we need to smooth out our wavefunctions $\psi^m$ along such coordinates, and more precisely we need to impose that $\p_z \Psi = \p_{\bar{z}} \Psi = 0$. Hence, under such assumption the operators ${D}_z$, ${D}_{\bar{z}}$ in (\ref{matrixDirac:ap}) reduce to those in (\ref{matrixDirac}), and so solutions to (\ref{Dirac9KK}) are also solutions to (\ref{Dirac9}) and (\ref{ferm7}).

Finally, applying twice (\ref{Dirac9KK}) we obtain
\be
{\bf D_A}^\dag {\bf D_A}\, \Psi\, =\, |m_\rho|^2 \Psi
\label{eigenferm:ap}
\ee
where
\be
{\bf D_A}^\dag\, =\, 
\left(
\begin{array}{cccc}
0 & {D}_{\bar{x}} & {D}_{\bar{y}} & {D}_{\bar{z}} \\
-{D}_{\bar{x}} & 0 & -{D}_{z} & {D}_{y} \\
-{D}_{\bar{y}} & {D}_{z} & 0 & -{D}_{x} \\
-{D}_{\bar{z}} & -{D}_{y} & {D}_{x} & 0
\end{array}
\right)
\label{Ddag}
\ee
and so by solving such eigenvalue equation we can also obtain the wavefunctions for the D9-brane massive replicas of the previous zero modes. Again, in order to translate such result to a D7-brane wavefunction we need to assume that $\Psi$ is $(z,\bar{z})$-independent.

As in the main text, in order to solve for the most general eigenmode of (\ref{eigenferm:ap}) it is useful to notice that
\be
{\bf D_A}^\dag {\bf D_A}\, =\, - \Delta \Id_4\, - i {\bf F}
\quad \quad \quad 
{\bf F}\, =\,
\left(
\begin{array}{cccc}
\sig_{+++} & {F}_{\bar{y}\bar{z}} & {F}_{\bar{z}\bar{x}} & {F}_{\bar{x}\bar{y}} \\
{F}_{zy} & \sig_{+--} & {F}_{y\bar{x}} & {F}_{z\bar{x}} \\
{F}_{xz} & {F}_{x\bar{y}} & \sig_{-+-} &  {F}_{z\bar{y}} \\
{F}_{yx} & {F}_{x\bar{z}} & {F}_{y\bar{z}} & \sig_{--+}
\end{array}
\right) 
\label{Lapfer:ap}
\ee
where $\Delta$ and $\sigma_{\eps_x\eps_y\eps_z}$ are defined as in (\ref{Lapsym}), (\ref{Lapsig}), respectively. If the entries of {\bf F} are constant, then the operators $\Delta \Id_4$ and {\bf F} commute and one can diagonalize them simultaneously. Indeed, imposing the F-term condition $F^{(0,2)} = 0$, one finds that the solutions for (\ref{eigenferm:ap}) are of the form\footnote{Imposing the D9-brane D-term condition $\tr F^2 \wedge J = 0$ implies that $\sigma_{+++} = 0 $ and so $\lam_1 + \lam_2 + \lam_3 = 0$.}
\begin{subequations}
\label{spec:ap}
\begin{align}
\label{spec:ap1}
 & |m_\rho|^2 = \rho^2 - i \sig_{+++}, \
\Psi =
\left(
\begin{array}{c}
1 \\ 0 \\ 0 \\ 0
\end{array}
\right) \psi_\rho
 &  
|m_\rho|^2 = \rho^2 + \lam_1, \
\Psi\, =
J  \left(
\begin{array}{c}
0 \\ 1 \\ 0 \\ 0
\end{array}
\right) \psi_\rho \\
\label{spec:ap2}
& |m_\rho|^2 = \rho^2 + \lam_2, \
\Psi =
J \left(
\begin{array}{c}
0\\ 0 \\ 1 \\ 0
\end{array}
\right) \psi_\rho
\quad &
|m_\rho|^2 = \rho^2 + \lam_3, \
\Psi = J
\left(
\begin{array}{c}
0\\ 0 \\ 0 \\ 1
\end{array}
\right) \psi_\rho
\end{align}
\end{subequations}
where $-\Delta \psi_\rho = \rho^2 \psi_\rho$, and the $4\times 4$ unitary matrix $J$ is defined by
\be
J\, =\, 
\left(
\begin{array}{cc}
1 & 0 \\ 
0 & j
\end{array}
\right)\quad \quad {\rm such\ that\ } \quad \quad
J^{-1} {\bf F} J \, =\, i
\left(
\begin{array}{cccc}
-i \sig_{+++} \\ 
& \lam_1 \\
& & \lam_2 \\
& & & \lam_3
\end{array}
\right)
\label{defJ}
\ee
In particular, the fermionic zero mode satisfying the Dirac equation (\ref{Dirac9KK}) for $m_{\rho} = 0$ will be of this form. Let us for instance assume that it has of the form of the second solution in (\ref{spec:ap2}). Then it is easy to see that the corresponding Laplace eigenfunction $\psi_0$ is such that $\rho^2 = -\lam_3$ and 
\be
\CD_{\bar{x}} \psi_0\, =\, \CD_{\bar{y}} \psi_0\, =\, \CD_z \psi_0\, =\, 0
\label{annihil:ap}
\ee
where the operators $\CD_m$ are linear combinations of the  covariant derivatives ${D}_m$ defined by 
\be
\left(
\begin{array}{c}
\CD_x \\ \CD_y \\ \CD_z
\end{array}
\right)
\, =\, j^t 
\left(
\begin{array}{c}
{D}_x \\ {D}_y \\ {D}_z
\end{array}
\right)
\quad \quad {\rm and} \quad \quad
\left(
\begin{array}{c}
\CD_{\bar{x}} \\ \CD_{\bar{y}} \\ \CD_{\bar{z}}
\end{array}
\right)
\, =\, j^{-1} 
\left(
\begin{array}{c}
{D}_{\bar{x}} \\ {D}_{\bar{y}} \\ {D}_{\bar{z}}
\end{array}
\right)
\label{rotated:ap}
\ee
To obtain all possible eigenvalues of $\Delta$ we can use its commutation relations with the $\CD_m$ to show that $\CD_x$, $\CD_y$ and $\CD_{\bar{z}}$ act as creation operators whereas $\CD_{\bar{x}}$, $\CD_{\bar{y}}$ and $\CD_z$ are annihilation operators as implied by (\ref{annihil:ap}). The final result is that the eigenfunctions of $\Delta$ are all of the form
\be
\psi^{mnl}\, =\, (\CD_x)^m (\CD_y)^n (\CD_{\bar{z}})^l\, \psi_0
\label{KKtower:ap}
\ee

Note that in the above discussion we have suppressed the gauge indices of $\psi_{\rho}$, which in practice will be a matrix of the form (\ref{sectors}) with only one non-vanishing entry or sector $\a^\pm$. As the worldvolume flux  ${F}_{n\bar{m}} \sim [{F}_{n\bar{m}}, \cdot]$ acts in the adjoint for such gauge indices, each sector (i.e., each matter curves in the T-dual D7-brane picture) will feel a different Laplace operator $\Delta$, a different matrix {\bf F} and a different rotation matrix $J$. Rather than giving a description of these sectors for a general D9-brane model, we will illustrate them for the particular case of the $U(3)$ model introduced in subsection \ref{toy}.

\subsection{Massive modes for the $U(3)$ model}

\subsubsection*{Sector $a$}

Similarly to section \ref{general}, the matrix {\bf F} for the sectors $a^{\pm}$ reads 
\be
{\bf F}_{a^\pm} \, =\, \pm i
\left(
\begin{array}{cccc}
\oh({M}_x + {M}_y) & 0 & 0 & 0\\
0 & \oh({M}_x - {M}_y) & 0 & -iM_{\Phi_x}\\
0 & 0 & -\oh({M}_x - {M}_y) &  0 \\
0 & iM_{\Phi_x}& 0 & -\oh({M}_x + {M}_y)
\end{array}
\right)
\label{Fa}
\ee
Note that in order to reproduce eq.(\ref{Lapa}) we just need to impose $M_{\Phi_x} = m_\Phi^2$.
The rotation matrix that satisfies (\ref{defJ}) is given by
\be
J_{a}\, =\, 
\left(
\begin{array}{cc}
1 & 0 \\ 
0 & j_{a}
\end{array}
\right), \quad \quad {\rm where\ } \quad \quad
j_{a}\, =\,
\left(
\begin{array}{ccc}
{\rm cos\, } \phi_a & 0 & -i {\rm sin\, } \phi_a \\ 
 0 & 1 & 0\\
-i {\rm sin\, } \phi_a & 0 & {\rm cos\, } \phi_a
\end{array}
\right)
\label{Ja}
\ee
where we have defined
\be
{\rm cos\, }\phi_a \, =\, \frac{1}{\sqrt{(\tilde{\lam}_{a,-})^2+ 1}} \quad \quad  \quad {\rm sin\, }\phi_a \, =\, \frac{\tilde{\lam}_{a,-}}{\sqrt{(\tilde{\lam}_{a,-})^2+ 1}}
\label{anglea}
\ee
and
\be
\lam_{a, \pm}\, =\, \frac{{M}_{x}}{2} \pm \sqrt{\left(\frac{{M}_{x}}{2}\right)^2 + M_{\Phi_x}^{2}} \quad \quad \quad  \tilde{\lam}_{a,\pm} \, =\, \frac{\lam_{a,\pm}}{M_{\Phi_x}}
\label{lambda}
\ee
Indeed, it is easy to check that 
\be
\label{rotatedD:ap}
{\bf {\cal D}_A}\,=\, J_{a}^{\, t} \cdot {\bf D_A} \cdot J_{a}\, =\,
\left(
\begin{array}{cccc}
0 & \CD_x & \CD_y & \CD_z \\
-\CD_x & 0 & -\CD_{\bar{z}} & \CD_{\bar{y}} \\
-\CD_y & \CD_{\bar{z}} & 0 & -\CD_{\bar{x}} \\
-\CD_z & -\CD_{\bar{y}} & \CD_{\bar{x}} & 0
\end{array}
\right)
\ee
where, as in (\ref{rotated:ap}), we have 
\begin{subequations}
\label{rotatedsa:ap}
\begin{align}
& \left(
\begin{array}{c}
\CD_x \\ \CD_y \\ \CD_z
\end{array}
\right)
\, =\, j_a^{t}
\left(
\begin{array}{c}
{D}_x \\ {D}_y \\ {D}_z
\end{array}
\right)
\, =\,
\left(
\begin{array}{c}
c_a {D}_x -is_a {D}_z \\ {D}_y \\ c_a {D}_z -i s_a {D}_x
\end{array}
\right)
\, =\,
\left(
\begin{array}{c}
c_a \left(\p_x \mp   \bar{x} \lam_{a,+} \right) \\  \p_y \mp   \bar{y} M_y  \\ -is_a\left(\p_x \mp   \bar{x} \lam_{a,-} \right)
\end{array}
\right)
\\
& \left(
\begin{array}{c}
\CD_{\bar{x}} \\ \CD_{\bar{y}} \\ \CD_{\bar{z}}
\end{array}
\right)
\, =\, j_a^{-1} 
\left(
\begin{array}{c}
{D}_{\bar{x}} \\ {D}_{\bar{y}} \\ {D}_{\bar{z}}
\end{array}
\right)
\, =\,
\left(
\begin{array}{c}
c_a {D}_{\bar{x}} +is_a {D}_{\bar{z}} \\ {D}_{\bar{y}} \\ c_a {D}_{\bar{z}} +i s_a {D}_{\bar{x}}
\end{array}
\right)
\, =\,
\left(
\begin{array}{c}
{c_a}\left(\p_{\bar{x}} \mp   x \lam_{a,-} \right) \\  \p_{\bar{y}}  \\ {is_a}\left(\p_{\bar{x}} \mp   x \lam_{a,+} \right)
\end{array}
\right)
\end{align}
\end{subequations}
in the semi-holomorphic gauge (\ref{sholgD9}) and for the $a^\pm$ sectors respectively, and where we have taken the abbreviations $c_a = {\rm cos\, } \phi_a$, $s_a = {\rm sin\, } \phi_a$. One may also check that in this rotated basis the operator (\ref{Lapfer:ap}) becomes
\be
{\bf \CD_A}^\dag{\bf \CD_A}\,=\, J^{-1}_{a} {\bf D_A}^\dag {\bf D_A} J_{a} \, =\, 
-(\Delta_{a^\pm} \pm  {M}_{xy}) \Id_4\, \pm 
\left(
\begin{array}{cccc}
2{M}_{xy} & 0 & 0 & 0\\
0 & \lam_{a,+}  & 0 & 0 \\
0 & 0 & {M}_{y} &  0 \\
0 & 0 & 0 & \lam_{a,-} 
\end{array}
\right)
\label{Lapferrota:ap}
\ee
where ${M}_{xy} \equiv \oh({M}_x+{M}_y)$ and $\Delta$ may be expressed either by (\ref{Lapsym}) or by
\be
\Delta\, =\,  \{\CD_x, \CD_{\bar{x}}\} + \{\CD_y, \CD_{\bar{y}}\} + \{\CD_z, \CD_{\bar{z}}\}
\ee

From (\ref{Lapferrota:ap}) it is easy to see that the eigenvectors of ${\bf D_A}^\dag {\bf D_A}$ in the $a^\pm$ sector are indeed of the form (\ref{spec:ap}), with the replacement $J \raw J_a$. For ${M}_x<0<{M}_y$, the lowest eigenvalue in the $a^-$ sector will arise from the first eigenvector in (\ref{spec:ap2}) with lowest eigenvalue of $\Delta_{a^-}$, while in the $a^+$ sector it will arise from the second eigenvector in (\ref{spec:ap2}) with lowest eigenvalue of $\Delta_{a^+}$. Similarly to the discussion around eq.(\ref{zmreala}) one can see that only the sector $a^+$ contains zero modes, which in the semi-holomorphic gauge (\ref{sholgD9}) are given by 
\be
\Psi_{0, a^+}\, =\, 
J_a \left(
\begin{array}{c}
0\\ 0 \\ 0 \\ 1
\end{array}
\right) \psi_{0, a^+}, \quad \quad \psi_{0, a^+}\, =\, e^{ \lam_{a,-}   |x|^2}  f_{a}(y)
\label{zmhola:ap}
\ee
with $f_{a}$ a holomorphic function to be determined by boundary conditions. Since the wavefunction $\psi_{0, a^+}$ satisfies
\be
\CD_{\bar{x}} \psi_{0, a^+} \, =\, \CD_{\bar{y}} \psi_{0, a^+} \, =\, \CD_z \psi_{0, a^+} \, = \,0
\label{zmwfa}
\ee
for the operators $\CD_m$ of the $a^+$ sector, it is easy to check that (\ref{zmhola:ap}) satisfies the zero mode equation (\ref{Dirac9}). From (\ref{zmwfa}) and the commutation relations $[\CD_m, \CD_{\bar{m}}]$ one can also check that $\psi_{0, a^+}$ is an eigenfunction of the Laplacian such that
\be
\Delta_{a^+} \psi_{0, a^+}\, =\, - \frac{1}{2} \left( (\lam_{a,+} - \lam_{a,-}) + {M}_y  \right)\, \psi_{0, a^+}\, =\,\left( \lam_{a,-} - {M}_{xy}\right) \psi_{0, a^+}
\ee
as expected from the fact that (\ref{zmhola:ap}) is a zero mode of ${\bf D_A}^\dag {\bf D_A}$ as well. In fact, the commutation relations $[\CD_m, \CD_{\bar{m}}]$ allow to reproduce the whole spectrum of eigenvalues of $\Delta_{a^+}$. Indeed, let us first rewrite them as
\begin{subequations}
\label{comma+}
\begin{align}
& [\Delta_{a^+}, \CD_x]\, =\, -  \lam_{a,+} \CD_x &  & [\Delta_{a^+}, \CD_{\bar{x}}]\, =\,  \lam_{a,+} \CD_{\bar{x}} \\
&[\Delta_{a^+}, \CD_y]\, =\, -   {M}_y \CD_y &  & [\Delta_{a^+}, \CD_{\bar{y}}]\, =\,  {M}_y \CD_{\bar{y}}\\
&[\Delta_{a^+}, \CD_z]\, =\, -  \lam_{a,-}  \CD_z &  & [\Delta_{a^+}, \CD_{\bar{z}}]\, =\,  \lam_{a,-}  \CD_{\bar{z}}
\end{align}
\end{subequations}
from which it is easy to see that the remaining eigenfunctions are of the form 
\be
\psi_{mnl,a^+}\, =\, (\CD_x)^m (\CD_y)^n (\CD_{\bar{z}})^l\,  \psi_{0, a^+}
\label{apKKtower}
\ee
with eigenvalue $-(m\lam_{a,+} + n{M}_y - (l+1) \lam_{a,-} + {M}_{xy})$. The full set of eigenfunctions and eigenvalues of ${\bf D_A}^\dag {\bf D_A}$ for the $a^+$ sector is displayed in Table \ref{paired}.

\begin{table}[h!]
\begin{center}
\begin{tabular}{|c|c|}
\hline
 mass$^2$ & $a^+$  mode $H_{a^+}$ \\
 \hline \hline 
 $ m \lambda_{a, +} + n M_y + l| \lambda_{a, -}|$ & 
 $\left(
\begin{array}{c}
0 \\ -\displaystyle{\frac{i\lam_{a,-}}{M_{\Phi_x}}} \\ 0 \\ 1
\end{array}
\right)\psi_{mnl} $
  \\ 
  \hline 
$m \lambda_{a, +}  + (n + 1)  M_y +  (l+1) |\lambda_{a, -}|$ &
$\left(
\begin{array}{c}
0 \\ 0 \\ 1 \\ 0
\end{array}
\right) \psi_{mnl}$
\\     
\hline 
$ (m + 1)\lambda_{a, +} + n M_y +  (l+1) |\lambda_{a, -}|$ &
$\left(
\begin{array}{c}
0 \\ 1  \\ 0 \\  -\displaystyle{\frac{i\lam_{a,-}}{M_{\Phi_x}}}
\end{array}
\right)\psi_{mnl} $
\\
\hline 
$ m \lambda_{a, +} + n M_y +  (l+1) |\lambda_{a, -}|+2M_{xy}$ &
$\left(
\begin{array}{c}
1 \\ 0 \\ 0 \\ 0
\end{array}
\right) \psi_{mnl}$
\\
\hline
\end{tabular}
\end{center}
\caption{Massive modes in the $a^+$ sector. Here $\psi_{mnl} = \psi_{mnl, a^+}$  is constructed as in (\ref{apKKtower}).}
\label{paired}
\end{table}

\subsubsection*{Sector $b$}

For sectors $b^\pm$ we have that the matrix {\bf F} reads
\be
{\bf F}_{b^\pm} \, =\, \pm i
\left(
\begin{array}{cccc}
-{M}_{xy} & 0 & 0 & 0\\
0 & -{M}_x  + {M}_{xy} & 0 & 0 \\
0 & 0 & -{M}_y + {M}_{xy} & iM_{\Phi_y} \\
0 & 0 & -iM_{\Phi_y} & {M}_{xy}
\end{array}
\right)
\label{Fb}
\ee
and so
\be
J_{b}\, =\, 
\left(
\begin{array}{cc}
1 & 0 \\ 
0 & j_{b}
\end{array}
\right), \quad \quad {\rm where\ } \quad \quad
j_{b}\, =\,
\left(
\begin{array}{ccc}
1 & 0 & 0 \\ 
 0 & {\rm cos\, } \phi_b & i {\rm sin\, } \phi_b\\
0 & i {\rm sin\, } \phi_b & {\rm cos\, } \phi_b
\end{array}
\right)
\label{Jb}
\ee
where now the rotation angle is specified by
\be
c_b ={\rm cos\, }\phi_b \, =\, \frac{1}{\sqrt{(\tilde{\lam}_{b,-})^2+ 1}} \quad \quad  \quad s_b ={\rm sin\, }\phi_b \, =\, \frac{\tilde{\lam}_{b,-}}{\sqrt{(\tilde{\lam}_{b,-})^2+1}}
\label{angleb}
\ee
and
\be
\lam_{b, \pm}\, =\, - \frac{{M}_{y}}{2} \pm \sqrt{\left(\frac{{M}_{y}}{2}\right)^2 + M_{\Phi_y}^{2}} \quad \quad \quad \tilde{\lam}_{b, \pm}\, =\, { \lam_{b, \pm} \over M_{\Phi_y}}
\label{lambdb}
\ee
matching the rotation angle in the $a^{\pm}$ sectors for ${M}_x = - {M}_y$ and $M_{\Phi_y} = m_\Phi^2$. Again, we can define a rotated Dirac operator $\CD_A$ as in (\ref{rotatedD:ap}), with the replacement $J_a \raw J_b$. The entries of such operator are then given by 
\begin{subequations}
\label{rotatedsb:ap}
\begin{align}
& \left(
\begin{array}{c}
\CD_x \\ \CD_y \\ \CD_z
\end{array}
\right)
\, =\, j_b^t
\left(
\begin{array}{c}
{D}_x \\ {D}_y \\ {D}_z
\end{array}
\right)
\, =\,
\left(
\begin{array}{c}
 {D}_x  \\ c_b {D}_y+ i s_b {D}_z \\ c_b {D}_z + i s_b {D}_x
\end{array}
\right)
\, =\,
\left(
\begin{array}{c}
  \p_x \pm   \bar{x} M_x  \\ {c_b}\left(\p_y \mp   \bar{y} \lam_{b,+} \right) \\ is_b\left(\p_y \mp   \bar{y} \lam_{b,-} \right)
\end{array}
\right)
\\
& \left(
\begin{array}{c}
\CD_{\bar{x}} \\ \CD_{\bar{y}} \\ \CD_{\bar{z}}
\end{array}
\right)
\, =\, j_b^{-1} 
\left(
\begin{array}{c}
{D}_{\bar{x}} \\ {D}_{\bar{y}} \\ {D}_{\bar{z}}
\end{array}
\right)
\, =\,
\left(
\begin{array}{c}
{D}_{\bar{x}}  \\ c_b {D}_{\bar{y}} - i s_a  {D}_{\bar{z}} \\ c {D}_{\bar{z}} - i s {D}_{\bar{y}}
\end{array}
\right)
\, =\,
\left(
\begin{array}{c}
 \p_{\bar{x}}  \\  c_b\left(\p_{\bar{y}} \mp   y \lam_{b,-} \right) \\ is_b\left(\p_{\bar{y}} \mp   y \lam_{b,+} \right)
\end{array}
\right)
\end{align}
\end{subequations}
in the semi-holomorphic gauge (\ref{sholgD9}) and for the $b^\pm$ sectors respectively. In this rotated basis we have
\be
{\bf \CD_A}^\dag{\bf \CD_A}\,=\, J^{-1}_{b} {\bf D_A}^\dag {\bf D_A} J_{b} \, =\, 
-(\Delta_{b^\pm} \mp  {M}_{xy}) \Id_4\, \pm
\left(
\begin{array}{cccc}
-2{M}_{xy} & 0 & 0 & 0\\
0 & -{M}_x   & 0 & 0 \\
0 & 0 & \lam_{b,+} &  0 \\
0 & 0 & 0 & \lam_{b,-} 
\end{array}
\right)
\label{Lapferrotb:ap}
\ee
and so the eigenvalues of ${\bf D_A}^\dag {\bf D_A}$ in the $b^\pm$ sector are also of the form (\ref{spec:ap}), now with the replacement $J \raw J_b$. For ${M}_x<0<{M}_y$, the zero mode arises in the $b^+$ sector, and it is given by
\be
\Psi_{0, b^+}\, =\, 
J_b \left(
\begin{array}{c}
0\\ 0 \\ 0 \\ 1
\end{array}
\right) \psi_{0, b^+}, \quad \quad \psi_{0, b^+}\, =\, e^{   \lam_{b,-}   |y|^2}  f_{b}(x)
\label{zmholb:ap}
\ee
with $f_b$ holomorphic. Note that again, in the $b^+$ sector
\be
\CD_{\bar{x}} \psi_{0, b^+} \, =\, \CD_{\bar{y}} \psi_{0, b^+} \, =\, \CD_z \psi_{0, b^+} \, = \,0
\label{zmwfb}
\ee
and so the spectrum of eigenvectors can be obtained by using an algebra of creation an annihilation operators. Such algebra is codified in the commutation relations 
\begin{subequations}
\label{commb}
\begin{align}
& [\Delta_{b^\pm}, \CD_x]\, =\, \pm   {M}_x  \CD_x &  & [\Delta_{b^\pm}, \CD_{\bar{x}}]\, =\,  \mp   {M}_x \CD_{\bar{x}} \\
&[\Delta_{b^\pm}, \CD_y]\, =\, \mp   \lam_{b,+} \CD_y &  & [\Delta_{b^\pm}, \CD_{\bar{y}}]\, =\, \pm   \lam_{b,+} \CD_{\bar{y}}\\
&[\Delta_{b^\pm}, \CD_z]\, =\, \mp    \lam_{b,-}  \CD_z &  & [\Delta_{b^\pm}, \CD_{\bar{z}}]\, =\, \pm    \lam_{b,-}  \CD_{\bar{z}}
\end{align}
\end{subequations}
that show that the wavefunction 
\be
\psi_{mnl,b^+}\, =\, (\CD_x)^m (\CD_y)^n (\CD_{\bar{z}})^l\,  \psi_{0, b^+}
\ee
is an eigenfunction of $\Delta_{b^+}$ with eigenvalue $  (m{M}_x - n \lam_{b,+}  + (l+1) \lam_{b,-} + {M}_{xy})$. From this we can construct the massive modes in the $b^+$ sector, following the same steps as in the sector $a^+$.

\subsubsection*{Sector $c$}

For this sector we have
\be
{\bf F}_{c^\pm} \, =\, \pm i 
\left(
\begin{array}{cccc}
0 & 0 & 0 & 0\\
0 & 0 & 0 &  iM_{\Phi_x} \\
0 & 0 & 0 & -iM_{\Phi_y} \\
0 & -iM_{\Phi_x} & iM_{\Phi_y} & 0
\end{array}
\right)
\label{Fc}
\ee
and so
\be
J_{c}\, =\, 
\left(
\begin{array}{cc}
1 & 0 \\ 
0 & j_{c}
\end{array}
\right), \quad \quad {\rm where\ } \quad \quad
j_{c}\, =\,\frac{1}{\sqrt{2}}
\left(
\begin{array}{ccc}
c_c & - c_c & i s_c \sqrt{2} \\ 
- s_c & s_c & i c_c \sqrt{2}\\ 
i  & i  & 0
\end{array}
\right)
\label{Jc}
\ee
\be
c_c  \, =\, \frac{1}{\sqrt{(M_{\Phi_y}/M_{\Phi_x})^2+ 1}} \quad \quad  \quad s_c  \, =\, \frac{1}{\sqrt{(M_{\Phi_x}/M_{\Phi_y})^2+ 1}}
\label{anglec}
\ee
Note that for $M_{\Phi_x}=M_{\Phi_y}$ we have $c_c = s_c = 1/\sqrt{2}$, and so the rotation matrix (\ref{Jc}) becomes flux-independent.
Similarly to the previous sectors, we now have that
\begin{subequations}
\label{rotatedsc:ap}
\begin{align}
& \left(
\begin{array}{c}
\CD_x \\ \CD_y \\ \CD_z
\end{array}
\right)
\, =\, j_c^t
\left(
\begin{array}{c}
{D}_x \\ {D}_y \\ {D}_z
\end{array}
\right)
\, =\, \frac{1}{\sqrt{2}} 
\left(
\begin{array}{c}
c_c \p_{x} - s_c \p_{y} \pm \sqrt{2}   M_\Phi (c_c \bar{x} - s_c\bar{y})  \\ 
- c_c \p_{x} + s_c \p_{y} \pm \sqrt{2}   M_\Phi (c_c \bar{x} - s_c\bar{y})  \\ 
i \sqrt{2}  (s_c \p_{x} + c_c \p_{y})
\end{array}
\right)
\\
& \left(
\begin{array}{c}
\CD_{\bar{x}} \\ \CD_{\bar{y}} \\ \CD_{\bar{z}}
\end{array}
\right)
\, =\, j_c^{-1} 
\left(
\begin{array}{c}
{D}_{\bar{x}} \\ {D}_{\bar{y}} \\ {D}_{\bar{z}}
\end{array}
\right)
\, =\, 
\frac{1}{\sqrt{2}} 
\left(
\begin{array}{c}
 c_s \p_{\bar{x}} - s_c \p_{\bar{y}} \mp \sqrt{2}  M_\Phi   (c_c{x} - s_c{y})   \\ 
- c_c \p_{\bar{x}} + s_c \p_{\bar{y}} \mp\sqrt{2}  M_\Phi   (c_c{x} - s_c{y}) \\ 
-i \sqrt{2}  (s_c \p_{\bar{x}} + c_c \p_{\bar{y}})
\end{array}
\right)
\end{align}
\end{subequations}
for the $c^\pm$ sectors respectively, and where we have defined $M_\Phi = \sqrt{M_{\Phi_x}^2 + M_{\Phi_y}^2}/\sqrt{2}$. Finally,
\be
{\bf \CD_A}^\dag{\bf \CD_A}\,=\, J^{-1}_{c} {\bf D_A}^\dag {\bf D_A} J_{c} \, =\, 
-\Delta_{c^\pm}  \Id_4\, \pm \sqrt{2}   M_{\Phi} 
\left(
\begin{array}{cccc}
0 & 0 & 0 & 0\\
0 & -1   & 0 & 0 \\
0 & 0 & 1 &  0 \\
0 & 0 & 0 & 0 
\end{array}
\right)
\label{Lapferrotc:ap}
\ee

Unlike the previous sectors, the sector $c$ is non-chiral, meaning that one finds zero mode solutions both in the $c^+$ and the $c^-$ sectors.  Indeed, such zero modes are given by
\be
\Psi_{0, c^+}\, =\, 
J_c \left(
\begin{array}{c}
0\\ 1 \\ 0 \\ 0
\end{array}
\right) \psi_{0,c}, \quad \quad 
\Psi_{0, c^-}\, =\, 
J_c \left(
\begin{array}{c}
0\\ 0 \\ 1 \\ 0
\end{array}
\right) \psi_{0,c},  \quad \quad \psi_{0, c}\, =\, e^{-\sqrt{2}   M_\Phi|c_c x-s_c y|^2  }
\label{zmholc:ap}
\ee
with the scalar wavefunction $\psi_{0, c}$ satisfying 
\be
\CD_{x} \psi_{0, c}\, =\, \CD_{\bar{y}} \psi_{0, c}\, =\, \CD_{\bar{z}} \psi_{0, c}\, =\, 0
\ee
in the $c^+$ sector and 
\be
\CD_{y} \psi_{0, c}\, =\, \CD_{\bar{x}} \psi_{0, c}\, =\, \CD_{\bar{z}} \psi_{0, c}\, =\, 0
\ee
in the $c^-$ sector. The commutation relations of these sectors are given by
\begin{subequations}
\label{commc}
\begin{align}
& [\Delta_{c^\pm}, \CD_x]\, =\, \pm  \sqrt{2} M_{\Phi}  \CD_x &  & [\Delta_{c^\pm}, \CD_{\bar{x}}]\, =\,  \mp   \sqrt{2} M_{\Phi} \CD_{\bar{x}} \\
&[\Delta_{c^\pm}, \CD_y]\, =\, \mp   \sqrt{2} M_{\Phi} \CD_y &  & [\Delta_{c^\pm}, \CD_{\bar{y}}]\, =\, \pm   \sqrt{2} M_{\Phi} \CD_{\bar{y}}\\
&[\Delta_{c^\pm}, \CD_z]\, =\, 0  &  & [\Delta_{c^\pm}, \CD_{\bar{z}}]\, =\, 0
\end{align}
\end{subequations}
so that we can build the eigenfunctions of, e.g., $\Delta_{c^+}$ by applying the creator operators $\CD_{\bar{x}}$ and $\CD_y$,  and so construct the spectrum of massive modes in the $c^+$ sector up to the Kaluza-Klein excitations along the direction $x+y$.

\subsection{Fermionic conventions\label{fconv}}

In order to describe explicitly fermionic wavefunctions we take the following representation for $\G$-matrices in flat 10d space
\beq
\G^{\ul{\mu}}\, = \, \g^\mu \otimes \Id_2 \otimes \Id_2 \otimes \Id_2 \quad \quad  \quad \G^{\ul{m}}\, =\, \g_{(4)} \otimes \tilde{\g}^{m-3}
\label{ulG:ap}
\eeq
where $\mu = 0, \dots, 3$, labels the 4d Minkowski coordinates, whose gamma matrices are
\beq
\g^0\, =\,
\left(
\begin{array}{cc}
 0 & -\Id_2 \\ \Id_2 & 0
\end{array}
\right)
\quad
\g^i\, =\,
\left(
\begin{array}{cc}
 0 & \sig_i \\ \sig_i & 0
\end{array}
\right)
\eeq
$m = 4, \dots, 9$ labels the extra $\mathbb{R}^6$ coordinates
\beq
\begin{array}{lll}
\tilde{\g}^{1}\, = \, \sig_1 \otimes  \Id_2 \otimes \Id_2 & \quad \quad &  \tilde{\g}^{4}\, = \,  \sig_2 \otimes  \Id_2 \otimes \Id_2 \\
\tilde{\g}^{2}\, = \, \sig_3 \otimes  \sig_1 \otimes \Id_2 & \quad \quad &  \tilde{\g}^{5}\, = \,  \sig_3 \otimes  \sig_2 \otimes \Id_2 \\
\tilde{\g}^{3}\, = \,  \sig_3 \otimes \sig_3 \otimes \sig_1 & \quad \quad &  \tilde{\g}^{6}\, = \,  \sig_3 \otimes \sig_3 \otimes \sig_2
\end{array}
\label{tilgamma}
\eeq
and $\sig_i$ indicate the usual Pauli matrices. The 4d chirality operator is then given by
\beq
\G_{(4)}\, = \, \g_{(4)} \otimes  \Id_2 \otimes \Id_2 \otimes \Id_2
\label{4dgamma}
\eeq
where $\g_{(4)} = i \g^0\g^1\g^2\g^3$, and the 10d chirality operator by
\beq
\G_{(10)}\, = \, \g_{(4)} \otimes \g_{(6)}\, =\, \left(
\begin{array}{cc}
 \Id_2 & 0 \\ 0 & -\Id_2
\end{array}
\right) \otimes  \sig_3 \otimes \sig_3 \otimes \sig_3
\eeq
with $\g_{(6)} = -i \tilde{\g}^1\tilde{\g}^2\tilde{\g}^3\tilde{\g}^4\tilde{\g}^5\tilde{\g}^6$. Finally, in this choice of representation a Majorana matrix is given by
\beq
\label{ap:Maj}
\mathcal{B}\, =\, \G^{\ul{2}}\G^{\ul{7}}\G^{\ul{8}}\G^{\ul{9}}\, =\,
\left(
\begin{array}{cc}
 0 & \sig_2 \\ -\sig_2 & 0
\end{array}
\right)
\otimes \sig_2 \otimes i\sig_1 \otimes \sig_2 \,= \, \mathcal{B}_4 \otimes \mathcal{B}_6
\eeq
which indeed satisfies the conditions $\mathcal{B}\mathcal{B}^* = \Id$ and $\mathcal{B}\, \G^{\ul{M}} \mathcal{B}^* = \G^{\ul{M}*}$. Notice that the 4d and 6d Majorana matrices $\mathcal{B}_4 \equiv \g^2 \g_{(4)}$ and $\mathcal{B}_6 \equiv \tilde{\g}^4 \tilde{\g}^5 \tilde{\g}^6$ satisfy analogous conditions $\mathcal{B}_4\mathcal{B}_4^* = \mathcal{B}_6\mathcal{B}_6^* = \Id$ and $\mathcal{B}_4\, \g^{\mu} \mathcal{B}_4^* = \g^{\mu*}$, $\mathcal{B}_6\, \g^{m} \mathcal{B}_6^* = -  \g^{m*}$.

In the text we mainly work with 10d Majorana-Weyl spinors of negative chirality, meaning those spinors $\theta$ satisfying $\theta = - \G_{(10)} \theta = \mathcal{B}^*\theta^*$. In the conventions above this means that we have spinors of the form
\begin{subequations}
\label{basisMW}
\begin{align}
\theta^0\, =\,
\psi^0 \,
\left(
\begin{array}{c}
\xi_+ \\ 0
\end{array}
\right) \otimes \chi_{---}
+ i (\psi^0)^*\,
\left(
\begin{array}{c}
0 \\ \sig_2\xi_+^*
\end{array}
\right) \otimes
\chi_{+++}\\
\theta^1\, =\,
\psi^1 \,
\left(
\begin{array}{c}
\xi_+ \\ 0
\end{array}
\right) \otimes \chi_{-++}
- i (\psi^1)^*\,
\left(
\begin{array}{c}
0\\ \sig_2\xi_+^*
\end{array}
\right) \otimes
\chi_{+--}\\
\theta^2\, =\,
\psi^2 \,
\left(
\begin{array}{c}
\xi_+ \\ 0
\end{array}
\right) \otimes \chi_{+-+}
+ i (\psi^2)^*\,
\left(
\begin{array}{c}
0 \\ \sig_2\xi_+^*
\end{array}
\right) \otimes
\chi_{-+-}\\
\theta^3\, =\,
\psi^3 \,
\left(
\begin{array}{c}
\xi_+ \\ 0
\end{array}
\right) \otimes \chi_{++-}
- i (\psi^3)^*\,
\left(
\begin{array}{c}
0 \\ \sig_2\xi_+^*
\end{array}
\right) \otimes
\chi_{--+}
\end{align}
\end{subequations}
where $\psi^j$ is the spinor wavefunction, $(\xi_+ \ 0)^t$ is a 4d spinor of positive
chirality and $\chi_{\epsilon_1\epsilon_2\epsilon_3}$ is a basis of 6d spinors of such that
\beq
\chi_{---}\, =\,
\left(
\begin{array}{c}
0\\1
\end{array}
\right) \otimes
\left(
\begin{array}{c}
0\\1
\end{array}
\right) \otimes
\left(
\begin{array}{c}
0\\1
\end{array}
\right)
\quad \quad
\chi_{+++}\, =\,
\left(
\begin{array}{c}
1\\0
\end{array}
\right) \otimes
\left(
\begin{array}{c}
1\\0
\end{array}
\right) \otimes
\left(
\begin{array}{c}
1\\0
\end{array}
\right)
\label{spinorbasis:ap}
\eeq
etc. Note that these basis elements are eigenstates of the 6d chirality operator $\g_{(6)}$, with eigenvalues $\epsilon_1\epsilon_2\epsilon_3$. In (\ref{basisMW}), we need to replace $(\psi^m)^* \raw (\psi^m)^\dag$ if the spinor $\theta$ transforms in the adjoint of a gauge group.

Finally, let us recall that to dimensionally reduce a 10d fermionic action, one has to simultaneously diagonalize two Dirac operators: $\slashed{\p}_{\IR^{1,3}}$ and $\slashed{D}^{\text{int}}$, built from $\G^{\ul{\mu}}$ and $\G^{\ul{m}}$, respectively. However, as these two set of $\G$-matrices do not commute, nor will $\slashed{\p}_{\IR^{1,3}}$ and $\slashed{D}^{\text{int}}$, and so we need instead to construct these Dirac operators from the alternative $\G$-matrices
\beq
\tilde{\G}^{\ul{\mu}}\, = \, \G_{(4)} \G^{\ul{\mu}}\, = \, \G_{(4)}\g^\mu \otimes \Id_2 \otimes \Id_2 \otimes \Id_2 \quad \quad  \quad  \tilde{\G}^{\ul{m}}\, =\, \G_{(4)} \G^{\ul{m}}\, =\, \Id_4 \otimes \tilde{\g}^{m-3}
\label{commG}
\eeq
following the common practice in the literature.

\section{Computing wavefunctions in the commutative formalism\label{betawf}}

In this appendix we address the computation of the corrected zero mode wavefunctions in the commutative 8d formalism described in susbsection \ref{8dapp}. In particular, we will look for a corrected wavefunction of the form (\ref{expwf}), with $\{\Psi_\lam\}$ the tower of unperturbed massive modes at the matter curve (i.e., those wavefunctions satisfying eq.(\ref{eigenferm})).

\subsection{$\b$-deformed wavefunctions and massive modes}

As already discussed in section \ref{corrwf}, the zero mode wavefunctions in the presence of a non-perturbative effect or a $\beta$-deformation can be expressed in terms of the expansion
\be
\Psi\, =\, \Psi^{(0)} + \eps \Psi^{(1)} + \eps^2 \Psi^{(2)} + \dots  
\label{ap:expwf2}
\ee
where $\Psi^{(0)} = \Psi_0$ is the wavefunction in the absence of the non-perturbative effect/$\beta$-deformation, satisfying the classical equation (\ref{Dirac9b}). The equation for the first order correction $\Psi^{(1)}$ is given by
\be
{\bf D_A} \Psi^{(1)}\, =\, {\bf K}^{(1)} \Psi_0
\label{ap:premaster}
\ee
where ${\bf K}^{(1)}$ is the following linear operator
\bea
\label{ap:K1K2}
{\bf K}^{(1)} & = & -
\left(
\begin{array}{cc}
0 & 0 \\
0 & \Theta ({\bf \tilde{K}} + i {\bf \tilde{A}})^{(0)}  \Theta
\end{array}
\right) \\ \nonumber
& &  -({\bf \tilde{K}} + i{\bf \tilde{A}}) \, = \,
\left(
\begin{array}{ccc}
0 & \CK_{\bar{z}} + i \CA_{\bar{z}} & -(\CK_{\bar{y}} + \frac{i}{2}\CA_{\bar{y}}) \\
-(\CK_{\bar{z}} + i \CA_{\bar{z}}) & 0 & \CK_{\bar{x}} + \frac{i}{2}\CA_{\bar{x}} \\
\CK_{\bar{y}} + \frac{i}{2}\CA_{\bar{y}} & -(\CK_{\bar{x}} +\frac{i}{2}\CA_{\bar{x}}) &0
\end{array}
\right)
\eea
\bea
 \nonumber
& & \CK_{\bar{m}} \, = \, \{F_{y\bar{m}}^{\, \th}, D_x\cdot\} -  \{F_{x\bar{m}}^{\, \th}, D_y \cdot\} \quad \quad \CA_{\bar{m}}\, =\, [\langle F_{y\bar{m}}^{\, \th} A_x -F_{x\bar{m}}^{\, \th} A_y \rangle,\cdot]
\\ \nonumber
& & F_{l\bar{n}}^{\, \th} = \langle F_{l\bar{n}} \rangle \quad \quad 
F_{l\bar{z}}^{\, \th} \, =\, \p_x(\theta \langle \Phi_{xy} \rangle) \quad \quad 
l,n \, =\, x, y
\eea

From the results of subsection \ref{4dapp}, one expects that the solution to (\ref{ap:premaster}) is given by a linear combination of massive modes of the unperturbed system. That is, $\Psi^{(1)}$ should be of the form
\be
\Psi^{(1)}\, =\, \sum_\lam c_\lam \Psi_\lam
\label{ap:massivexp}
\ee
with $c_\lam \in \IC$ and each $\Psi_\lam$ satisfying
\be
{\bf D_A}^\dag {\bf D_A}\, \Psi_\lam\, =\, |m_\lam|^2 \Psi_\lam
\label{eigenfermlam}
\ee
Hence, using the eigenmode decomposition (\ref{ap:massivexp}) we obtain that eq.(\ref{ap:premaster}) can be expressed as
\be
 {\bf D_A}^\dag {\bf K}^{(1)} \Psi_0\, =\, \sum_{\lam} c_{\lam} |m_{\lam}|^2  \Psi_{\lam} 
\label{ap:master}
\ee
for some set of coefficients $c_\lam$ to be found for each different $\Psi_0$. The corrected wavefunction will then be given by
\be
\Psi\, =\, \Psi_0 + \eps \sum_{\lam} c_{\lam} \Psi_{\lam} + \CO(\e^2)
\label{expwf3}
\ee
In the following, we will illustrate the above method by computing the corrected wavefunctions for the $U(3)$ toy model of section \ref{toy}, for the particular case of constant $\theta$.

\subsection{The $U(3)$ model\label{u3corrwf}}

Let us again consider the $U(3)$ model of subsection \ref{toy}, together with its full spectrum of massive modes worked out in appendix \ref{apD9KK}. In particular, recall from that appendix that the classical massive modes $\Psi_\lam$ of a model with constant fluxes are given by (\ref{spec:ap}), for some specific rotation matrix $J$ that differs for each sector, and that the scalar wavefunctions $\psi_\rho$ are related to the zero mode wavefunction $\psi_0$ by means of the creation operators $\CD_m$ defined in (\ref{rotated:ap}). Hence, one may solve eq.(\ref{ap:master}) by expressing ${\bf D_A}^\dag {\bf K}^{(1)}$ in terms of such creator operators. We will show how to do so for each of the chiral sectors of the $U(3)$ model, restricting for simplicity to the case where $\theta$ is a constant.

\subsubsection*{sector $a^+$}

For this sector and for constant $\theta$ it is easy to see that
\be
\Theta {\bf \tilde{K}} \Theta\, =\, \frac{i\theta}{6} 
\left(
\begin{array}{ccc}
 0 & im_\Phi^2( D_y + 2 D_x) &  - M_y D_x \\
-  im_\Phi^2( D_y + 2 D_x)& 0 & - M_x D_y \\
 M_y D_x & M_x D_y &0
\end{array}
\right)
\label{K1a}
\ee
\be
i \Theta {\bf \tilde{A}} \Theta\, =\, \frac{i\theta}{6}
\left(
\begin{array}{ccc}
0 & - 2m^2_{\Phi}\bar{A_x} &  -iM_y \bar{A_x} \\
2m^2_{\Phi}\bar{A_x} & 0 & -iM_x \bar{A_y} \\
i M_y \bar{A_x} & iM_x \bar{A_y} &0
\end{array}
\right)
\label{K2a}
\ee
where $\bar{A}_x$ and $\bar{A}_y$ are defined as in (\ref{holg}). Since for the sector $a^+$ we have that  $D_x = \p_x - i\bar{A_x}$ and $D_y = \p_y - i\bar{A_y}$, we obtain
\be
\label{K12a}
{\bf K}^{(1)}_{a^+} \,  =\, 
\frac{i\theta}{6} \left(
\begin{array}{cccc}
0 & 0 & 0 & 0\\
0 &  0 & -im_\Phi^2( D_y + 2 \p_x) &  M_y \p_x \\
0 & im_\Phi^2( D_y + 2 \p_x)& 0 &  M_x \p_y \\
0 & -M_y \p_x & -M_x \p_y &0
\end{array}
\right)
\ee

Recall that the classical zero modes for the sector $a^+$ in the holomorphic gauge can be expressed either as (\ref{zmhola}) or as (\ref{zmhola:ap}). One then deduces that
\be
{\bf D_A}^\dag{\bf K}^{(1)}_{a^+} \Psi_{0, a^+}\, =\, \CP_a  \Psi_{0, a^+}\, =\, 
\left(
\begin{array}{c}
0\\ -i\frac{\lam_a^-}{m_\Phi^2} \\ 0 \\ 1
\end{array}
\right) 
\CP_a
\psi_{0, a^+}
\label{mastera}
\ee
with the differential operator $\CP_a$ given by
\be
\CP_a\, =\, 
\frac{i\theta}{6}
\left[ (2s_a^2\lam_a^-) \CD_x^2 + \left(c_a (\lam_a^- + 2 M_{xy})- M_y/c_a\right)\CD_x\CD_y +ic_aM_x \bar{A_y} \CD_x\right]
\label{Pa}
\ee
and where, following the definitions of appendix \ref{apD9KK}, $\lam_a^{\pm}$ is defined as in (\ref{eigenlam}) and
\bea
\CD_x\, =\, c_a \left( \p_x - \lam_a^+\bar{x}\right) \quad & & \quad  \CD_y\, =\, \p_y - M_y \bar{y} \\
c_a \, =\, \frac{m_\Phi^2}{\sqrt{({\lam}_a^{-})^2+ m_\Phi^4}} \quad & &   \quad s_a \, =\, \frac{{\lam}_a^{-}}{\sqrt{({\lam}_a^{-})^2+ m_\Phi^4}}
\eea
Notice that in (\ref{mastera}) the terms proportional to $\psi_{200, a^+} \equiv \CD_x^2 \psi_{0, a^+}$ and $\psi_{110, a^+} \equiv \CD_x \CD_y \psi_{0, a^+}$ are already of the form (\ref{ap:master}), for these are eigenfunctions of the Laplace operator $\Delta$. The same is not true for the term proportional to $\bar{A_y}\CD_x$. Indeed, this term is difficult to express purely in terms of creation operators $\CD_m$, since $\bar{A_y}$ depends on the coordinate $\bar{y}$ along the matter curve. In order to circumvent this difficulty one may choose a gauge for $\langle A \rangle$ slightly different from the one implicit in (\ref{order1}). Such gauge $\langle A' \rangle$ is defined by
\be
\langle A' \rangle\, =\, \langle A \rangle + {\rm d} \Omega\quad \raw \quad \Psi' = e^{i\Omega} \Psi \quad \quad \quad \Omega\, =\, \frac{\eps \theta}{2} \langle A_{x} A_{y} \rangle^{(0)}
\label{minigauge}
\ee
In this new gauge we have that (\ref{K2a}) is replaced by
\be
i \Theta {\bf \tilde{A}}' \Theta\, =\, \frac{i\theta}{6}
\left(
\begin{array}{ccc}
0 & -2m^2_{\Phi}\bar{A_x} &  -2iM_y \bar{A_x} \\
2m^2_{\Phi}\bar{A_x} & 0 & 0 \\
2 i M_y \bar{A_x} & 0 &0
\end{array}
\right)
\label{K2apr}
\ee
and so instead of (\ref{mastera}) we have
\be
{\bf D_A}^\dag ({\bf K}^{(1)}_{a^+})' \Psi_{0, a^+}'\, =\, \CP_a'  \Psi_{0, a^+}'
\label{masterapr}
\ee
where
\be
\CP_a' = 
\frac{i\theta}{6} s_am_\Phi^2 
\left[ (2s_a^2/c_a) \CD_x^2 + \left(1 + \frac{2M_{xy}c_a}{m_\Phi^2s_a}- \frac{2M_ys_a}{m_\Phi^2c_a}  \right)\CD_x\CD_y\right]
\ee
Hence, in this gauge we have that
\be
{\bf D_A}^\dag ({\bf K}^{(1)}_{a^+})' \Psi_{0, a^+}'\, =\, \frac{i\theta}{6} m_\Phi^2 s_a \left[ \frac{2s_a^2}{c_a} \Psi_{200,a^+} +\left(1 + \frac{2M_{xy}c_a}{m_\Phi^2s_a}- \frac{2M_ys_a}{m_\Phi^2c_a}  \right) \Psi_{110,a^+}\right]
\ee
where $\Psi_{200,a^+} = \CD_x^2 \Psi_{0,a^+}$ and $\Psi_{110,a^+}= \CD_x\CD_y \Psi_{0,a^+}$ are massive replicas of the chiral mode $\Psi_{0,a^+}$, of masses $|m_{200}|^2 = 2 \lam_a^+ = -2 m_\Phi^2 c_a/s_a$ and $|m_{110}|^2 = (M_y + \lam_a^+) = (2 M_{xy} - m_\Phi^2 s_a/c_a)$, respectively (see table \ref{paired} in appendix \ref{apD9KK}). By plugging the above expression into (\ref{expwf3}) we obtain that the coefficients $c_\lam$ are
\be
c_{200}\, =\, - \frac{i\theta}{6} \frac{s_a^4}{c_a^2} \quad \quad {\rm and} \quad \quad c_{110}\, =\, - \frac{i\theta}{6} \left( \frac{2Ms_a}{m_\Phi^2} + c_a\right)
\ee
where for simplicity we have set $M \equiv M_x = -M_y$. We finally conclude that the corrected wavefunction for the $a^+$ sector is given by
\bea\nonumber
\Psi_{a^+} & = & 
 \Psi_{0, a^+} - \frac{i\eps\theta}{6} \left[\frac{s_a^4}{c_a^2} \Psi_{200,a^+} + \left( 2Ms_a/m_\Phi^2 + c_a\right) \Psi_{110,a^+} - \bar{A_x}\bar{A_y} \Psi_{0,a^+} \right] + \CO(\eps^2)
 \\ 
& = & \left(
\begin{array}{c}
0\\ -i\frac{\lam_a^-}{m_\Phi^2} \\ 0 \\ 1
\end{array}
\right) 
 \left[ 1 -\frac{i\eps\theta}{6} \left( \frac{s_a^4}{c_a^2} \CD_x^2 +  \left( 2Ms_a/m_\Phi^2 + c_a\right)\CD_x\CD_y  - \bar{A_x}\bar{A_y} \right)\right] \psi_{0,a^+}
\label{finalwfa}
\eea
where we have performed the gauge transformation
\be
\Psi_{a^+}\, =\, 
e^{-i\Omega}\Psi_{a^+}'
\, =\, e^{\frac{i\eps\theta}{6} \bar{A_x}\bar{A_y}} \Psi_{a^+}'
\ee
and expanded it to first order in $\e$.

\subsubsection*{sector $b^+$}

For this sector we have
\be
\label{K12b}
{\bf K}^{(1)}_{b^+}\,  =\, 
\frac{i\theta}{6} \left(
\begin{array}{cccc}
0 & 0 & 0 & 0\\
0 &  0 & im_\Phi^2( D_x + 2 \p_y) &  M_y \p_x \\
0 & -im_\Phi^2( D_x + 2 \p_y)& 0 &  M _x\p_y \\
0 & -M_y \p_x & -M_x \p_y &0
\end{array}
\right)
\ee
where for this sector we have $D_x = \p_x + i\bar{A_x}$ and $D_y = \p_y + i\bar{A_y}$. Just like in the sector $a^+$ we have that eq.(\ref{ap:master}) has the form
\be
{\bf D_A}^\dag {\bf K}^{(1)}_{b^+} \Psi_{0, b^+}\, =\, \CP_b  \Psi_{0, b^+}
\label{masterb}
\ee
with
\be
\CP_b\, =\, 
\frac{i\theta}{6}
\left[ (2s_b^2\lam_b^-) \CD_y^2 + \left(c_b (\lam_b^- - 2 M_{xy}) + M_x/c_b\right)\CD_x\CD_y -ic_bM_y \bar{A_x} \CD_y\right]
\label{Pb}
\ee
\be
\CD_x\, =\,  \p_x + M_x \bar{x}   \quad \quad \quad  \CD_y\, =\, c_b \left( \p_y - \lam_b^+\bar{y}\right)
\ee
where $\lam_b^-$ is given by (\ref{simlams}) and $s_b$, $c_b$ are defined as $s_a$, $c_a$ with the replacement $\lam_a^- \raw \lam_b^-$. Note that the operators (\ref{Pa}) and (\ref{Pb}) are basically related by the exchange $x \leftrightarrow y$, and so one may already guess that the same will be true for the corrected wavefunctions.

Just like in the sector $a^+$, expressing the term $\bar{A_x} \CD_y$ in term of creation operators is difficult, so one may take a different gauge for $\langle A \rangle$. The appropriate choice is again (\ref{minigauge}) but with the replacement $\Omega \raw -\Omega$. We then find
\be
{\bf D_A}^\dag ({\bf K}^{(1)}_{b^+})' \Psi_{0, b^+}' = \frac{i\theta}{6} m_\Phi^2 s_b \left[ \frac{2s_b^2}{c_b} \Psi_{020,b^+} + \left(1 - \frac{2M_{xy}c_b}{m_\Phi^2s_b} + \frac{2M_xs_b}{m_\Phi^2c_b}  \right) \Psi_{110,b^+}\right]
\ee
where $\Psi_{020,b^+} = \CD_y^2 \Psi_{0,b^+}$ and $\Psi_{110,b^+}= \CD_x\CD_y \Psi_{0,b^+}$ correspond to massive replicas with masses $|m_{020}|^2 = 2 \lam_b^+ = -2 m_\Phi^2 c_b/s_b$ and $|m_{110}|^2 = \lam_b^+-M_x = -(2M_{xy} + m_\Phi^2 s_b/c_b)$, respectively. The coefficients $c_\lam$ in (\ref{ap:master}) now read
\be
c_{020}\, =\, - \frac{i\theta}{6} \frac{s^4_b}{c^2_b} \quad \quad {\rm and} \quad \quad c_{110}\, =\, - \frac{i\theta}{6} \left( \frac{2Ms_b}{m_\Phi^2} + c_b\right)
\ee
where again we have set $M \equiv M_x = -M_y$. Hence
\bea\nonumber
\Psi_{b^+} & = &  
 \Psi_{0, b^+} - \frac{i\eps\theta}{6} \left[\frac{s_b^4}{c_b^2} \Psi_{020,b^+} + \left( 2Ms_b/m_\Phi^2 + c_b\right) \Psi_{110,a^+} - \bar{A_x}\bar{A_y} \Psi_{0,b^+} \right] + \CO(\eps^2)
 \\ 
& = & \left(
\begin{array}{c}
0\\ 0 \\ i\frac{\lam_-}{m_\Phi^2} \\ 1
\end{array}
\right) 
 \left[ 1 -\frac{i\eps\theta}{6} \left( \frac{s^4_b}{c^2_b} \CD_y^2 +  \left( 2Ms_b/m_\Phi^2 + c_b\right)\CD_x\CD_y  - \bar{A_x}\bar{A_y} \right)\right] \psi_{0,b^+}
\label{finalwfb}
\eea
where now
\be
\Psi_{b^+}\, =\, 
e^{i\Omega}\Psi_{b^+}'
\, =\, e^{\frac{i\eps\theta}{6} \bar{A_x}\bar{A_y}} \Psi_{b^+}'
\ee

\subsubsection*{sector $c^+$}

For this sector we have
\be
\Theta {\bf \tilde{K}} \Theta\, =\, \frac{i\theta}{6} 
\left(
\begin{array}{ccc}
 0 & -im_\Phi^2( \p_x - \p_y) &  2M_y \p_x \\
 im_\Phi^2( \p_x - \p_y)& 0 &  2M_x \p_y \\
 -2M_y \p_x & -2M_x \p_y &0
\end{array}
\right)
\label{K1c}
\ee
\be
\Theta {\bf \tilde{A}} \Theta\, =\, \frac{i\theta}{6}
\left(
\begin{array}{ccc}
0 & 2m^2_{\Phi}(\bar{A_x} +\bar{A_y}) & 0 \\
-2m^2_{\Phi}(\bar{A_x} +\bar{A_y}) & 0 & 0 \\
0 & 0 &0
\end{array}
\right)
\label{K2c}
\ee
from which we deduce that
\be
{\bf D_A}^\dag {\bf K}^{(1)}_{c^+} \Psi_{0, c^+}\, =\, \CP_c  \Psi_{0, c^+}
\label{masterc}
\ee
with
\be
\CP_c\, =\, 
\frac{i\theta}{6}
\left[ \left( M_x - M_y -\frac{m_\Phi^2}{\sqrt{2}}\right) \CD_y^2 + m_\Phi^2 M_{xy} (\bar{x}+\bar{y}) \CD_y\right]
\label{Pc}
\ee
\be
\CD_y\, =\, \oh \left( -\p_x + \p_y + \sqrt{2}  m_\Phi^2 (\bar{x} - \bar{y})\right)
\ee

Note that we also find a non-trivial dependence of $\CP_c$ along the coordinate $\bar{x}+\bar{y}$ along the matter curve $\Sigma_c$. This time, however, such term vanishes if we set $M \equiv M_x = -M_y$, and then the correction only involves the massive mode $\Psi_{020,c^+} = \CD_y^2 \Psi_{0,c^+}$, of mass $|m_{020}| = 2\sqrt{2} m_\Phi^2$. In particular, from (\ref{Pc}) we can deduce the coefficient $c_{020}$ and so
\bea\nonumber
\Psi_{c^+} & = &  
 \Psi_{0, c^+} - \frac{i\eps\theta}{6} \left[\frac{1}{4} - \frac{M}{\sqrt{2}m_\Phi^2} \right] \Psi_{020,c^+} + \CO(\eps^2)
 \\ 
& = & \left(
\begin{array}{c}
0\\ -\frac{i}{\sqrt{2}} \\ \frac{i}{\sqrt{2}} \\ 1
\end{array}
\right) 
 \left[ 1 -\frac{i\eps\theta}{6} \left( \frac{1}{4} - \frac{M}{\sqrt{2}m_\Phi^2}\right)  \CD_y^2\right] \psi_{0,c^+} + \CO(\e^2)
\label{finalwfc}
\eea

\section{Equivalence of commutative and non-commutative actions\label{apncmap}}

As shown in the main text, computing non-perturbative effects on 7-brane wavefunctions is equivalent to compute wavefunctions for either 7-branes in a $\b$-deformed background, or for a non-commutative version of the 7-brane action, at least up to $\CO(\e^2)$ corrections. That the last two approaches are equivalent could have been guessed from the results of \cite{kapustin03,pestun06}, where it was argued that on a D-brane world-volume the effect of a $\b$-deformation can be seen as a non-commutative deformation of the gauge theory. This point was made more precise in \cite{mm09} where it was shown, via a Seiberg-Witten map, the equivalence between a $\beta$-deformed, commutative 7-brane action and a non-commutative 7-brane action. 

The purpose of this appendix is to reproduce in detail the derivation in \cite{mm09} of the equivalence of such commutative and non-commutative 7-brane actions, by means of a slightly simpler Seiberg-Witten map that will also be used in section \ref{ncmap}.

Let us consider a non-commutative 7-brane action with non-commutative parameter $\Theta^{xy} = - \Theta^{yx} = - \eps\, \theta$, with $\eps$ constant and $\theta$ holomorphic on $(x,y)$, and all the remaining components of $\Theta^{ij}$ vanishing. Such action is given by 
\be
\hat W\, =\, m_*^4 \int_S \tr(\hat\Phi\circledast \hat F )\, =\, m_*^4 \int_S \tr(\hat\Phi\wedge\hat F)+\CO(\e^2)
\label{ncsup}
\ee
Let us now define the Seiberg-Witten map 
\begin{subequations}
\label{apSWmap}
\begin{align}
\label{apSWmapA}
\hat A_{\bar{m}}&=\, A_{\bar{m}} - \frac12\Theta^{ij}\{A_i, (\p_j+D_j) A_{\bar{m}} \}+\CO(\eps^2)\\
\label{apSWmapphi}
\hat\Phi &=\, \Phi - \frac12 \{A_i, (\p_j+D_j)(\Theta^{ij}\Phi)\} +\CO(\eps^2)
\end{align}
\end{subequations}
Since $\Theta^{ij}$ is holomorphic it commutes with $\p_{\bar{n}}$ and so, just like in \cite{sw99}, we have that
\be
\label{apSWmapF}
 \hat F_{\bar{x}\bar{y}}\, =\, F_{\bar{x}\bar{y}} - \Theta^{ij} \left[\{F_{\bar{x}i} , F_{j\bar{y}}\} +\frac12 \{A_i,(\p_j+D_j) F_{\bar{x}\bar{y}} \right] +\CO(\e^2)
 \ee
which together with (\ref{apSWmapphi}) reproduces eq.(15) of \cite{mm09} for the case $\theta =$ const. Plugging these two expressions into (\ref{ncsup}) we obtain 
\bea
m_*^{-4}\, \hat W  & =  &\int_S \tr(\Phi \wedge F) 
- \int_S \Theta^{ij}\tr \left(\Phi_{xy}  \{F_{\bar{x}i}, F_{j\bar{y}}\}\right)  {\rm d }{\rm vol}_S
\\ \nonumber
& & - \oh \int_S \Theta^{ij}\tr \left(\Phi_{xy} \{ A_i, (D_j + \p_j) F_{\bar{x}\bar{y}}\}\right) {\rm d }{\rm vol}_S 
 \\ \nonumber
 & &  - \oh \int_S \tr \left( \{A_i,  (D_j+ \p_j)(\Theta^{ij} \Phi_{xy})\} F_{\bar{x}\bar{y}} \right)  {\rm d }{\rm vol}_S+\CO(\e^2)
\eea
We now have that
\bea
& & \oh \tr \left(\Theta^{ij}\Phi_{xy} \{ A_i, (D_j + \p_j) F_{\bar{x}\bar{y}}\} + \{A_i,  (D_j+ \p_j)(\Theta^{ij} \Phi_{xy})\} F_{\bar{x}\bar{y}} \right) \cr 
& & = \Theta^{ij} \tr\, (\Phi_{xy} \{F_{ij}, F_{\bar{x}\bar{y}}\}) +  2 \p_j \left[ \Theta^{ij} \tr\, (A_i \{\Phi_{xy}, F_{\bar{x}\bar{y}}\})\right]
\eea
and so
\be
\hat W\,  =\, m_*^4\left[ \int_S \tr\, (\Phi \wedge F) + \frac{\e}{2}  \int_S \theta\, \tr\, ( \Phi_{xy} F \wedge F) + \int_S \p_i\tr \left[\Theta^{ij} \left(A_j \Phi \wedge F \right)\right] \right]+\CO(\e^2)
\ee
Since $S$ has no boundary and the last term is a total derivative it must vanish, and so we recover the $\beta$-deformed superpotential (\ref{supototal}).

\end{document}